\documentclass[JHEP,12pt]{article}
\usepackage{subcaption}
\usepackage{amssymb,amsmath}
\usepackage{xcolor}
\usepackage{bm}
\usepackage{bbm}
\usepackage{braket}
\usepackage{jheppub}
\usepackage{endnotes}
\usepackage{comment} 
\usepackage[most]{tcolorbox}
\definecolor{block-gray}{gray}{0.85}
\newtcolorbox{oldversion}{colback=block-gray,grow to right by=0mm,grow to left by=0mm,boxrule=0pt,boxsep=0pt,breakable}

\def\be{\begin{equation}}
\def\ee{\end{equation}}

\renewcommand{\[}{\begin{equation}\begin{aligned}}
\renewcommand{\]}{\end{aligned}\end{equation}}

\newcommand{\Rop}{\mathbbm{R}}

\newcommand{\Pop}{\mathbbm{P}}
\newcommand{\Kop}{\mathbb{K}}

\usepackage[normalem]{ulem}

\author[1,2]{Alessandra Buonanno,}
\author[1,2]{Mohammed Khalil,}
\author[3]{Donal O'Connell,}
\author[4]{Radu Roiban,}
\author[5]{Mikhail P. Solon,}
\author[3]{Mao Zeng}

\affiliation[1]{Max Planck Institute for Gravitational Physics (Albert Einstein Institute), Am M\"uhlenberg~1, Potsdam, 14476, Germany}
\affiliation[2]{Department of Physics, University of Maryland, College Park, MD 20742, USA}
\affiliation[3]{Higgs Centre for Theoretical Physics, University of Edinburgh, James Clerk Maxwell Building, Peter Guthrie Tait Road, Edinburgh, EH9 3FD, UK}
\affiliation[4]{Department of Physics, Pennsylvania State University, University Park, PA 16802, USA}
\affiliation[5]{Mani L. Bhaumik Institute for Theoretical Physics, University of California at Los Angeles, Los Angeles, CA 90095, USA}

\begin{document}

\

\title{Snowmass White Paper: Gravitational Waves and Scattering Amplitudes}


\abstract{
We review recent progress and future prospects for harnessing powerful tools from theoretical high-energy physics, such as scattering amplitudes and effective field theory, to develop a precise and systematically improvable framework for calculating gravitational-wave signals from binary systems composed of black holes and/or neutron stars. This effort aims to provide state-of-the-art predictions that will enable high-precision measurements at future gravitational-wave detectors. In turn, applying the tools of quantum field theory in this new arena will uncover theoretical structures that can transform our understanding of basic phenomena and lead to new tools that will further the cycle of innovation. While still in a nascent stage, this research direction has already derived new analytic results in general relativity, and promises to advance the development of highly accurate waveform models for ever more sensitive detectors.}

\maketitle

\section{Executive Summary}
\begin{figure}[t]
\begin{center}
\includegraphics[scale=0.53]{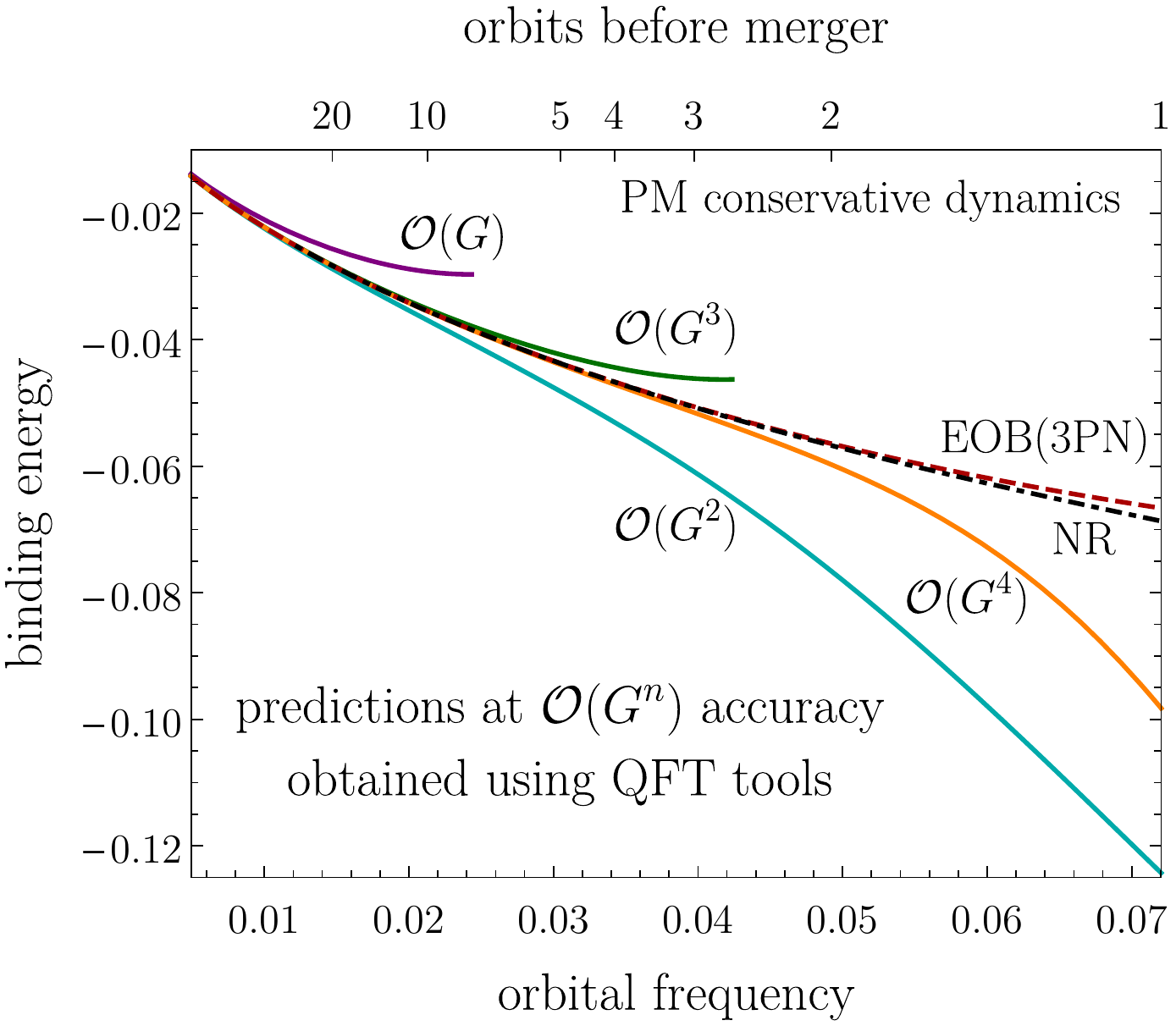}
~
\includegraphics[scale=0.53]{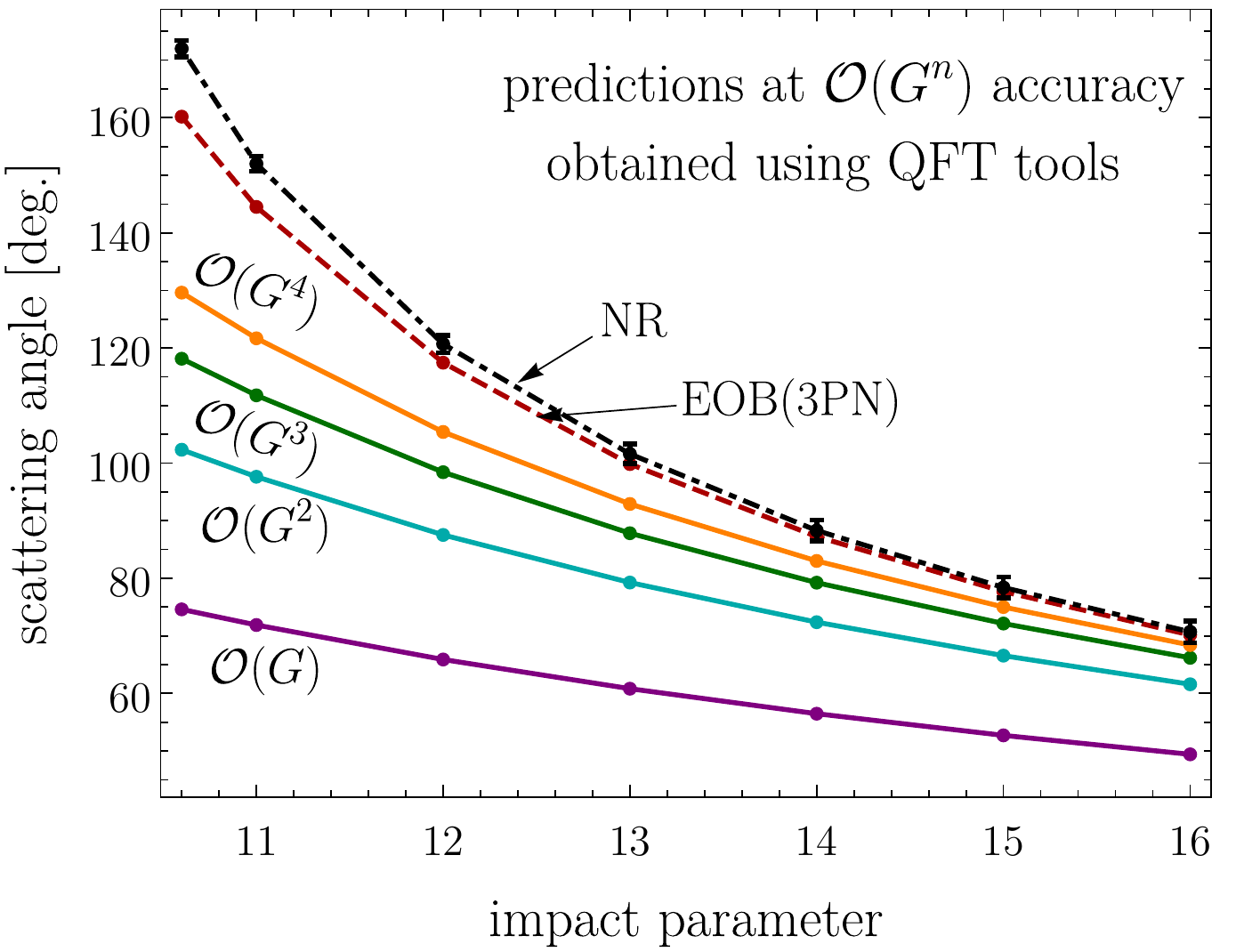}
\end{center}
\vspace{-0.5cm}
\caption{ \textbf{(left)} The binding energy (in units of the reduced mass) versus the orbital frequency of an equal-mass non-spinning binary black hole following an adiabatic quasi-circular orbit towards merger. 
The horizontal axis is also given as the number of orbits before merger.
\textbf{(right)} The scattering angle versus impact parameter of two equal-mass non-spinning black holes following hyperbolic trajectories with initial relative velocity $v=0.4$. These plots are adapted from~\cite{Khalil:2022}, where the authors compared predictions for post-Minkowskian (PM) conservative dynamics obtained using QFT tools (solid lines of increasing accuracy in Newton’s constant $G$, i.e., in the PM approximation) with NR~\cite{Damour:2014afa,Ossokine:2017dge} and effective-one-body (EOB) results, here at third post-Newtonian (PN) order, which are the benchmarks for building waveform models in the LVK collaboration. 
 }
\label{fig:binding_and_angle}
\end{figure}

The ambitious future of gravitational-wave (GW) science calls for
invigorating the theoretical framework for precision calculations of
GW signals, thus galvanizing a new approach that
harnesses the cutting-edge tools of theoretical high-energy physics
such as on-shell methods, double copy, advanced multiloop integration,
and effective field theory (EFT). These are the engines that drive modern
calculations of scattering amplitudes in particle theory, and
integrating them together for application to GWs has
recently led to new results in the perturbative, analytic solution of the two-body problem in general relativity. Theorists developing waveform models for the
LIGO-Virgo-KAGRA (LVK) collaboration~\cite{LIGOScientific:2014pky,VIRGO:2014yos,KAGRA:2020agh} have performed initial studies~\cite{Antonelli:2019ytb,Khalil:2022} of these early
results, see Figure~\ref{fig:binding_and_angle}, and have strongly encouraged further developments. Indeed, if these calculations are pushed to higher orders, and are extended to include all physical effects (i.e., spins and tides), they can
be used, in combination with other analytic methods~\cite{Regge:1957td,Zerilli:1970se,Vishveshwara:1970cc,Teukolsky:1973ha,Buonanno:1998gg,Buonanno:2000ef}\footnote{See also \cite{Blanchet:2013haa,Porto:2016pyg,Schafer:2018kuf,Barack:2018yvs} for recent reviews.} and with numerical-relativity (NR) simulations~\cite{Pretorius:2005gq,Campanelli:2005dd,Baker:2005vv,Foucart:2022iwu}, to provide highly accurate
waveform models of binary systems composed of black holes and/or neutron stars. Another aspect of this program that has drawn significant interest from both the high-energy 
physics and general relativity communities is the exploration
of theoretical structures that emerge in the classical limit of
scattering amplitudes.

This new research direction in theoretical high-energy physics is an
opportunity to deploy the classic and modern tools of quantum field theory (QFT)  
in a new arena, thereby impacting an important experimental frontier and
uncovering rich theoretical structure that can lead to new tools. The
program is in a nascent stage, and significant progress will come in
the next several years, building towards the vision that QFT tools
will advance the computation of gravitational waveforms. In particular, 
they will address the need for high precision in upcoming LVK runs, in space-based detectors such as LISA~\cite{LISA:2017pwj}, and in future ground-based detectors such as LIGO-India~\cite{Saleem:2021iwi}, Cosmic Explorer~\cite{Reitze:2019iox} and Einstein Telescope~\cite{Punturo:2010zz}. High-precision waveform models will be crucial for maximizing the discovery potential and extracting the best science with GW observations of ever more sensitive future detectors~\cite{LISA:2017pwj,Sathyaprakash:2019yqt,Maggiore:2019uih,Kalogera:2021bya,Berti:2022wzk}.

\section{Introduction}

The emergence of 
GW science~\cite{Abbott:2016blz, TheLIGOScientific:2017qsa,LIGOScientific:2021qlt} has already transformed multiple domains of astronomy, cosmology, and particle physics, yet this represents only a small fraction of its future potential~\cite{LISA:2017pwj,Kalogera:2021bya}. Space- and ground-based observatories of the coming decade will map out and characterize millions of merger events per year with sensitivity well beyond that of current LIGO/Virgo facilities (see, e.g., ~\cite{Favata:2013rwa,Samajdar:2018dcx,Purrer:2019jcp,Huang:2020pba,Gamba:2020wgg}). One of the key challenges will be advancing the theoretical modeling of compact binary coalescences to produce accurate gravitational waveforms.

Theoretical modeling of GW
sources is challenging due to multiple physical scales that are nonlinearly coupled through general relativity. Notably, these complications — multiple scales and nonlinearity — are exactly what drove breakthroughs in QFT in the last few decades, leading to the modern scattering amplitudes program. Scattering amplitudes have revealed mathematical structures in gauge theory and gravity, leading to new physical insights, efficient methods for computation, and the seeds for even bolder ideas.

The Parke-Taylor formula~\cite{Parke:1986gb} reduced pages of Feynman calculus to a half-line expression describing gluon scattering, famously heralding the enormous value of understanding theoretical structures lying at the heart of scattering amplitudes. In recent decades, major advances have been driven by two parallel developments. First are new methods that formulate QFT without explicit quantum fields, thus focusing on physical quantities. These “on-shell methods”, reinvigorated by twistor string ideas~\cite{Witten:2003nn, Roiban:2004yf, Gukov:2004ei, Cachazo:2004kj}, have become efficient mainstream tools for tree-level \cite{Britto:2005fq} and loop-level \cite{Bern:1994zx, Fusing, TripleCuteeJets, BCFUnitarity} calculations in gauge and gravity theories; see \cite{BDKUniarityReview, ElvangHuangReview, Elvang:2015rqa, JJHenrikReview, BernHuangReview} for reviews. 
Second, is a radically new perspective on gravity: gravitational scattering amplitudes ${\cal M}_{\rm gravity}$ can be realized as a “double copy” of gauge theory amplitudes ${\cal M}_{\rm gauge}$~\cite{BCJ, BCJLoop},
\be
{\cal M}_{ \rm gauge} \times {\cal M}_{ \rm gauge}  \sim {\cal M}_{\rm gravity} \,.
\ee
This structure builds on the relation between tree-level open and closed string scattering~\cite{KLT}, its field 
theory limit~\cite{Bern:1993wt}, and structures gleaned in explicit higher-loop calculations~\cite{CompactThree, Bern:2009kd}. It extends vanilla examples of related gauge and gravity theories to a veritable web of theories that share common building blocks, and has been applied to the exploration of a number of new directions, such as the ultraviolet properties of supergravity theories up to five loops~\cite{GravityThree, Bern:2012cd, N4GravFourLoop,N5GravFourLoop, UVFiveLoops}, and the nonperturbative structure of classical solutions of Einstein's equations with sources such as Schwarzschild~\cite{Monteiro2014cda}; see~\cite{BCCJRReview} for a recent review and also the dedicated Snowmass White Paper~\cite{doublecopyWhitePaper2022}.

\begin{figure}[t]
\begin{center}
\includegraphics[scale=.45]{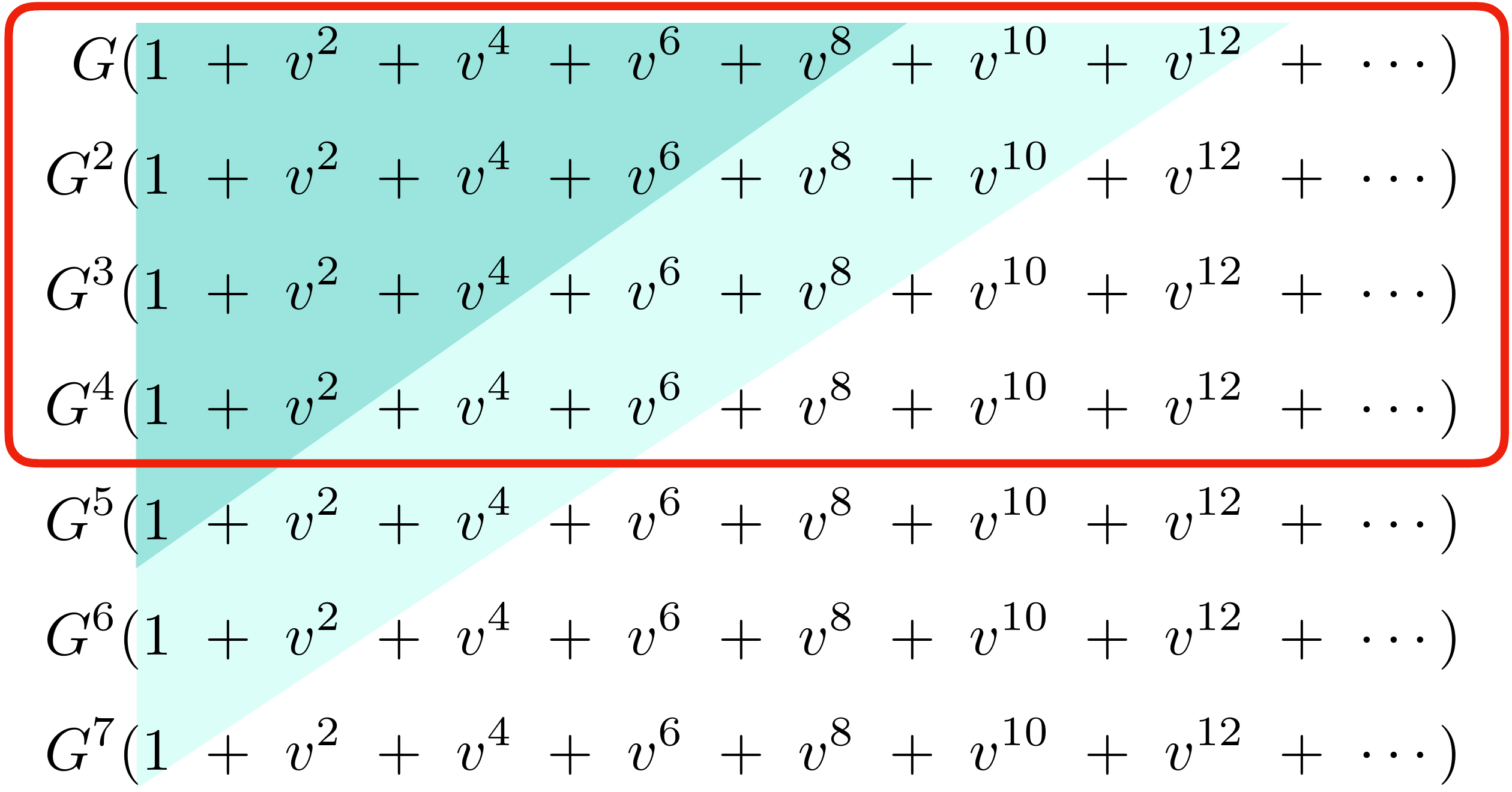}
\end{center}
\vspace{-0.5cm}
\caption{Map of perturbative corrections to Newton's potential, where $G$ is Newton's constant and $v$ is the relative velocity of the binary constituents. New results through $O(G^4)$ were recently obtained using QFT tools (red box). They are valid to all orders in velocity, and overlap with the state-of-the-art from the PN expansion (dark triangle) and the contributions required by future detectors (light triangle) (see, e.g., ~\cite{Favata:2013rwa, Samajdar:2018dcx,Purrer:2019jcp,Huang:2020pba,Gamba:2020wgg}).}
\label{fig:mapPT}
\end{figure}

In the past few years there has been a flurry of activity in applying both on-shell methods and the double copy, in combination with advanced multiloop integration techniques and EFT, to develop new tools for state-of-the-art predictions of 
GW signals. This development was encouraged by the general relativity community~\cite{Damour:2017zjx}, and has led to a number of new results in general relativity; for example see Figures~\ref{fig:binding_and_angle} and~\ref{fig:mapPT}.

The new approach based on tools from theoretical high-energy physics aims to complement, and has greatly benefited from, decades of successful work using traditional approaches to solve the relativistic two-body problem, including the post-Newtonian (PN) approximation~\cite{Einstein:1938yz, Einstein:1940mt, Ohta:1973je, Jaranowski:1997ky, Damour:1999cr, Blanchet:2000nv, Damour:2001bu, Damour:2014jta, Jaranowski:2015lha}, the gravitational self-force formalism~\cite{Mino:1996nk, Quinn:1996am}, the effective--one-body (EOB) 
formalism~\cite{Buonanno:1998gg,Buonanno:2000ef}, the nonrelativistic general-relativity (NRGR) formalism~\cite{Goldberger:2004jt}, the post-Minkowskian (PM) approximation~\cite{Bertotti:1956pxu, Kerr:1959zlt, Bertotti:1960wuq, Portilla:1979xx, Westpfahl:1979gu, Portilla:1980uz, Bel:1981be, Westpfahl:1985tsl, Damour:2016gwp, Damour:2017zjx}, and NR~ \cite{Pretorius:2005gq, Campanelli:2005dd, Baker:2005vv} (see, e.g.,~\cite{Blanchet:2013haa, Porto:2016pyg, Schafer:2018kuf, Barack:2018yvs, Levi:2018nxp} for recent reviews). In particular, the seminal work~\cite{Goldberger:2004jt} introduced nonrelativistic EFT ideas from particle physics to the worldline approach to binary dynamics, and has led to a number of landmark results; for recent developments see the dedicated 
Snowmass White Paper~\cite{NRGRWhitePaper2022} on this subject.

\begin{figure}[t]
\begin{center}
\includegraphics[scale=.3]{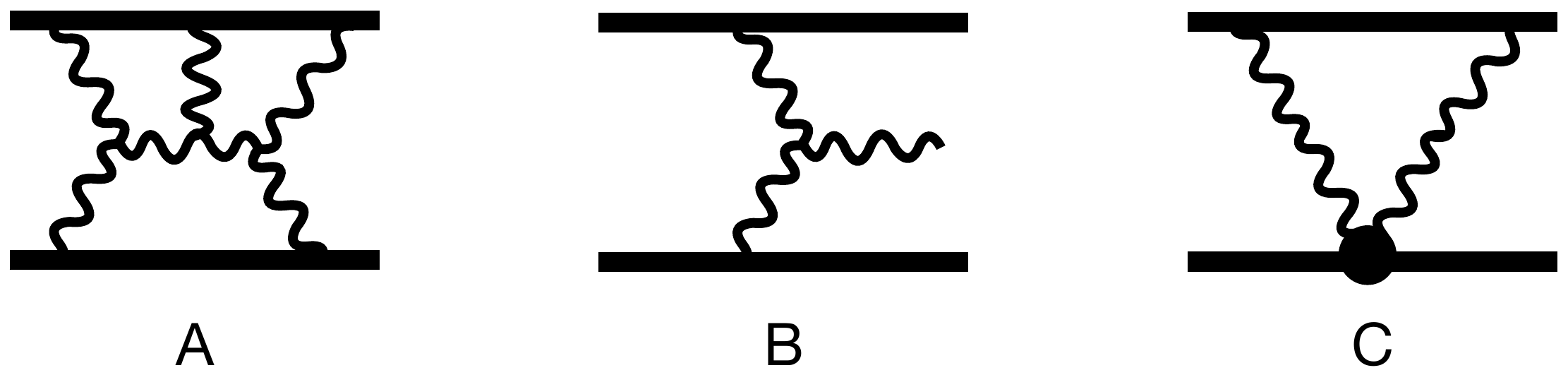}
\\[10pt]
\includegraphics[scale=.4]{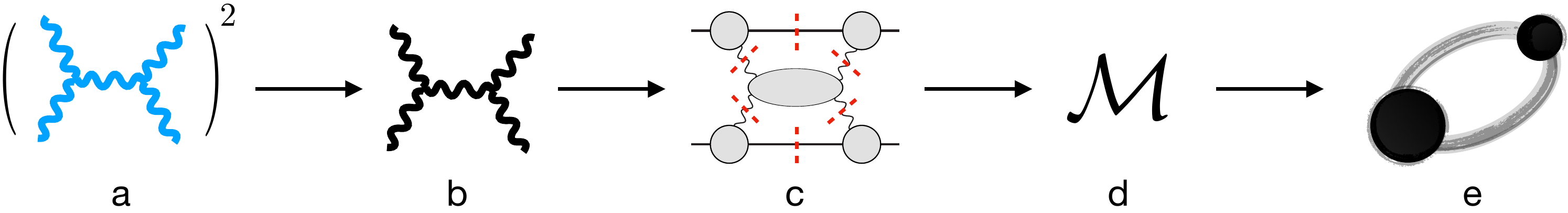}
\end{center}
\vspace{-0.7cm}
\caption{\textbf{(top)} Classical binary dynamics is encoded in scattering amplitudes of massive particles (thick lines) interacting through gravitons (wavy lines):  (A) four-point scattering encodes higher-order corrections to conservative binary dynamics, (B) five-point scattering encodes radiative effects due to graviton emission, and (C) higher-dimension operators (solid circle) encode tidal deformation of neutron stars. Spinning black holes can be described by higher-spin representations in QFT. \hspace{0.15cm} \textbf{(bottom)} A sample calculational pipeline using the tools of theoretical high-energy physics: Starting from tree-level gauge theory amplitudes (a), corresponding gravitational amplitudes (b) are obtained using the double-copy. These are then fused into loop amplitudes (c) using generalized unitarity. The integrated amplitude (d) is obtained using advanced multiloop integration methods developed in high-energy physics, in combination with EFT. The amplitude can then be mapped, using a variety of methods, to the EOB Hamiltonian used for producing waveforms (e). The classical limit is applied at every stage, leading to vast simplifications.}
\label{fig:method}
\end{figure}

The tools of theoretical particle physics have been honed in a wide variety of intricate quantum calculations such as the Higgs gluon-fusion cross section at N$^{3}$LO~\cite{Anastasiou:2015vya}, 
NNLO corrections to $e^+e^-$ event shapes \cite{Gehrmann-DeRidder:2007vsv}, cusp anomalous dimension at four loops~\cite{Henn:2019swt}, electron $g-2$ at four loops~\cite{Laporta:2017okg}, ${\cal N}=8$ supergravity four-point amplitude at five loops~\cite{Bern:2018jmv}, and ${\cal N}=4$ super-Yang-Mills  four-point amplitude at six loops~\cite{Carrasco:2021otn}. Leveraging these tools for 
GW physics yields several advantages. First, the structure of perturbation theory is vastly simplified by special relativity, on-shell methods, and the double copy, leading to compact expressions that make theoretical structures manifest. Second, the technology and deep knowledge base for integration of loops in QFT, which have seen decades of development for collider physics applications, are directly transferred, including integration-by-parts systems~\cite{Chetyrkin:1981qh, Laporta:2000dsw, Smirnov:2008iw} and differential equations~\cite{Kotikov:1990kg,Bern:1993kr,Remiddi:1997ny,Gehrmann:1999as,Henn:2013pwa,Henn:2013nsa,Parra-Martinez:2020dzs}. Finally, EFT efficiently and systematically targets contributions to various processes in the classical limit. Figure~\ref{fig:method} illustrates these tools in action.

Aside from providing state-of-the-art predictions, the new approach using particle physics tools aims to explore theoretical structures that emerge in the classical limit of scattering amplitudes. These structures may be familiar from particle physics, but are not manifest in traditional approaches to the two-body problem in general relativity, or they may be rooted in the classical regime and are yet to be explored through the lens of QFT. Examples include universality in the high-energy limit, the interplay between conservative and dissipative effects, nonperturbative connections to classical solutions, and perturbation theory in curved backgrounds.

The upshot is that scattering amplitudes are highly effective tools for understanding and precise modeling of 
GW sources, in ways similar to their application for interactions of fundamental particles. The recent progress and future goals of this program revolve around three objectives, broadly defined:
\begin{itemize}
\item Provide state-of-the-art predictions for the dynamics and gravitational radiation of compact binaries composed of black holes and neutron stars, which can be used, in combination with other analytic methods and NR simulations, to produce highly accurate waveforms. This includes modeling spin, tidal, and radiative effects.
\item Develop the mathematical and physical tools of theoretical high-energy physics for application to 
GWs. This involves tailoring existing tools to improve scalability, and formulating new tools and approaches to the 
binary problem.
\item Explore the theoretical landscape of scattering amplitudes in the classical regime. Examples include the universality of high-energy scattering, nonperturbative relations between scattering amplitudes and classical dynamics, and connections to exact spacetime geometries.
\end{itemize}
In the rest of this white paper, we will discuss the classical limit, and then describe each of the three topics above, summarizing recent progress. Then we will describe future directions.


\section{Classical Limit}\label{classical}

The correspondence principle states that classical physics emerges from the quantum theory in the limit of macroscopic conserved charges such as masses, electric charges, spins, 
orbital angular momenta, etc. Perturbations around such a configuration, which are subleading in the large charges, can be systematically included and have the natural interpretation 
of quantum corrections.\footnote{This perspective was used to great effect in string theory tests of the asymptotic Bethe ansatz for the anomalous dimensions of single-trace operators 
in ${\cal N}=4$ super-Yang-Mills theory, see e.g. \cite{Gubser:2002tv, Kruczenski:2004kw, Giombi:2009gd} and \cite{Beisert:2010jr} for a review.}

For the application to 
GWs, the classical limit of scattering amplitudes is defined by two properties that distinguish compact binaries from their quantum counterparts:
\begin{itemize}
\item Bound compact objects have large angular momentum $J \gg \hbar$, as opposed to $J \sim \hbar \equiv 1$ for quantum bound states.
\item Compact objects, such as black holes or neutron stars, have large gravitational charges $M_{\odot} / M_{\rm Planck} \sim 10^{38}$, as opposed to $e/Q_{\rm Planck} \sim 10^{-1}$ for electric charges of elementary particles. Here $M_{\odot}$ and $e$ denote the solar mass and electron charge, while $M_{\rm Planck}$ and $Q_{\rm Planck}$ are the Planck mass and charge, respectively. 
\end{itemize}

In other words, for a system of two gravitationally interacting spinless bodies with masses $m_1$ and $m_2$, the classical regime emerges in the limit $J \gg \hbar$ and $m_1, m_2 \gg M_{\rm Planck}$. This limit also coincides with the limit in which the de Broglie wave length $\lambda$ of the particles is much smaller than the particles' separation $|\bm b|$, which is conjugate to the momentum transfer $\bm q$. Thus, the kinematic regime of classical physics is when the momentum transfer is much smaller than the incoming momenta, $|\bm p|\gg |\bm q|$, similar to the Regge limit. Other charges that may characterize classical particles also have a similar scaling in the classical limit; for example, the spin $S$ and finite size $R$ scale as $|\bm q|R\sim {\cal O}(1)$, $|\bm q||\bm S|\sim m_{1,2}$. These scalings are consistent with the classical nature of Newton's potential, $|\bm b|\gg Gm$, and fixes the general form of four-point amplitudes as well as of generating functions of observables. 

Another perspective on the classical limit was taken in~\cite{Kosower:2018adc}, which together with the associated framework is usually referred to as the KMOC formalism.  
Quantum mechanical computation of observables yields the classical result
in the correspondence limit which is arranged in two steps: (1) a suitable restoration of Planck's constant $\hbar$, which effectively plays a role similar to the momentum transfer $\bm q$, and expansion at small $\hbar$, and (2) incorporation 
from the outset that from a quantum-mechanical perspective, particles are described by wavepackets having finite width in both position- and momentum-space 
rather than plane waves. This width must be negligible in the classical limit, leading to inequalities which constrain the parameters of the scattering, analogous to 
those coming from the large-$J$ limit. 

One implication of the classical limit is that loop amplitudes involving massive particles may contain classical contributions. The classical limit can be taken at the earliest stages of calculations, yielding vast simplifications prior to integration. This enables the calculation of loop amplitudes that would otherwise be beyond the reach of current technology if evaluated including quantum effects. Properties of amplitudes in an expansion around the classical limit were previously unexplored systematically. The application to 
GW physics provides the motivation to fill this gap, and have led to interesting results~\cite{Cristofoli:2021jas,Britto:2021pud}.


\section{State-of-the-Art Predictions \label{predictions} }

Waveform models for the inspiral stage of a binary 
system are built from the conservative and dissipative two-body 
dynamics.
Therefore a direct path for amplitudes methods to have an
impact is to compute state-of-the-art conservative two-body
potentials, including the effects of spin, and tidal deformation for
modeling neutron stars~\cite{Blanchet:2013haa,Punturo:2010zz,LISA:2017pwj,Reitze:2019iox}. For instance, the general relativity community had urged amplitudes
experts to compute the ${\cal O}(G^3)$ contributions to conservative
binary dynamics for spinless compact objects~\cite{Damour:2017zjx}.

Potentials are naturally encoded in four-point scattering amplitudes, and can be extracted from the latter via a variety of methods such as an amplitude matching calculation~\cite{CRS}, direct mapping to an EOB parameterization~\cite{Damour:2019lcq,Damgaard:2021rnk}, using the eikonal phase as a generating functional~\cite{Amati:1987wq, tHooft:1987vrq, Muzinich:1987in, Amati:1987uf}, and analytic continuation~\cite{Cho:2018upo, Kalin:2019rwq,Kalin:2019inp, Cho:2021arx}. However, at ${\cal O}(G^4)$, the conservative two-body Hamiltonian is no longer universal to both bound and unbound trajectories. 
The effect of back-scattered radiation on the conservative dynamics --- in particular the so-called tail effects~\cite{Thorne:1980ru, Blanchet:1987wq, Blanchet:1993ec} --- leads to nonlocal-in-time contributions to the Hamiltonian that depend on the specific trajectory. Understanding the mapping between bound and unbound orbits in the presence of radiation is an important open problem that needs to be solved to maximally leverage scattering amplitudes for deriving classical bound state dynamics.
\bigskip

\noindent \textbf{Higher-Order Corrections to Conservative Binary Dynamics.} 
Conservative binary dynamics at ${\cal O}(G)$ and ${\cal O}(G^2)$ and to all orders in velocity\footnote{The ${\cal O} (G^n)$ result to all orders in velocity is also referred to as the $n$-th order post-Minkowskian result or $n$PM result. In this White Paper we will often simply refer to this as the ${\cal O} (G^n)$ result, following the conventional nomenclature of perturbative orders in particle physics.} were derived from scattering amplitudes in Ref.~\cite{CRS}. While these results were known in general relativity using classical methods, the method employed in Ref.~\cite{CRS} laid a basic path for extending to higher orders using scattering amplitudes.

\begin{figure}
\begin{center}
\includegraphics[scale=.70]{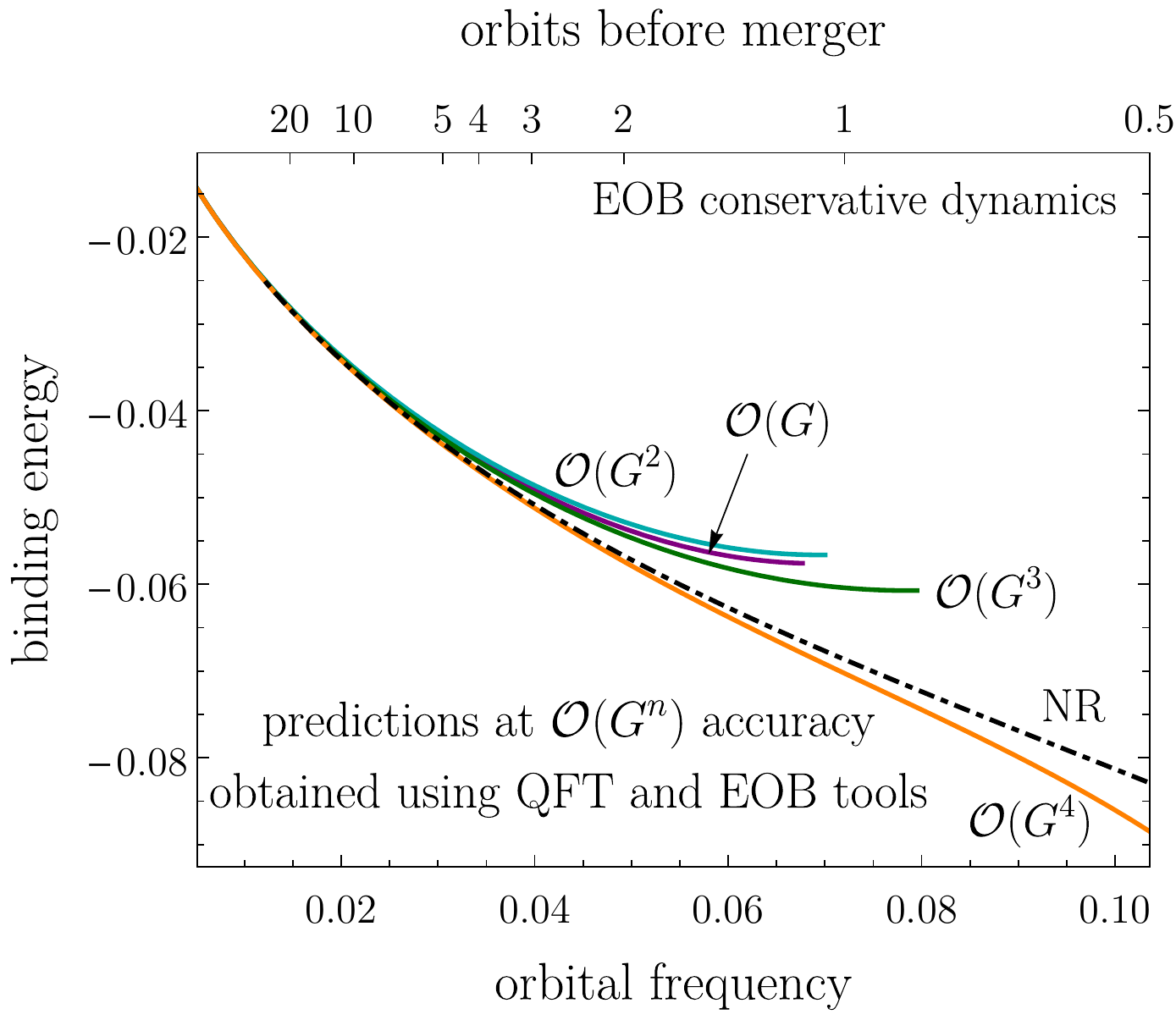}
\end{center}
\caption{The binding energy (in units of the reduced mass) versus the orbital frequency of an equal-mass non-spinning binary black hole following an adiabatic quasi-circular orbit towards merger. 
The horizontal axis is also given as the number of orbits before merger.
 The plot is adapted from~\cite{Khalil:2022}, where the authors compared predictions for post-Minkowskian (PM) conservative dynamics obtained using QFT tools (solid lines of increasing accuracy in Newton’s constant $G$, i.e., in the PM approximation) to NR results~\cite{Ossokine:2017dge}. In contrast to Figure~\ref{fig:binding_and_angle}, the predictions shown here are obtained by incorporating the QFT results into the EOB formalism, resulting in better agreement with NR towards merger.
}
\label{bindingenergyEOBPM}
\end{figure}

Conservative binary dynamics at ${\cal O}(G^3)$ was computed in Refs.~\cite{3PM,3PMLong} using amplitudes techniques, and has since been verified using a variety of methods (see, e.g.,~\cite{Blumlein:2020znm,Cheung:2020gyp,Kalin:2020mvi}). The result of Refs.~\cite{3PM,3PMLong} for conservative binary dynamics has been extended to include dissipative effects in a number of studies~\cite{DiVecchia:2020ymx, Damour:2020tta,DiVecchia:2021bdo,Bjerrum-Bohr:2021din,Damgaard:2021ipf,Brandhuber:2021eyq,Herrmann:2021lqe,Herrmann:2021tct,DiVecchia:2021ndb,Heissenberg:2021tzo}.%

Conservative binary dynamics at ${\cal O}(G^4)$ was computed in Refs.~\cite{Bern:2021dqo,Bern:2021yeh} using amplitudes techniques. Radiative contributions to the conservative dynamics first appear at ${\cal O}(G^4)$, and complicates the analysis due to difficulties in precisely defining conservative processes in the presence of radiation. This represents one of the main obstacles for advancing the theoretical modeling of 
GW sources to higher orders. Interestingly, some of the issues are sidestepped in the QFT-based approach where everything follows from well-established properties of scattering amplitudes. In particular, four-point scattering amplitudes should fully capture two-body conservative dynamics 
for hyperbolic orbits and is equivalent to using time-symmetric graviton propagators~\cite{Wheeler:1949hn, Damour:1995kt, Damour:2016gwp}.

The result of Refs.~\cite{Bern:2021dqo,Bern:2021yeh} reproduces the sixth-order PN result in Ref.~\cite{Bini:2021gat}. Moreover, due to Lorentz invariance and dimensional analysis, the result exhibits a simple mass dependence, confirming arguments made in~\cite{Damour:2019lcq}. This mass dependence puts strong constraints on radiative contributions to conservative binary dynamics~\cite{Bini:2021gat}. For instance, the result of Refs.~\cite{Bern:2021dqo,Bern:2021yeh} also agrees with the fifth-order PN result of Ref.~\cite{Blumlein:2021txe} up to a single term that does not have the expected mass dependence. The origin of such contributions requires further study. The result of Refs.~\cite{Bern:2021dqo,Bern:2021yeh} to all orders in velocity has also been partially verified~\cite{Dlapa:2021npj,Dlapa:2021vgp}. Initial studies of these results were performed by theorists developing waveform models for the LVK collaboration with encouraging conclusions; see Figure~\ref{fig:binding_and_angle}. Quite interestingly, the results derived from amplitudes and EFT can be included in the EOB framework in a way that improves the comparison with NR towards merger; see Figure~\ref{bindingenergyEOBPM}. While these results at 4PM order are in an early stage with assumptions~\footnote{The results at 4PM order, i.e. at ${\cal O}(G^4)$, have been obtained from the Hamiltonian computed for scattering orbits in~\cite{Bern:2021dqo,Bern:2021yeh,Dlapa:2021npj,Dlapa:2021vgp}.} that require further investigation, the key message is that results from scattering amplitudes are promising and further developments are strongly encouraged.

\bigskip

\noindent \textbf{Spin Effects.}
From a theoretical perspective, inclusion of the spin of the binary components in amplitudes methods faces the difficulty that fields of arbitrary spins are needed while no-go theorems~\cite{Shamaly:1972zu, Hortacsu:1974bm, Deser:2000dz, Porrati:2008gv, Camanho:2014apa, Afkhami-Jeddi:2018apj} show that under certain assumptions such field theories have unphysical features.
Five proposals~\cite{Arkani-Hamed:2017jhn, Bern:2020buy, Chiodaroli:2021eug, Aoude:2020onz,Jakobsen:2021zvh} have 
been put forth 
to construct classical spin-dependent two-body Hamiltonians.
Their results are consistent with each other, and also with results obtained through 
standard general-relativity methods, when the latter are available.

The stress tensor of a Kerr black hole~\cite{Vines:2017hyw} was obtained from higher-spin interactions in~\cite{Guevara:2019fsj} and~\cite{Bern:2020buy}, the latter also obtaining the stress tensor of more general spinning compact objects~\cite{Levi:2015msa}. 
The all-orders-in-spin ${\cal O}(G)$ and quartic-in-spin ${\cal O}(G^2)$ scattering angle for 
an aligned-spin configuration (i.e., spins that are aligned or anti-aligned with the orbital angular momentum) was found in~\cite{Guevara:2018wpp}. The ${\cal O}(G^2)$ Hamiltonians expected to describe Kerr black hole binaries for general spin configurations through fourth power of the spin were 
derived in~\cite{Bern:2020buy, Kosmopoulos:2021zoq, Chung:2019duq, Chung:2020rrz, Chen:2021qkk}, and a general direct 
relation between amplitudes and observables was conjectured in~\cite{Bern:2020buy}.  
The change in momentum and in the spin at ${\cal O}(G^3)$ to quadratic order in spin have been computed 
in~\cite{Jakobsen:2022fcj}, together with the radiation-reaction effects due to the radiated momentum at 2PM order \cite{Jakobsen:2021smu}.

Proceeding past fourth order in spin at ${\cal O}(G^2)$ had been an important open problem for standard general-relativity methods.
The neutron star Hamiltonian quintic in spin was predicted in \cite{spin5} for arbitrary spin orientation, and the Kerr scattering amplitude to eighth order in spin in~\cite{spin5competition}. Refs.~\cite{spin5, spin5competition} proposed that the Kerr black hole corresponds to a particular subclass of spin structures that can appear in a general scattering amplitude, which was explained in~\cite{spin5}
through a shift symmetry of the spin vector reminiscent of reparametrization invariance in heavy-particle EFT~\cite{Luke:1992cs, Heinonen:2012km}. 

Relatedly, the dynamics of spin is technically very similar to the dynamics of color in classical (infrared-free) Yang-Mills theories; using similar techniques, 
the 2PM Hamiltonian for color-charged matter was found in~\cite{delaCruz:2020bbn,delaCruz:2021gjp}.

\bigskip
\noindent \textbf{Tidal Effects.} 
Recent detections of 
GWs from the merger of neutron stars already constrain the equation of state of matter at nuclear densities~\cite{TheLIGOScientific:2017qsa, LIGOScientific:2020aai}, and more accurate measurements at future detectors strongly motivate calculations at higher precision~\cite{Favata:2013rwa, Samajdar:2018dcx,Purrer:2019jcp,Huang:2020pba,Gamba:2020wgg}. Similar to traditional approaches using classical methods~\cite{Goldberger:2004jt,Bini:2020flp,Kalin:2020mvi}, the approach based on scattering amplitudes uses higher-dimension operators to model the rigidity of the body and the susceptibility of its shape to change in response to a tidal potential~\cite{Cheung:2020sdj}. The ${\cal O}(G^3)$ contributions from the leading tidal operators were computed using scattering amplitudes in~\cite{Cheung:2020sdj}, and was verified using a worldline approach in~\cite{Kalin:2020lmz}. Other amplitudes-based calculations of tidal effects have been pursued in Refs.~\cite{Haddad:2020que, Aoude:2020ygw, Bern:2020uwk,Cheung:2020gbf}, including results for infinite classes of tidal operators as well as all orders in $G$ results in the probe limit.

Aside from calculating predictions for tidal effects for neutron stars, tidal effects for black holes have also been the subject of many recent studies within the particle physics community~\cite{Chia:2020yla,Hui:2020xxx,Charalambous:2021mea}. Black holes have vanishing static, conservative tidal responses, and a symmetry explanation for this has recently been put forth~\cite{Charalambous:2021kcz,Hui:2021vcv}.

\bigskip
\noindent \textbf{Beyond General Relativity.}
Two-body Hamiltonians capturing specific models of physics beyond general relativity have also been computed using scattering amplitudes methods~\cite{Emond:2019crr, Cristofoli:2019ewu, AccettulliHuber:2019jqo, AccettulliHuber:2020dal, Bern:2020uwk, 
Cheung:2020sdj, AccettulliHuber:2020oou, Carrillo-Gonzalez:2021mqj}.

\bigskip

\noindent \textbf{Radiative and Absorptive Effects.}
As mentioned above, waveform models rely on both the conservative and dissipative dynamics. Calculations of radiative effects have so far focused on hyperbolic scattering trajectories, in particular the scattering angle~\cite{DiVecchia:2020ymx, DiVecchia:2021bdo,Bjerrum-Bohr:2021din,Damgaard:2021ipf,Brandhuber:2021eyq,Herrmann:2021lqe,Herrmann:2021tct,DiVecchia:2021ndb,Heissenberg:2021tzo}, as well as the loss of linear~\cite{Herrmann:2021tct, Herrmann:2021lqe,Riva:2021vnj} and angular momentum~\cite{Damour:2020tta,Bini:2021gat,Gralla:2021qaf,Jakobsen:2021smu,Jakobsen:2021lvp,Mougiakakos:2021ckm,Manohar:2022dea}, 
and the energy flux from higher-order tail effects~\cite{Bini:2021qvf, Edison:2022cdu}. A proposal for deriving radiation-reaction forces from scattering amplitudes was recently given in~\cite{Manohar:2022dea}.

So far, absorptive effects have not been incorporated in a QFT framework, see \cite{Goldberger:2019sya} for a worldline EFT approach.


\section{Theoretical Structures \label{structures} }

In this section we give an overview of mathematical and physical structures that are relevant in the classical regime of scattering amplitudes. Some of these structures, such as nonperturbative properties of amplitudes and their direct connections to classical solutions, were exposed through explicit calculations of different observables at higher orders. Others, such as the eikonal phase and universality of high-energy scattering, have been studied in the context of theoretical high energy physics but only recently applied for 
GWs. We highlight examples where theoretical structures bring insight to the phenomenology of 
GWs, and can be leveraged to develop calculational tools.
\bigskip

\noindent \textbf{High Energy Limit.}
Even for classical scattering, the high energy (ultra-relativistic) limit exposes interesting structures such as the interplay between exclusive and inclusive observables~\cite{Bloch:1937pw,Kinoshita:1962ur,Lee:1964is}, connections to soft graviton theorems~\cite{DiVecchia:2021ndb}, and universality among gravitational theories with and without supersymmetry~\cite{tHooft:1987vrq, DiVecchia:2020ymx, Parra-Martinez:2020dzs}. 

For example, the four-point classical potential scattering amplitude at ${\cal O}(G^3)$ develops a singularity in the high-energy limit, $s \to \infty$:
\begin{align}\label{eq:high}
{\cal M} \to - 8 \pi G^3 s^2 \log(-t) \log\left( m_1 m_2 \over s \right)\,,
\end{align}
where $s$ and $t$ are Mandelstam variables, and $m_{i}$ are the masses of the scattering black holes. On general grounds, such singularities cannot exist in complete amplitudes~\cite{Akhoury:2011kq}. The finite inclusive observable, including contributions from radiation modes, can be derived using soft graviton theorems to describe graviton emission~\cite{DiVecchia:2021ndb} or by a linear response analysis~\cite{Damour:2020tta}.  Moreover, the double logarithmic structure can be understood from the Regge limit, and is given by the cusp anomalous dimension~\cite{Henn:2012qz}, or by the heavy-heavy current anomalous dimension in Heavy Quark Effective Theory~\cite{Manohar:2000dt}. 

These connections illustrate that amplitudes in simplifying limits, even in supersymmetric versions of gravity, can carry useful information about classical binary dynamics. More generally, there is a wealth of knowledge on four- and five-point amplitudes in theoretical high energy physics, and many of the tools and physical insights can be transplanted to classical binary dynamics. For example, quantum electrodynamics (QED) as well as supersymmetric cousins of gravity, such as ${\cal N}=8$ supergravity, offer controlled laboratory settings for sharpening tools and dissecting basic phenomena.
\bigskip

\noindent \textbf{Eikonal Phase.} It has long been known that in the classical limit the eikonal phase provides a good description of elastic four-point amplitudes
in gauge and gravity theories, which effectively exponentiate as a consequence of unitarity~ \cite{Levy:1969cr, Abarbanel:1969ek, Cheng:1969eh, Amati:1990xe, 
Kabat:1992tb, Saotome:2012vy, Akhoury:2013yua} 
\begin{equation}
i{\cal M} = e^{i\delta} - 1 \ .
\label{eq:Eikonal}
\end{equation}
It is moreover expected that the dominant contributions 
at high energies to the eikonal $\delta$ come from the exchange of the highest-spin state in the theory~\cite{tHooft:1987vrq, Amati:1990xe, Muzinich:1987in}, and therefore that it is universal in gravitational theories~\cite{Bellini:1992eb}. 

Modern evidence for these properties was obtained in recent work on massive theories~\cite{Parra-Martinez:2020dzs, DiVecchia:2020ymx,Bjerrum-Bohr:2021din}, as well as massless theories with various amounts of supersymmetry, including general relativity~\cite{Bern:2020gjj}. This led to the idea that inclusion of all contributions from exchanged gravitons with momenta of the order of the transferred momentum is a necessary (and possibly sufficient) condition for universality of the eikonal in the ultrarelativistic limit and for a smooth interpolation to the nonrelativistic limit~\cite{DiVecchia:2020ymx}. A self-contained treatment of the real and imaginary parts of the eikonal in the entire soft region at ${\cal O}(G^3)$ in massive ${\cal N}=8$ and in general relativity was presented in \cite{DiVecchia:2021bdo} and confirmed in \cite{Bjerrum-Bohr:2021vuf}, giving another demonstration of the universality properties.

The eikonal form of the S matrix is quite similar to that of the amplitude-radial action relation~\eqref{eq:AA}, but they differ in the definition of the iteration terms~\cite{Bern:2021dqo}. Similar to the radial action, the eikonal is a generating 
function of scattering observables~\cite{Amati:2007ak}, such as the scattering angle or the time delay. See Sec.~\ref{tools} for further discussion.
\bigskip

\noindent \textbf{Eikonal with Spin.}
The unitarity argument for the exponentiation of elastic amplitudes holds also for spinning external states. This was demonstrated explicitly for spin-1/2 particles in~\cite{Haddad:2021znf} and for arbitrary spin in the classical limit from the perspective of minimization of the spin variance in~\cite{Cristofoli:2021jas}.

The eikonal continues to be a generating function for observables when scattering spinning particles. This was verified explicitly
through ${\cal O}(S_1S_2)$ in~\cite{Bern:2020buy} where the change in momentum (also referred to as the ``impulse")  and the change in spin (also referred to as
the ``spin kick") obtained from the eikonal were compared with the results of Hamilton's equations. A conjecture was also put forth~\cite{Bern:2020buy}, connecting eikonal and scattering observables to all orders in spin. This relation was successfully tested through fourth order in spin~\cite{Kosmopoulos:2021zoq, Chen:2021qkk} at ${\cal O}(G^2)$.
It is quite interesting and surprising that a single function can capture the complicated three-dimensional dynamics of scattering of spinning particles.
\bigskip

\noindent \textbf{Nonperturbative Structures.} EFT methods can be used to connect scattering amplitudes and classical binary dynamics through a matching calculation between gravity and a low energy EFT that describes particles interacting through an effective potential.
An outcome of this matching calculation is a beautiful relation between the classical limit of the scattering amplitude ${\cal M}$ in impact parameter space and the radial action $I_r$ of Hamilton-Jacobi theory: 
\begin{align}\label{eq:AA}
i {\cal M} = e^{i I_r} - 1  .
\end{align}
This means that the scattering amplitude directly determines the radial action $I_r$, which in turn determines orbital trajectories.
This equation was first derived in~\cite{Bern:2021dqo}, and is dubbed the ``amplitude-action relation". The exponential form of the amplitude-action relation is reminiscent of the eikonal phase~\eqref{eq:Eikonal}, and has an important practical implication for calculations: a large class of integrals come from exponentiation of lower-order contributions and can be systematically dropped prior to integration. The amplitude-action relation was studied to all orders in perturbation theory for a probe in the Schwarzschild or pure-NUT 
gravitational backgrounds as well as for a probe interacting with point-charges and monopoles in~\cite{Kol:2021jjc}.

\begin{figure}[t]
\begin{center}
\includegraphics[scale=.49]{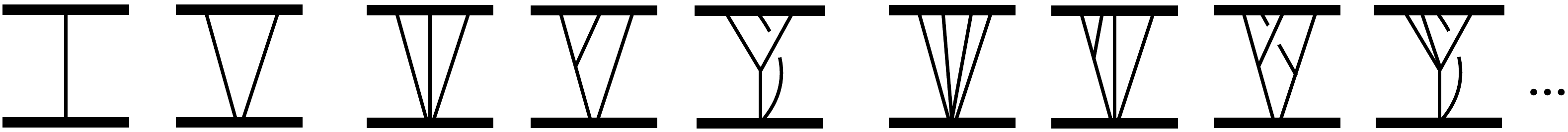}
\end{center}
\vspace{-0.5cm}
\caption{The resummation of these diagrams to all orders in $G$ yields the amplitude for a probe particle in a Schwarzschild background.}
\label{fig:geodesic}
\end{figure}

The amplitude-action relation in Eq.~(\ref{eq:AA}) is an example of remarkable nonperturbative structures that appear in the classical limit. 
Another example is the following relation between the classical scattering amplitude in position space ${\cal M}(r)$ and the local center-of-mass momentum in a hyperbolic orbit,
\begin{align}
{\cal M}(r) = {p(r)^2 - p(\infty)^2 \over 2E} \,.
\end{align}
Here $p(r)^2$ is the squared center-of-mass momentum at position $r$, and $E$ is the total center-of-mass energy. This was first noticed and used in~\cite{3PM, 3PMLong}, and further developed and formalized in~\cite{Kalin:2019rwq} (where it was dubbed ``impetus formula") and also~\cite{Bjerrum-Bohr:2019kec}. This relation can be used to derive amplitudes to all orders in $G$. For instance, the diagrams in Figure~\ref{fig:geodesic} describe a probe particle in a Schwarzschild background, and can be resummed by determining $p(r)$ from geodesic motion. One can also derive nonperturbative amplitudes involving higher-dimensional operators that describe tidal effects~\cite{Cheung:2020gbf}. 

Another nonperturbative relation builds on the Newman-Janis~\cite{Newman:1965tw} relation between the Schwarzschild and Kerr solutions of Einstein's equations.  This relation, employing a certain complex shift, has been used in~\cite{Arkani-Hamed:2019ymq} to obtain the impulse for spinning particles from that of spinless particles at leading post-Minkowskian order. It remains an intriguing open question whether the relation holds at higher orders and how to exploit it~\cite{Guevara:2020xjx}. 

\bigskip

\noindent \textbf{Gravitational Self-Force.} Extreme-mass-ratio inspirals are binary systems consisting of compact bodies with masses $m_1$ and $m_2$ with $m_2 \gg m_1$. The limiting case is described by a probe particle orbiting in a background spacetime such as Schwarzschild or Kerr. Beyond this, the particle interacts with its own gravitational field, giving rise to an effective ``self-force", which is computed as an expansion in $m_1/m_2$ but to all orders in $G$. 
The gravitational self-force for generic bound geodesics in Schwarzschild and Kerr spacetimes  were found in \cite{Barack:2010tm} and \cite{vandeMeent:2017bcc}, respectively.
The precision of LISA will require the second-order self force in the conservative and dissipative dynamics, i.e., corrections of ${\cal O}(m_1^2/m_2^2)$ and all orders in $G$, which is not completely solved. However, see Ref.~\cite{Pound:2019lzj} for significant recent progress.
 
Interestingly, scattering amplitudes have revealed a connection between perturbative corrections to binary dynamics and self-force corrections. The mass dependence of the $n$-loop scattering amplitude in the classical limit follows simply from dimensional analysis, and implies that it can probe the ${\cal O}(G^{n+1})$ contribution to the $\lfloor \frac{n}{2} \rfloor$-th order self-force correction. In other words, the $n$-loop scattering amplitude contains contributions of the form $\sim G^{n+1} (m_1/m_2)^{\lfloor \frac{n}{2} \rfloor}$. 
This structure is leveraged in a powerful new method, dubbed ``Tutti-Frutti", for extracting perturbative corrections to binary dynamics from self-force calculations~\cite{BDG1,BDG2,BDG3,Damour:2019lcq,Antonelli:2020aeb,Antonelli:2020ybz,Khalil:2021fpm}. This mass dependence, which ultimately traces back to Lorentz invariance, also has strong implications for the radiative contribution to conservative dynamics~\cite{Damour:2019lcq}. In particular, it implies nontrivial cancellations among contributions from source multipole moments and imposes strong consistency checks on perturbative calculations. Notably, there are now numerical and theoretical efforts to extend gravitational self-force calculations to hyperbolic trajectories in order to make contact with results from scattering amplitudes; see, e.g.,~\cite{Gralla:2021qaf}.
\bigskip

\noindent \textbf{Coherent States and Waveforms.} 
The gravitational waveform during a scattering event can itself be directly computed from amplitudes~\cite{Cristofoli:2021vyo}. At lowest perturbative order
it is given by an integral of a five-point tree scattering amplitude. The relation to the intuitive picture, in which a 
GW consists of a large number of gravitons, is interesting: 
GWs are described by coherent states (with very large occupation numbers) of the gravitational field.
Gravitons building up this state are emitted independently~\cite{Cristofoli:2021jas, Britto:2021pud}, and each emission is 
described by the same five-point amplitude.

Coherent states also play a role in understanding the emergence of classical spin and color from QFT~\cite{Bern:2020buy, delaCruz:2020bbn,Aoude:2021oqj}, and in understanding the emergence of classical physics from quantum mechanics quite generally~\cite{Yaffe:1981vf}.

Eikonal methods have been generalized to include such coherent outgoing radiation~\cite{Cristofoli:2021jas}, building on earlier work of~\cite{Ciafaloni:2018uwe} and~\cite{Cristofoli:2021vyo}. In particular, Ref.~\cite{Cristofoli:2021jas} argued that the eikonal is effectively extended by a coherent state operator, which creates arbitrarily many outgoing massless particles and that the nonperturbative radiation field is described by the waveshape parameter defining the coherent state, which in turn is determined by the five-point amplitude.
\bigskip

\noindent \textbf{Soft Gravitons.} 
Recently it has become clear that there is a beautiful relation between the soft limit in quantum theories and memory effects in classical dynamics~\cite{Strominger:2014pwa}. This can be understood in the KMOC formalism (see Sec.~\ref{tools}) by studying the radiated momentum. In the long-wavelength limit, the scattering amplitude involved in the radiation simplifies as a soft factor times a lower point amplitude, recovering the impulse~\cite{Bautista:2019tdr}. Classically, this impulse is the ``memory'' of the step-change in the field described by its very low frequency Fourier components. More generally, the KMOC formalism reveals a rich interplay between classical physics, soft or low-frequency radiation, and scattering amplitudes~\cite{Laddha:2018rle,Laddha:2018myi,Sahoo:2018lxl,Manu:2020zxl,Bautista:2021llr}.
\bigskip

\noindent \textbf{Analytic Continuation and Time Nonlocality.}
The integrability of the two-body equations of motion with Newton's potential guarantees that their solutions are uniquely specified 
by integrals of motion -- the total energy and the orbital angular momentum. Hyperbolic and elliptic motions are mapped into each 
other by analytic continuation of the boundary conditions.
While integrability is not known to exist for two-body conservative Hamiltonians even at 
2PM order~\cite{Caron-Huot:2018ape}, Refs.~\cite{Cho:2018upo,Kalin:2019rwq, Kalin:2019inp, Cho:2021arx}
argued that bound state observables can be obtained through a suitable analytic continuation in energy and in angular momentum 
from scattering observables. This procedure was dubbed ``Boundary-to-Bound" or ``B2B" and formalizes the fact 
that such Hamiltonians can be constructed by matching, e.g. the scattering angle, and then subsequently used for bound motion 
by changing the boundary conditions. Physically, this can be understood as a consequence of the time locality of the 
potential generated by potential-region gravitons. 

However, this approach fails for conservative radiation-reaction effects, beginning at ${\cal O}(G^4)$ with the tail effects~\cite{Cho:2021arx}. 
Given a scattering angle, it is always possible to construct a local Hamiltonian that reproduces it. Fundamentally however, because it captures 
effects of radiation modes propagating over long periods before being reabsorbed by the binary, the ``off-shell" Hamiltonian has both 
an instantaneous component and a non-local in time one~\cite{Damour:2014jta}. 
While the analytic continuation of the local and universal (logarithmic) part of the nonlocal Hamiltonian is straightforward, obtaining the non-universal 
part of the bound Hamiltonian from the unbound one is an important open problem.
\bigskip

\noindent \textbf{Exploring Structure in Simpler Theories.} Even in the classical limit gravitational interactions are complicated. Before proceeding to full-fledged gravitational calculations, it is useful to test ideas and tools and search for theoretical structures in much simpler settings such as gauge theory. Since color in the classical limit becomes essentially abelian, it suffices to study QED. Calculations designed to explore some of the subtleties that appear at ${\cal O}(G^3)$ in gravitational calculations were carried out in gauge theories, both using scattering amplitudes~\cite{delaCruz:2020bbn, Bern:2021xze}, and classical gravity methods~\cite{Saketh:2021sri,Buonanno:2000qq}.

Gravitational theories with additional symmetries, like supersymmetry are another possibility. As discussed above, results obtained in supergravity theories confirmed universality properties of the eikonal function and led to an understanding of the role of radiative corrections in classical scattering.
\bigskip

\noindent \textbf{Relations Between Amplitude Fragments.} 
The essential difference between classical and quantum measurements is the variance: it is nonzero for quantum measurements and must become negligible 
in the classical limit. Defining measurements in terms of expectation values of operators, Ref.~\cite{Cristofoli:2021jas} showed that the condition of zero variance (effectively requiring factorization of expectation values of products of operators) implies an infinite set of relations between different amplitude ``fragments"
(that is different orders in an amplitude's expansion in momentum transfer) with different numbers of loops and legs. 
For four-point amplitudes they are the relations required for eikonal exponentiation. 
At five points, higher-loop fragments are related to lower-loop five-point and four-point amplitude fragments.
Among the implications of the zero-variance condition are detailed predictions for the momentum transfer dependence of amplitudes and a novel understanding of the eikonal in the presence of outgoing radiation mentioned above.


\section{Developing Tools \label{tools} }

In this section we describe the tools that enable state-of-the-art calculations, such as those summarized in Sec.~\ref{predictions}, as well as, recently developed tools that were inspired by new theoretical structures or by the challenges of pushing the cutting edge. These tools are forged by streamlining existing calculations, testing on toy theories such as QED or supersymmetric cousins of gravity, and direct forays into new calculations in Einstein's gravity. We highlight synergies between scattering amplitudes and other particle physics tools such as EFT and advanced multiloop integration techniques, as well as, with methods in general relativity such as gravitational self-force and the worldline formalism. 
\bigskip

\noindent \textbf{Effective Field Theory.} Compact binaries share many of the essential characteristics of bound states of elementary particles such as positronium, hydrogen, or quarkonia. Therefore, the nonrelativistic EFT techniques developed for QED and QCD~\cite{Caswell:1985ui,Isgur:1989vq,Bodwin:1994jh,Beneke:1997zp,Manohar:2000dt,Manohar:2000hj,Manohar:2006nz} are well-suited for application to compact binaries. This is underscored by the many landmark calculations that were enabled by the NRGR framework~\cite{Goldberger:2004jt}, see the dedicated 
Snowmass White Paper~\cite{NRGRWhitePaper2022} for recent developments on this subject. 
These physical systems are characterized by scales that define the following modes or regions:
\setlength{\tabcolsep}{10pt}
\renewcommand{\arraystretch}{1}
\begin{align}
\begin{tabular}{|c|c|}
\hline
hard &  $(m,m)$ \\ \hline
soft &  $\left({mv \over J}, {mv \over J} \right)$ \\[1pt] \hline
potential  & $\left( {mv^2 \over J}, {mv \over J} \right)$ \\[2pt] \hline
radiation  & $\left( {mv^2 \over J}, {mv^2 \over J} \right)$\\[2pt]
  \hline
\end{tabular}
\end{align}
Here $(E,p)$ denotes the scalings of the energy and three-momentum of the modes, $m$ is the mass of the binary constituents, $v$ is the relative velocity, and $J$ is the angular momentum.
As discussed in Sec.~\ref{classical}, the classical limit corresponds to the limit of large angular momentum. 

The new approach based on scattering amplitudes adapts these EFT tools, and closely related methods such as the method of regions~\cite{Beneke:1997zp}, to efficiently extract classical contributions from the various modes described above. In particular, power-counting and factorization allows contributions from potential and radiation modes to be systematically and separately considered in scattering amplitudes, and then resummed to all orders in the velocity $v$ using differential equations. 

Similar to applications in QED and QCD~\cite{Manohar:2000hj}, another basic role of EFT is to connect scattering amplitudes and classical binary dynamics through a matching calculation between gravity and a low energy EFT that describes particles interacting through an effective potential~\cite{Neill:2013wsa, Vaidya:2014kza, CRS}. For the case of conservative dynamics, the basic framework has now been applied and extended for application at higher orders~\cite{CRS, 3PM, 3PMLong, Bern:2021dqo, Bern:2021yeh}, for including spin and tidal effects~\cite{Cheung:2020sdj,Bern:2020buy,Kosmopoulos:2021zoq,Haddad:2020que,Cheung:2020gbf,Bern:2020uwk,Aoude:2020ygw}, and for gravitational theories with supersymmetry and in arbitrary dimensions~\cite{Caron-Huot:2018ape,Parra-Martinez:2020dzs,Cristofoli:2020uzm}. In Ref.~\cite{Bern:2021dqo}, this EFT matching procedure was used to derive the amplitude-action relation~\eqref{eq:AA}.
\bigskip

\noindent \textbf{Advanced Multiloop Integration.} 
The quest for high-precision calculations in particle physics, e.g. for colliders and for theoretical explorations of various field theories, has led to the development of advanced technology for integration of loop contributions. This can be transplanted for application to 
GWs, where the evaluation of integrals is a challenge common to all field-theory--based approaches, including traditional approaches in general relativity. 

Integration by parts reduction as implemented in automated programs, such as AIR~\cite{Anastasiou:2004vj},
FIRE~\cite{Smirnov:2008iw, Smirnov:2013dia}, Kira~\cite{Maierhoefer:2017hyi}, Reduze~\cite{Studerus:2009ye},  and LiteRED~\cite{Lee:2012cn, Lee:2013mka},
can be used to obtain a set of master integrals, which are then evaluated using the method of differential equations~\cite{Kotikov:1990kg, Bern:1993kr, Remiddi:1997ny, Gehrmann:1999as}, perhaps in canonical form~\cite{Henn:2013pwa, Henn:2013nsa}. The results are given in terms of 
multiple polylogarithms, their elliptic generalizations and perhaps more complicated functions. A judicious choice of variables is crucial~\cite{Landshoff:1969yyn, Parra-Martinez:2020dzs}. Integrals that appear in general-relativity--based approaches to PN calculations are also amenable to these methods, as was demonstrated in~\cite{Bini:2020uiq, Bini:2020rzn}. These methods mesh well with the expansion around the classical limit, and extraction of potential and radiation contributions as implemented through the method of regions~\cite{Beneke:1997zp}. 
\bigskip

\noindent \textbf{Generating Functions.} The radial action $I_r$, defined as the integral of the radial momentum along the trajectory, contains the entire classical 
central-field dynamics. Observables such as the scattering angle, redshift, and time delay, can be derived from the radial action via thermodynamic-type relations, e.g. 
$dI_r = \frac{\theta}{2\pi} dL + \tau dE + \sum_a \langle z_a\rangle dm_a$, where $\theta$ is the scattering angle, $L$ is the orbital angular 
momentum, $\tau$ is the time delay, and $\langle z\rangle$ is the redshift. Ref.~\cite{Bern:2021dqo} demonstrated a direct relation \eqref{eq:AA} between elastic four-point scattering amplitudes and the radial action for the scattering trajectory of the two particles. The radial action can thus be directly obtained from particular contributions to the four-point scattering amplitude~\cite{Bern:2021dqo, Bjerrum-Bohr:2021wwt}. Ref.~\cite{Damgaard:2021ipf} argued that the radial action constructed this way is related to the WKB approximation to the scattering process, which establishes a connection with the eikonal exponentiation of amplitudes~\eqref{eq:Eikonal}. The amplitude-action relation has also been used to explore non-perturbative results in the probe limit~\cite{Kol:2021jjc}.

From a field theory point of view, the eikonal function provides another generating function for observables.  
Indeed, the kinematics that selects the classical part of a 
four-point amplitude, $s\gg t = -q^2$, is the same as the Regge kinematics, so properties of amplitudes in this regime, such as its exponentiation 
in terms of the eikonal function, can be used to gain insight into classical physics. 
The original framework has been extended to capture the
exponentiation of amplitudes of spinning particles and of five-point amplitudes, under suitable assumptions regarding the 
coherence of the emitted gravitons~\cite{Cristofoli:2021jas}. While structurally similar, the radial action and the eikonal functions differ in the details of the definition of the exponential or, equivalently, in the definition of the iteration terms. 

\bigskip

\noindent \textbf{KMOC -- A Formalism for Observables from Amplitudes.}
Another approach to extracting classical physics from amplitudes focuses on determining physical observables which are well defined in 
both the classical and quantum theories. This observables-based approach was first discussed in~\cite{Kosower:2018adc}, and is sometimes 
referred to as the KMOC formalism. It is a quantum-first treatment of observables, which incorporates key aspects of classical physics from 
the beginning.
By arranging a system to be under the purview of the correspondence principle,  then quantum effects are negligible and 
it must be the case that the quantum mechanical computation of the observables will yield the classical result.

In the KMOC formalism observables are evaluated as expectation values of operators in the quantum field theory, $\langle \psi|{\cal O}|\psi\rangle $.  
Scattering amplitudes enter through time evolution which, given an initial state $\ket{\psi}$ in the far past, produces a final state $S \ket{\psi}$  in the 
far future, and scattering amplitudes are the matrix elements of $S$.
This strategy is somewhat reminiscent to QCD event shapes~\cite{Basham:1978bw, Basham:1978zq, Richards:1983sr, Korchemsky:1999kt, Dokshitzer:1999sh, Lee:2006fn, Bauer:2008dt}, such as the energy or charge flow~\cite{Belitsky:2001ij} and energy-energy correlators~\cite{Basham:1978bw, Basham:1978zq}. 

A whole range of observables can be computed using the KMOC formalism.  
A particularly simple example is the impulse.
The impulse is the change in the expectation value of the momentum operator $\Pop^\mu$ of the quantum field describing the massive particle and
is closely related to the classical potential. 
There is a direct connection between the impulse and the scattering angle: once the final momentum is known, it is straightforward to determine the 
direction of motion relative to the initial momentum. On the other hand if the scattering angle is known the final momentum can be reconstructed, 
assuming no energy is radiated away.

The amount of momentum radiated is itself another simple observable, arising as the expectation value of the momentum operator $\Kop^\mu$
of whatever field is transporting the momentum.  For example, in gravitational scattering, it is the momentum operator of the gravitational field.
Because conservation of momentum is built into the underlying quantum field theory,  the KMOC formalism naturally incorporates the 
effects of radiation reaction.
It also has natural generalizations that include the spin of particles ~\cite{Maybee:2019jus,Guevara:2019fsj,Arkani-Hamed:2019ymq} and 
capture the dynamics of color in the context of classical (infrared-free) Yang-Mills theories~\cite{delaCruz:2020bbn,delaCruz:2021gjp}. 

The gravitational waveform emitted during a scattering event follows directly from amplitudes~\cite{Cristofoli:2021vyo}; the observable of interest 
is the Riemann curvature operator $\Rop_{\mu\nu\rho\sigma}(x)$ in linearized gravity. The linearized approximation is appropriate as the 
GWs travel over extremely large spatial distances and retains only the leading term in inverse distance.
The waveform itself is an integral of an appropriate component of the expectation $\bra{\psi} S^\dagger \Rop_{\mu\nu\rho\sigma}(x) S \ket{\psi}$.
At lowest perturbative order, it is an integral of a five-point tree scattering amplitude. The relation with the intuitive picture in which a wave is composed of a large number of gravitons is established by describing the wave as a coherent state of the gravitational field.
Interestingly, vanishing variance in the classical limit \cite{Cristofoli:2021jas} implies 
that correlators of products of operators, 
${}_\text{out}\langle \psi| {\cal O}_1{\cal O}_2|\psi\rangle_\text{out}$ factorize as ${}_\text{out}\langle \psi| {\cal O}_1|\psi\rangle_\text{out}
{}_\text{out}\langle \psi|{\cal O}_2|\psi\rangle_\text{out}$; they nevertheless contain new classical information as demonstrated in~\cite{Manohar:2022dea}.
\bigskip

\noindent \textbf{Amplitude Building-Blocks in the Classical Limit.}
Through generalized unitarity, tree-level amplitudes are the building blocks for all higher-order calculations. The application of scattering amplitudes to 
GWs has motivated reorienting the structure of these building blocks to take advantage of simplifications in the classical limit.
A number of approaches are based on heavy-particle effective theory~\cite{Damgaard:2019lfh}, and the double-copy in such theories~\cite{Brandhuber:2021kpo,Brandhuber:2021eyq,Brandhuber:2021bsf}, which build on ongoing efforts to understand the kinematic algebra behind color/kinematics duality, see e.g.~\cite{Chen:2019ywi, Chen:2021chy}. Other approaches focus on reorganizing the soft expansion to reduce the number of master integrals by combining terms related by permutations of graviton legs~\cite{Bjerrum-Bohr:2021vuf}. 

Another approach uses generalized unitarity to first construct gauge-theory amplitudes and then applies the double-copy to obtain the corresponding gravitational amplitudes~\cite{Carrasco:2020ywq}. Various methods are available to project out the dilaton and axion~\cite{Johansson:2014zca, Luna:2017dtq} that naturally appear in this approach. It will be interesting to explore how to take the classical limit before the double-copy, and use gauge theory tree-level amplitudes adapted for this purpose.

Many developments focused on amplitudes for higher-spin particles. The amplitudes obtained in the classical limit avoid the strong constraints imposed by no-go theorems~\cite{Shamaly:1972zu, Hortacsu:1974bm, Deser:2000dz, Porrati:2008gv, Camanho:2014apa, Afkhami-Jeddi:2018apj}. 
Ref.~\cite{Arkani-Hamed:2017jhn} constructed three-point amplitudes for arbitrary-spin massive particles through standard amplitudes methods and, by demanding good high-energy properties, identified the so-called ``minimal amplitudes" which correspond to the linear response of Kerr black holes to an external gravitational field~\cite{Vines:2017hyw}.
Using Lagrangian methods, Ref.~\cite{Bern:2020buy, spin5} constructed an EFT that described the classical interactions of higher-spin fields 
at large impact parameter, $|\bm b|\gg |\bm p|^{-1}$, which includes arbitrary spin-induced multipoles and whose three-point amplitudes exhibit a 
double-copy relation to higher-spin fields coupled with gluons.  
The heavy-particle approach in Ref.~\cite{Damgaard:2019lfh} was extended for higher-spin fields with double-copy properties in~\cite{Aoude:2020onz, Haddad:2020tvs, spin5competition}.
For suitable values of the parameters, the three-point amplitudes following from these Lagrangians reproduce the Kerr stress tensor~\cite{Vines:2017hyw}.
A gravitational Compton amplitude for spin-$5/2$ massive particles that is free of spurious poles was found in~\cite{Chiodaroli:2021eug}, 
demonstrating that the methods employed can produce consistent higher-point amplitudes for higher-spin particles.
Another construction~\cite{Bjerrum-Bohr:2020syg, Bjerrum-Bohr:2019nws} makes use of the scattering equation formalism to provide 
diagrammatic and recursive tools for finding covariant expressions for $D$-dimensional tree-level $n$-point amplitudes with pairs of spinning 
massive particles.

\bigskip

\noindent \textbf{Synergy with Traditional Approaches.} The new approach based on the tools of theoretical high-energy physics has benefited immensely from traditional methods, in terms of both conceptual guidance from existing frameworks and practical guidance from explicit results. In turn, the emergence of new results derived from scattering amplitudes has spurred new developments within traditional methods, as well as, hybrid approaches that meld traditional methods and particle-physics tools. Such an extensive exchange between various approaches in general relativity and particle physics has led to the rapid advance of recent years, and will continue to drive breakthroughs in the future.

For instance, one of the main tools for describing binary dynamics within the PN approach to general relativity is the worldline approach~\cite{Goldberger:2004jt, Foffa:2019yfl, Foffa:2019rdf, Blumlein:2019zku, Foffa:2019hrb, Foffa:2011ub, Gilmore:2008gq, Levi:2018nxp, Kol:2007bc}. Recently, inspired by amplitudes methods, this has been extended to a PM framework~\cite{Kalin:2020mvi} to derive results to all orders in velocity. Calculations in this approach share many similarities with amplitudes-based methods, such as the use of integration technology imported from particle physics. Other worldline-based approaches having aspects of scattering amplitudes have also been recently developed for calculations of tail effects in a nonrelativistic EFT for a binary~\cite{Edison:2022cdu} and of PM observables~\cite{Jakobsen:2022fcj, Jakobsen:2021smu, Mogull:2020sak}, including spin~\cite{Jakobsen:2022fcj, Jakobsen:2021zvh, Cho:2021mqw} and radiative~\cite{Jakobsen:2022fcj, Jakobsen:2021lvp} effects.

As discussed in Sec.~\ref{structures}, new results from scattering amplitudes have exposed a number of theoretical structures. One feature that was hidden in traditional non-relativistic calculations was the simple mass dependence of scattering amplitudes, or equivalently the radial action and scattering angle. Indeed, it was noticed~\cite{Vines:2018gqi}, and then rigorously proved~\cite{Damour:2019lcq}, that the scattering angle exhibits a particular dependence in the symmetric mass ratio $m_1m_2/(m_1+m_2)^2$, leading to the so-called ``good mass polynomiality" rule. The latter is the basis  of the ``Tutti-Frutti" method \cite{BDG1,BDG2,BDG3,Damour:2019lcq}, which 
establishes the overlap between the self-force expansion and the PM expansion. This powerful method has derived a number of new results, which have in turn provided important guidance for new amplitudes-based calculations. The ``Tutti-Frutti" method has also been extended for systems with spins~\cite{Antonelli:2020aeb,Antonelli:2020ybz,Khalil:2021fpm}.

The B2B map is another instance in which patterns revealed from explicit amplitudes calculations~\cite{3PM, 3PMLong} were combined with the application of analytic continuation~\cite{Cho:2018upo} to develop efficient new tools~\cite{Kalin:2019rwq, Kalin:2019inp, Cho:2021arx}.
%


\section{Future directions and challenges}

In this section we summarize several broad directions that are as essential for long-term progress and fruitful synergy between particle and 
GW physics.
\bigskip

\noindent
{\bf New Perspectives on Observables.} 
Scattering amplitudes are naturally defined by data at asymptotic infinity and can offer insights on this class of observables in general relativity, which can be subtle due to general coordinate invariance. Of particular significance are questions regarding radiative contributions, which are not only of theoretical interest, but also of practical importance for obtaining high-precision waveforms. 

One open question is: how are observables in bound and unbound orbits generally related? In the absence of radiative effects, both types of dynamics can be straightforwardly derived from the same Hamiltonian and are related by analytic continuation. However, this is no longer true in the presence of radiative effects, which introduce correlations over arbitrary time intervals, e.g., due to tail effects~\cite{Thorne:1980ru,Blanchet:1987wq, Blanchet:1993ec}. This leads to a nonlocal-in-time Hamiltonian that depends on the particular trajectory~\cite{Damour:2014jta} (see also~\cite{Foffa:2011np} for a more recent perspective). This non-universality prevents a simple analytic continuation~\cite{Cho:2021arx}. 

It is imperative to address this question in order to fully leverage scattering amplitudes, which are naturally associated with unbound motion, for describing the dynamics of bound compact objects. A complete framework for connecting bound and unbound trajectories will not only apply to conservative dynamics but also to the radiated energy, momentum, and angular momentum, and perhaps to the waveforms themselves. Interestingly, there is evidence for double copy in 
GW emission in both bound and unbound systems~\cite{Goldberger:2016iau, Goldberger:2017vcg, Shen:2018ebu}.

Relatedly, it would be useful to carefully formulate important concepts such as conservative vs. dissipative dynamics, as well as inclusive vs. exclusive or IR safe vs. unsafe observables. For instance, recent differences in the conservative dynamics at ${\cal O}(G^4)$ are due to difficulties in precisely defining conservative dynamics in the presence of radiative effects. To this end, it may be fruitful to compute new observables, beyond those traditionally considered in binary dynamics, that can serve as toy models for understanding these issues. 
Such new observables may also be of phenomenological interest. 
For example, it would be interesting to detect waveforms from hyperbolic encounters. Although such events are expected to be rare and have a short-duration signal, their theoretical predictions are well-understood and can be obtained directly with scattering amplitude methods without need of analytic continuation.

Observables and unexpected structures may be further revealed by formulations of gravitational interactions without direct reference to geometry, in analogy with structure exposed by formulations of quantum field theory without quantum fields. The classical double copy~\cite{Monteiro2014cda} provides an example, as do recent attempts to define the Kerr black hole from a QFT perspective \cite{spin5, spin5competition}. While currently perturbative, such theoretical tools may help fully exploit 
GW observations in the quest to understand gravitating compact objects and black-hole horizons.

\bigskip

\noindent
{\bf Still-Higher Perturbative Orders.} The precision demands of future experiments~\cite{KAGRA:2013rdx,Saleem:2021iwi,Reitze:2019iox,Punturo:2010zz,LISA:2017pwj} are quite challenging since, depending on the source, GWs from binary systems will be observed with a signal-to-noise ratio that is one or two orders of magnitude higher than with current detectors~\cite{LIGOScientific:2014pky,VIRGO:2014yos,KAGRA:2020agh}. To keep the modeling systematics below the expected statistical errors~\cite{Favata:2013rwa, Samajdar:2018dcx,Purrer:2019jcp,Huang:2020pba,Gamba:2020wgg}, the accuracy of current state-of-the-art waveform models~\cite{Hinderer:2016eia,Dietrich:2017aum,Nagar:2018plt,Nagar:2018zoe,Cotesta:2018fcv,Varma:2018mmi,Dietrich:2019kaq,Varma:2019csw,Ossokine:2020kjp,Pratten:2020ceb,Thompson:2020nei,Matas:2020wab,Estelles:2021gvs,Gamba:2021ydi}, which are built from combining analytic and numerical relativity, needs to be improved by, say, two orders of magnitude. This would require up to ${\cal O}(G^7)$ and/or ${\cal O}(v^{12})$, as shown in Figure~\ref{fig:mapPT} for the potential. Moreover, higher-order calculations are needed for the dissipative sector, the multipolar waveforms, and other physical effects such as spins, tides, and eccentricity.

While amplitude-based methods are powerful, achieving this will still require a significant build-up of technology over several years of cycling between performing state-of-the-art calculations and developing new tools to solve bottlenecks.
For example, current pipelines can be significantly improved by streamlining the construction of classical scattering amplitudes and efficiently removing iteration terms that are already determined by lower-order calculations. 
In the same spirit, it will be important to exploit the nontrivial interplay between different graph topologies that lead to vast simplifications in the integrand, as demonstrated in~\cite{Akhoury:2013yua}
and further explored in the context of 
GW physics in~\cite{Brandhuber:2021eyq, Brandhuber:2021kpo}.
Similarly, it will be interesting to understand and utilize the consequences of the infinite number of relationships between multi-point, multi-loop amplitudes 
in the classical limit~\cite{Cristofoli:2021jas}, which suggest that various terms in higher-point and/or higher loop amplitudes 
can be predicted from lower loop, lower point amplitudes. It also would be interesting to examine whether this phenomenon could be useful in contexts 
outside of 
GW physics, for example in collider experiments.

Apart from applications to precision 
GW physics, higher-order calculations also bring new perspectives on known phenomena. For example, the calculation of conservative binary dynamics at ${\cal O}(G^5)$ including radiative effects can give a better understanding of the role of 
GW memory in binary dynamics.

Higher-order calculations contain a wealth of information about the analytic structure of the 
two-body Hamiltonian and of observables. Up to rational functions, they are given by the nontrivial loop Feynman integrals that 
can appear at a given order in perturbation theory.  A classification of the functionally independent integrals can directly constrain observables from a general knowledge of their analytic properties or their behavior in various limits.
Moreover, these functions can be matched to NR simulations, which, for the time being, provide the most accurate description of the final moments of a binary inspiral. This fitting procedure may offer further analytic insight into this period of the binary's evolution and, conversely, on analytic approaches to nonperturbative properties of gravitational interactions. 
\bigskip

\noindent
{\bf Integration Challenges.}
Evaluation of higher-loop Feynman-type integrals is a challenge shared by QFT applications to both particle physics and to gravitational 
and 
GW physics.  

The integration-by-parts (IBP) reduction~\cite{Chetyrkin:1981qh, Laporta:2000dsw, Laporta:1996mq, Smirnov:2008iw, 
Maierhoefer:2017hyi, Studerus:2009ye, Lee:2012cn, Lee:2013mka} used to reduce the expanded Feynman integrals to 
basis of integrals, referred to as master integrals, becomes computationally demanding as the number of loops increase. 
Cutting edge developments that increase the efficiency of IBP reduction  \cite{Gluza:2010ws, Schabinger:2011dz, Ita:2015tya, 
Georgoudis:2016wff, Ita:2016oar, Zhang:2016kfo, Bern:2017gdk, Mastrolia:2018uzb, Frellesvig:2019uqt}, 
perhaps with further improvements related to the specific structure 
of the integrals that appear in the classical limit, will be vital for field theory-based approaches to reach high perturbative orders 
in the post-Minkowskian and the post-Newtonian expansion demanded by future observations. 
Further improvements may come from the development of an algorithmic choice of master integral basis tailored for the 
two-body general-relativity problem.

The master integrals relevant for the two-body problem depend only on the relative velocity of the two bodies and can be evaluated through a system of linear first-order 
ODEs~\cite{Kotikov:1990kg, Bern:1993kr, Remiddi:1997ny, Gehrmann:1999as} in this velocity. Expansion in the static limit, $v\rightarrow 0$, 
yields the integrals that appear in the post-Newtonian expansion and the leading terms serve as boundary conditions 
for the ODE system. The differential equations themselves simplify when a ``canonical basis'' of master integrals is 
chosen~\cite{Henn:2013pwa, Henn:2013nsa}. New algorithms  for finding such canonical bases, improving upon those 
in~\cite{Lee:2014ioa, Argeri:2014qva, Henn:2014qga, Gituliar:2016vfa, Gituliar:2017vzm, Prausa:2017ltv, Chicherin:2018old, Dlapa:2020cwj, Chen:2020uyk, Dlapa:2021qsl}, and the study of the special functions 
that solve the resulting differential equations \cite{Brown:2009ta, Brown:2009rc, BrownArxiv, Panzer:2014caa, Broedel:2017kkb, Bourjaily:2018yfy} will be important for further progress, see also the dedicated Snowmass White Paper~\cite{integrationWhitePaper2022}.
For specific practical applications it will also be profitable to explore other methods for evaluating master integrals, such as 
Mellin-Barnes representations~\cite{Smirnov:2004ym}, direct Feynman parameter integration~\cite{Panzer:2014caa, Bourjaily:2018aeq}, and difference equations 
from dimensional recurrence \cite{Tarasov:1996br, Lee:2009dh, Lee:2012te}.
These methods apply equally well to conservative calculations and to the calculation of observables involving outgoing gravitational radiation, 
such as the total energy radiated during a scattering event. The collider method of reverse unitarity \cite{Anastasiou:2002yz, Anastasiou:2002qz, 
Anastasiou:2003yy, Anastasiou:2015yha} allows systematic evaluation of the phase space integrals appearing in the KMOC formalism 
with techniques similar to those used for loop integrals, as demonstrated in~\cite{Herrmann:2021tct,Herrmann:2021lqe}.

\bigskip

\noindent
{\bf Higher-Order Structure, Resummation, and Beyond Perturbation Theory.}
Exact results in interacting theories, even in the classical regime, are rare.  Patterns and structure exposed by explicit higher perturbative orders 
may hold the key to understanding the structure of perturbation theory and eventually lead to its resummation, as illustrated e.g. by the BDS/ABDK 
conjectures \cite{Anastasiou:2003kj, Bern:2005iz} for the resummation of the planar MHV amplitudes in ${\cal N}=4$ super-Yang-Mills theory.
This is an important theoretical justification, which complements the practical one emphasized earlier, for further in-depth higher-order exploration.
The complete velocity dependence gives us access to the analytic structure of the Hamiltonian -- and of the observables derived from it -- and thus 
can inform on the structure at even higher orders, possibly providing sufficient information to resum the entire perturbation theory.

A proof of principle is the conjectured expression relating scattering observables of spinning bodies to the eikonal of the corresponding 
scattering amplitude \cite{Bern:2020buy}. While expected from the motion of spinless bodies, the existence a single function -- the eikonal, or
the radial action -- that potentially captures all conservative classical observables of such processes is surprising. 
Tests of such conjectures and the identification of novel, bolder ones may exploit special configurations of particles, new perturbative expansions 
which access different regions of parameter space, etc.

When present, symmetries are powerful means to extrapolate fixed-order properties of observables. The double copy hints at a relationship between the asymptotic symmetry group of Yang-Mills theory and of general relativity. Since the expectation value of the field itself can be computed (e.g., using scattering amplitudes and the KMOC formalism), it may be possible to test this idea directly.  Quantum mechanical deformations of the symmetry group, relating for example to infrared divergences, should be accessible by studying appropriate observables and may reveal all-order structures. 

Another path to exact results proceeds through enhancements of perturbative calculations.
For instance, the EOB approach~\cite{Buonanno:1998gg,Buonanno:2000ef} provides a framework that can incorporate not only perturbative results from PN, PM and gravitational self-force but also  exact results available in the probe limit. In general, it is critical that information from distinct
regions of parameter space and from different physical phenomena be included in the same formalism. 
The amount of physics contained in all-order results is remarkable and includes, among others, possible analytic access to horizon 
formation in the binary coalescence.

Yet another systematic strategy is the gravitational self-force approach~\cite{Mino:1996nk,Quinn:1996am, Barack:2018yvs}, in which one expands in the mass ratio of the binary. It makes contact with QFT in curved space, which we will return to shortly. It is a challenge for the future to develop the tools necessary to carry out such calculations with QFT  methods.

Apart from having been instrumental for driving our understanding of gravity, iconic classical solutions of general relativity such as the 
Schwarzschild and Kerr solutions also are at the foundation of the EOB theory and of the self-force approach. 
Scattering amplitudes and the KMOC formalism applied to three-point amplitudes~\cite{Monteiro:2020plf} can access these via analytic
continuation either to complex momenta or to spacetimes with an exotic signature. This led, for example, to a new understanding of the 
Newman-Janis shift~\cite{Arkani-Hamed:2019ymq} which relates the Kerr solution in a very simple way to the Schwarzschild solution.
Extending this observation beyond leading order should establish a connection between scattering amplitudes and higher-order terms 
in the self-force expansion to all orders in Newton's constant, beyond the ``good mass polynomiality" already manifest in perturbative 
scattering amplitudes and associated observables.
It could also be interesting to examine this connection in higher dimensions where the space of known classical solutions if far richer 
than in four dimensions.

\bigskip

\noindent
{\bf Many-Body Dynamics from Amplitudes.} 
While there is a large effort directed towards understanding compact binaries, systems of three or more massive bodies also occur in our Universe and may source 
GWs detectable in the future. These systems may probe scenarios of sequential and hierarchical mergers~\cite{Naoz:2012bx,Lim:2020cvm,Martinez:2020lzt,Fragione:2020gly}, and have qualitatively different dynamics~\cite{Boekholt:2021ifv}, including chaotic dynamics with interesting effects at relatively high velocities~\cite{Zwart:2021qxe}. The current state-of-the-art Hamiltonian is the 2PN order~\cite{SCHAFER1987336, Chu:2008xm, Will:2013cza}  obtained through general-relativity methods. A formal 2PM expression was obtained in~\cite{Loebbert:2020aos} using the worldline formalism and can be used to generate higher-order PN terms of the form $G^2 v^{2n}$. 

Scattering amplitude methods used for two-body dynamics can also be applied for $N$-body dynamics, and exploring this may uncover novel features of gravitational interactions. This will require extending tools for constructing $2N$-point matter amplitudes, identifying the classical limit, evaluation of integrals, etc. Further assumptions about the hierarchy of scales in $N$-body dynamics establish connections with multi-Regge kinematics in particle physics. 
\bigskip

\noindent {\bf Ringdown from Amplitudes.} The focus of the recent effort to understand and explore binary coalescence with particle physics methods has been the inspiral phase of the process,
which can be treated perturbatively around flat space. The 
GW signal contains information about the masses and spins of the constituents, the shape of the orbit (eccentricity and mean anomaly), 
and in late stages also about their tidal deformability. 
For binary black holes, the final phase of a merger, known as the ringdown, appears as a superposition of quasi-normal modes of the merged remnant~\cite{Vishveshwara:1970cc,Press:1971wr}. The frequency and decay time 
of each mode are determined in general relativity by the remnant’s mass and spin, while their relative amplitude 
contains information about the properties of the binary's components and geometry, including the spin orientations (see, e.g.,~\cite{Kamaretsos:2011um, Hughes:2019zmt, Li:2021wgz}).

This poses an interesting challenge and an equally interesting opportunity: to develop field theory methods to describe the emission of gravitons from 
the remnant of a binary merger. 
A naive expectation is that the relevant framework is QFT in a time-dependent curved space which is perhaps a 
classical double copy and that the relevant observables are final-state correlation functions akin to cosmological correlators. 
While still in its infancy, see e.g.~\cite{Adamo:2017sze, Adamo:2018mpq, Adamo:2020qru, Adamo:2020yzi, Adamo:2021rfq, Diwakar:2021juk, Cheung:2022pdk, Herderschee:2022ntr, Arkani-Hamed:2018kmz, Baumann:2020dch, Benincasa:2018ssx} and the Snowmass White Paper \cite{Baumann:2022jpr}, 
the generalization of the amplitudes program to curved space together with the gauge theory realization of classical solutions of general 
relativity \cite{Monteiro2014cda, Luna2015paa,  Luna2016due, Luna2016hge, CarrilloGonzalez:2019gof, 
CarrilloGonzalez:2018ejf, Carrillo-Gonzalez:2017iyj} may offer a new perspective on the ringdown phase, as well as, new means to explore
analytically the correlation between initial parameters and intrinsic properties of the remnant, including its tidal deformability, etc.
Clearly, a systematic development of these methods will have wider applications, establishing connections to other areas of physics, such as gauge/string duality.
Curved space methods will also make contact with the self-force approach to the two-body problem~\cite{Mino:1996nk,Quinn:1996am, Barack:2018yvs},  which can be interpreted as the back-reacted propagation of a light massive particle in the curved space generated by a heavy massive particle,  thus providing another path to analytic resummation of the two-body Hamiltonian or two-body observables. See \cite{Adamo:2022rmp} for a related application of the KMOC formalism.

\bigskip

\noindent
{\bf Probes of General Relativity, Quantum Gravity and New Physics.}
Quantum gravity remains one of the biggest mysteries of fundamental physics of this century. While indirect evidence for it abound, its precise formulation 
is elusive, and there is currently no observed deviations from predictions of classical general relativity in 
GW observations~\cite{LIGOScientific:2021sio}. Future more precise
observations with also a much larger number of events will provide a means for more efficient probes of physics beyond general relativity~\cite{Barack:2018yly}, or even of quantum gravity and 
horizon-scale physics (e.g.,~\cite{Mathur:2005zp, Mathur:2008nj, Almheiri:2012rt, Almheiri:2013hfa}); see also the 
Snowmass White Paper~\cite{Berti:2022wzk} on fundamental physics and beyond the standard model.
More accurate waveforms, covering long-time evolution, as well as various possibilities for new physics, is a step towards this goal, as are 
the identification of new observables tailored for such physics.
A complementary approach is the development of model-independent studies, covering both classical extensions of general relativity and its quantization. From an EFT perspective there is of course a certain overlap between them, so identifying methods to lift this degeneracy 
will be important. 

Further developments in amplitude methods, e.g. extending the work in~\cite{Carrasco:2021ptp, Bonnefoy:2021qgu, Chi:2021mio,Endlich:2017tqa} to systematically include and classify all possible higher-dimension/higher-derivative operators that can appear as counterterms in gravitational theories, including those discussed using standard GR methods in e.g.~\cite{deRham:2020ejn,Cano:2020cao,Cano:2021myl}, will aid both model-specific and model-independent studies of 
GW signals, both with regard to the inspiral and the ringdown phase, see also Snowmass White Paper~\cite{deRham:2022hpx} on UV constraints and IR physics. 


\section{Coda}

In its short history, the program to apply particle physics methods to gravitational physics in general and 
gravitational-wave 
physics in particular has achieved remarkable success by uncovering rich theoretical structures, developing powerful new tools, and producing predictions for future precision gravitational-wave detectors. 
Close synergy with the EFT and 
gravitational-wave 
communities has been essential, and 
provided crucial guidance regarding the theoretical needs for waveform modeling and the necessary tools 
to achieve them.

This white paper provides a snapshot of the current status of this vibrant field and of its progress 
to date.
There are many open questions and avenues to explore, ranging from the very formal to the very practical. Some
can be addressed with existing methods whose full potential is not yet realized, while others require novel ideas that may even open up completely unexpected directions that will further our knowledge of gravitational interactions, binary dynamics, black-hole physics and perhaps even quantum field theory.

\bigskip
\bigskip

\section*{Acknowledgements}
D.O.C. is supported by the U.K. Science and Technology Facilities Council (STFC) grant ST/P000630/1.
R.R. is supported by the US Department of Energy under Grant No. DE-SC00019066.
M.P.S. is grateful to the Mani L. Bhaumik Institute for Theoretical Physics for support.
M.Z. is supported by the U.K.\ Royal Society through Grant URF{\textbackslash}R1{\textbackslash}20109.

\bibliographystyle{JHEP}
\bibliography{bibliography}

\providecommand{\href}[2]{#2}\begingroup\raggedright\begin{thebibliography}{100}

\bibitem{LIGOScientific:2014pky}
{\scshape LIGO Scientific} collaboration, \emph{{Advanced LIGO}},
  \href{https://doi.org/10.1088/0264-9381/32/7/074001}{\emph{Class. Quant.
  Grav.} {\bfseries 32} (2015) 074001}
  [\href{https://arxiv.org/abs/1411.4547}{{\ttfamily 1411.4547}}].

\bibitem{VIRGO:2014yos}
{\scshape VIRGO} collaboration, \emph{{Advanced Virgo: a second-generation
  interferometric gravitational wave detector}},
  \href{https://doi.org/10.1088/0264-9381/32/2/024001}{\emph{Class. Quant.
  Grav.} {\bfseries 32} (2015) 024001}
  [\href{https://arxiv.org/abs/1408.3978}{{\ttfamily 1408.3978}}].

\bibitem{KAGRA:2020agh}
{\scshape KAGRA} collaboration, \emph{{Overview of KAGRA: Calibration, detector
  characterization, physical environmental monitors, and the geophysics
  interferometer}}, \href{https://doi.org/10.1093/ptep/ptab018}{\emph{PTEP}
  {\bfseries 2021} (2021) 05A102}
  [\href{https://arxiv.org/abs/2009.09305}{{\ttfamily 2009.09305}}].

\bibitem{Antonelli:2019ytb}
A.~Antonelli, A.~Buonanno, J.~Steinhoff, M.~van~de Meent and J.~Vines,
  \emph{{Energetics of two-body Hamiltonians in post-Minkowskian gravity}},
  \href{https://doi.org/10.1103/PhysRevD.99.104004}{\emph{Phys. Rev. D}
  {\bfseries 99} (2019) 104004}
  [\href{https://arxiv.org/abs/1901.07102}{{\ttfamily 1901.07102}}].

\bibitem{Khalil:2022}
M.~Khalil, A.~Buonanno, J.~Steinhoff and J.~Vines, \emph{{In preparation}}, .

\bibitem{Regge:1957td}
T.~Regge and J.~A. Wheeler, \emph{{Stability of a Schwarzschild singularity}},
  \href{https://doi.org/10.1103/PhysRev.108.1063}{\emph{Phys. Rev.} {\bfseries
  108} (1957) 1063}.

\bibitem{Zerilli:1970se}
F.~J. Zerilli, \emph{{Effective potential for even parity Regge-Wheeler
  gravitational perturbation equations}},
  \href{https://doi.org/10.1103/PhysRevLett.24.737}{\emph{Phys. Rev. Lett.}
  {\bfseries 24} (1970) 737}.

\bibitem{Vishveshwara:1970cc}
C.~V. Vishveshwara, \emph{{Stability of the schwarzschild metric}},
  \href{https://doi.org/10.1103/PhysRevD.1.2870}{\emph{Phys. Rev. D} {\bfseries
  1} (1970) 2870}.

\bibitem{Teukolsky:1973ha}
S.~A. Teukolsky, \emph{{Perturbations of a rotating black hole. 1. Fundamental
  equations for gravitational electromagnetic and neutrino field
  perturbations}}, \href{https://doi.org/10.1086/152444}{\emph{Astrophys. J.}
  {\bfseries 185} (1973) 635}.

\bibitem{Buonanno:1998gg}
A.~Buonanno and T.~Damour, \emph{{Effective one-body approach to general
  relativistic two-body dynamics}},
  \href{https://doi.org/10.1103/PhysRevD.59.084006}{\emph{Phys. Rev. D}
  {\bfseries 59} (1999) 084006}
  [\href{https://arxiv.org/abs/gr-qc/9811091}{{\ttfamily gr-qc/9811091}}].

\bibitem{Buonanno:2000ef}
A.~Buonanno and T.~Damour, \emph{{Transition from inspiral to plunge in binary
  black hole coalescences}},
  \href{https://doi.org/10.1103/PhysRevD.62.064015}{\emph{Phys. Rev. D}
  {\bfseries 62} (2000) 064015}
  [\href{https://arxiv.org/abs/gr-qc/0001013}{{\ttfamily gr-qc/0001013}}].

\bibitem{Blanchet:2013haa}
L.~Blanchet, \emph{{Gravitational Radiation from Post-Newtonian Sources and
  Inspiralling Compact Binaries}},
  \href{https://doi.org/10.12942/lrr-2014-2}{\emph{Living Rev. Rel.} {\bfseries
  17} (2014) 2} [\href{https://arxiv.org/abs/1310.1528}{{\ttfamily
  1310.1528}}].

\bibitem{Porto:2016pyg}
R.~A. Porto, \emph{{The effective field theorist\textquoteright{}s approach to
  gravitational dynamics}},
  \href{https://doi.org/10.1016/j.physrep.2016.04.003}{\emph{Phys. Rept.}
  {\bfseries 633} (2016) 1} [\href{https://arxiv.org/abs/1601.04914}{{\ttfamily
  1601.04914}}].

\bibitem{Schafer:2018kuf}
G.~Sch\"afer and P.~Jaranowski, \emph{{Hamiltonian formulation of general
  relativity and post-Newtonian dynamics of compact binaries}},
  \href{https://doi.org/10.1007/s41114-018-0016-5}{\emph{Living Rev. Rel.}
  {\bfseries 21} (2018) 7} [\href{https://arxiv.org/abs/1805.07240}{{\ttfamily
  1805.07240}}].

\bibitem{Barack:2018yvs}
L.~Barack and A.~Pound, \emph{{Self-force and radiation reaction in general
  relativity}}, \href{https://doi.org/10.1088/1361-6633/aae552}{\emph{Rept.
  Prog. Phys.} {\bfseries 82} (2019) 016904}
  [\href{https://arxiv.org/abs/1805.10385}{{\ttfamily 1805.10385}}].

\bibitem{Pretorius:2005gq}
F.~Pretorius, \emph{{Evolution of binary black hole spacetimes}},
  \href{https://doi.org/10.1103/PhysRevLett.95.121101}{\emph{Phys. Rev. Lett.}
  {\bfseries 95} (2005) 121101}
  [\href{https://arxiv.org/abs/gr-qc/0507014}{{\ttfamily gr-qc/0507014}}].

\bibitem{Campanelli:2005dd}
M.~Campanelli, C.~O. Lousto, P.~Marronetti and Y.~Zlochower, \emph{{Accurate
  evolutions of orbiting black-hole binaries without excision}},
  \href{https://doi.org/10.1103/PhysRevLett.96.111101}{\emph{Phys. Rev. Lett.}
  {\bfseries 96} (2006) 111101}
  [\href{https://arxiv.org/abs/gr-qc/0511048}{{\ttfamily gr-qc/0511048}}].

\bibitem{Baker:2005vv}
J.~G. Baker, J.~Centrella, D.-I. Choi, M.~Koppitz and J.~van Meter,
  \emph{{Gravitational wave extraction from an inspiraling configuration of
  merging black holes}},
  \href{https://doi.org/10.1103/PhysRevLett.96.111102}{\emph{Phys. Rev. Lett.}
  {\bfseries 96} (2006) 111102}
  [\href{https://arxiv.org/abs/gr-qc/0511103}{{\ttfamily gr-qc/0511103}}].

\bibitem{Foucart:2022iwu}
F.~Foucart, P.~Laguna, G.~Lovelace, D.~Radice and H.~Witek, \emph{{Snowmass2021
  Cosmic Frontier White Paper: Numerical relativity for next-generation
  gravitational-wave probes of fundamental physics}},
  \href{https://arxiv.org/abs/2203.08139}{{\ttfamily 2203.08139}}.

\bibitem{LISA:2017pwj}
{\scshape LISA} collaboration, \emph{{Laser Interferometer Space Antenna}},
  \href{https://arxiv.org/abs/1702.00786}{{\ttfamily 1702.00786}}.

\bibitem{Saleem:2021iwi}
M.~Saleem et~al., \emph{{The science case for LIGO-India}},
  \href{https://doi.org/10.1088/1361-6382/ac3b99}{\emph{Class. Quant. Grav.}
  {\bfseries 39} (2022) 025004}
  [\href{https://arxiv.org/abs/2105.01716}{{\ttfamily 2105.01716}}].

\bibitem{Reitze:2019iox}
D.~Reitze et~al., \emph{{Cosmic Explorer: The U.S. Contribution to
  Gravitational-Wave Astronomy beyond LIGO}}, {\emph{Bull. Am. Astron. Soc.}
  {\bfseries 51} (2019) 035}
  [\href{https://arxiv.org/abs/1907.04833}{{\ttfamily 1907.04833}}].

\bibitem{Punturo:2010zz}
M.~Punturo et~al., \emph{{The Einstein Telescope: A third-generation
  gravitational wave observatory}},
  \href{https://doi.org/10.1088/0264-9381/27/19/194002}{\emph{Class. Quant.
  Grav.} {\bfseries 27} (2010) 194002}.

\bibitem{Sathyaprakash:2019yqt}
B.~S. Sathyaprakash et~al., \emph{{Extreme Gravity and Fundamental Physics}},
  \href{https://arxiv.org/abs/1903.09221}{{\ttfamily 1903.09221}}.

\bibitem{Maggiore:2019uih}
M.~Maggiore et~al., \emph{{Science Case for the Einstein Telescope}},
  \href{https://doi.org/10.1088/1475-7516/2020/03/050}{\emph{JCAP} {\bfseries
  03} (2020) 050} [\href{https://arxiv.org/abs/1912.02622}{{\ttfamily
  1912.02622}}].

\bibitem{Kalogera:2021bya}
V.~Kalogera et~al., \emph{{The Next Generation Global Gravitational Wave
  Observatory: The Science Book}},
  \href{https://arxiv.org/abs/2111.06990}{{\ttfamily 2111.06990}}.

\bibitem{Berti:2022wzk}
E.~Berti, V.~Cardoso, Z.~Haiman, D.~E. Holz, E.~Mottola, S.~Mukherjee et~al.,
  \emph{{Snowmass2021 Cosmic Frontier White Paper: Fundamental Physics and
  Beyond the Standard Model}},  in \emph{{2022 Snowmass Summer Study}}, 3,
  2022, \href{https://arxiv.org/abs/2203.06240}{{\ttfamily 2203.06240}}.

\bibitem{Damour:2014afa}
T.~Damour, F.~Guercilena, I.~Hinder, S.~Hopper, A.~Nagar and L.~Rezzolla,
  \emph{{Strong-Field Scattering of Two Black Holes: Numerics Versus
  Analytics}}, \href{https://doi.org/10.1103/PhysRevD.89.081503}{\emph{Phys.
  Rev. D} {\bfseries 89} (2014) 081503}
  [\href{https://arxiv.org/abs/1402.7307}{{\ttfamily 1402.7307}}].

\bibitem{Ossokine:2017dge}
S.~Ossokine, T.~Dietrich, E.~Foley, R.~Katebi and G.~Lovelace, \emph{{Assessing
  the Energetics of Spinning Binary Black Hole Systems}},
  \href{https://doi.org/10.1103/PhysRevD.98.104057}{\emph{Phys. Rev. D}
  {\bfseries 98} (2018) 104057}
  [\href{https://arxiv.org/abs/1712.06533}{{\ttfamily 1712.06533}}].

\bibitem{Abbott:2016blz}
{\scshape LIGO Scientific, Virgo} collaboration, \emph{{Observation of
  Gravitational Waves from a Binary Black Hole Merger}},
  \href{https://doi.org/10.1103/PhysRevLett.116.061102}{\emph{Phys. Rev. Lett.}
  {\bfseries 116} (2016) 061102}
  [\href{https://arxiv.org/abs/1602.03837}{{\ttfamily 1602.03837}}].

\bibitem{TheLIGOScientific:2017qsa}
{\scshape LIGO Scientific, Virgo} collaboration, \emph{{GW170817: Observation
  of Gravitational Waves from a Binary Neutron Star Inspiral}},
  \href{https://doi.org/10.1103/PhysRevLett.119.161101}{\emph{Phys. Rev. Lett.}
  {\bfseries 119} (2017) 161101}
  [\href{https://arxiv.org/abs/1710.05832}{{\ttfamily 1710.05832}}].

\bibitem{LIGOScientific:2021qlt}
{\scshape LIGO Scientific, KAGRA, VIRGO} collaboration, \emph{{Observation of
  Gravitational Waves from Two Neutron Star\textendash{}Black Hole
  Coalescences}},
  \href{https://doi.org/10.3847/2041-8213/ac082e}{\emph{Astrophys. J. Lett.}
  {\bfseries 915} (2021) L5}
  [\href{https://arxiv.org/abs/2106.15163}{{\ttfamily 2106.15163}}].

\bibitem{Favata:2013rwa}
M.~Favata, \emph{{Systematic parameter errors in inspiraling neutron star
  binaries}}, \href{https://doi.org/10.1103/PhysRevLett.112.101101}{\emph{Phys.
  Rev. Lett.} {\bfseries 112} (2014) 101101}
  [\href{https://arxiv.org/abs/1310.8288}{{\ttfamily 1310.8288}}].

\bibitem{Samajdar:2018dcx}
A.~Samajdar and T.~Dietrich, \emph{{Waveform systematics for binary neutron
  star gravitational wave signals: effects of the point-particle baseline and
  tidal descriptions}},
  \href{https://doi.org/10.1103/PhysRevD.98.124030}{\emph{Phys. Rev. D}
  {\bfseries 98} (2018) 124030}
  [\href{https://arxiv.org/abs/1810.03936}{{\ttfamily 1810.03936}}].

\bibitem{Purrer:2019jcp}
M.~P\"urrer and C.-J. Haster, \emph{{Gravitational waveform accuracy
  requirements for future ground-based detectors}},
  \href{https://doi.org/10.1103/PhysRevResearch.2.023151}{\emph{Phys. Rev.
  Res.} {\bfseries 2} (2020) 023151}
  [\href{https://arxiv.org/abs/1912.10055}{{\ttfamily 1912.10055}}].

\bibitem{Huang:2020pba}
Y.~Huang, C.-J. Haster, S.~Vitale, V.~Varma, F.~Foucart and S.~Biscoveanu,
  \emph{{Statistical and systematic uncertainties in extracting the source
  properties of neutron star - black hole binaries with gravitational waves}},
  \href{https://doi.org/10.1103/PhysRevD.103.083001}{\emph{Phys. Rev. D}
  {\bfseries 103} (2021) 083001}
  [\href{https://arxiv.org/abs/2005.11850}{{\ttfamily 2005.11850}}].

\bibitem{Gamba:2020wgg}
R.~Gamba, M.~Breschi, S.~Bernuzzi, M.~Agathos and A.~Nagar, \emph{{Waveform
  systematics in the gravitational-wave inference of tidal parameters and
  equation of state from binary neutron star signals}},
  \href{https://doi.org/10.1103/PhysRevD.103.124015}{\emph{Phys. Rev. D}
  {\bfseries 103} (2021) 124015}
  [\href{https://arxiv.org/abs/2009.08467}{{\ttfamily 2009.08467}}].

\bibitem{Parke:1986gb}
S.~J. Parke and T.~R. Taylor, \emph{{An Amplitude for $n$ Gluon Scattering}},
  \href{https://doi.org/10.1103/PhysRevLett.56.2459}{\emph{Phys. Rev. Lett.}
  {\bfseries 56} (1986) 2459}.

\bibitem{Witten:2003nn}
E.~Witten, \emph{{Perturbative gauge theory as a string theory in twistor
  space}}, \href{https://doi.org/10.1007/s00220-004-1187-3}{\emph{Commun. Math.
  Phys.} {\bfseries 252} (2004) 189}
  [\href{https://arxiv.org/abs/hep-th/0312171}{{\ttfamily hep-th/0312171}}].

\bibitem{Roiban:2004yf}
R.~Roiban, M.~Spradlin and A.~Volovich, \emph{{On the tree level S matrix of
  Yang-Mills theory}},
  \href{https://doi.org/10.1103/PhysRevD.70.026009}{\emph{Phys. Rev. D}
  {\bfseries 70} (2004) 026009}
  [\href{https://arxiv.org/abs/hep-th/0403190}{{\ttfamily hep-th/0403190}}].

\bibitem{Gukov:2004ei}
S.~Gukov, L.~Motl and A.~Neitzke, \emph{{Equivalence of twistor prescriptions
  for superYang-Mills}},
  \href{https://doi.org/10.4310/ATMP.2007.v11.n2.a1}{\emph{Adv. Theor. Math.
  Phys.} {\bfseries 11} (2007) 199}
  [\href{https://arxiv.org/abs/hep-th/0404085}{{\ttfamily hep-th/0404085}}].

\bibitem{Cachazo:2004kj}
F.~Cachazo, P.~Svrcek and E.~Witten, \emph{{MHV vertices and tree amplitudes in
  gauge theory}},
  \href{https://doi.org/10.1088/1126-6708/2004/09/006}{\emph{JHEP} {\bfseries
  09} (2004) 006} [\href{https://arxiv.org/abs/hep-th/0403047}{{\ttfamily
  hep-th/0403047}}].

\bibitem{Britto:2005fq}
R.~Britto, F.~Cachazo, B.~Feng and E.~Witten, \emph{{Direct proof of tree-level
  recursion relation in Yang-Mills theory}},
  \href{https://doi.org/10.1103/PhysRevLett.94.181602}{\emph{Phys. Rev. Lett.}
  {\bfseries 94} (2005) 181602}
  [\href{https://arxiv.org/abs/hep-th/0501052}{{\ttfamily hep-th/0501052}}].

\bibitem{Bern:1994zx}
Z.~Bern, L.~J. Dixon, D.~C. Dunbar and D.~A. Kosower, \emph{{One loop n point
  gauge theory amplitudes, unitarity and collinear limits}},
  \href{https://doi.org/10.1016/0550-3213(94)90179-1}{\emph{Nucl. Phys. B}
  {\bfseries 425} (1994) 217}
  [\href{https://arxiv.org/abs/hep-ph/9403226}{{\ttfamily hep-ph/9403226}}].

\bibitem{Fusing}
Z.~Bern, L.~J. Dixon, D.~C. Dunbar and D.~A. Kosower, \emph{{Fusing gauge
  theory tree amplitudes into loop amplitudes}},
  \href{https://doi.org/10.1016/0550-3213(94)00488-Z}{\emph{Nucl. Phys.}
  {\bfseries B435} (1995) 59}
  [\href{https://arxiv.org/abs/hep-ph/9409265}{{\ttfamily hep-ph/9409265}}].

\bibitem{TripleCuteeJets}
Z.~Bern, L.~J. Dixon and D.~A. Kosower, \emph{{One-loop amplitudes for $e^+
  e^-$ to four partons}},
  \href{https://doi.org/10.1016/S0550-3213(97)00703-7}{\emph{Nucl. Phys.}
  {\bfseries B513} (1998) 3}
  [\href{https://arxiv.org/abs/hep-ph/9708239}{{\ttfamily hep-ph/9708239}}].

\bibitem{BCFUnitarity}
R.~Britto, F.~Cachazo and B.~Feng, \emph{{Generalized unitarity and one-loop
  amplitudes in ${\cal N}=4$ super-Yang-Mills}},
  \href{https://doi.org/10.1016/j.nuclphysb.2005.07.014}{\emph{Nucl. Phys.}
  {\bfseries B725} (2005) 275}
  [\href{https://arxiv.org/abs/hep-th/0412103}{{\ttfamily hep-th/0412103}}].

\bibitem{BDKUniarityReview}
Z.~Bern, L.~J. Dixon and D.~A. Kosower, \emph{{Progress in one loop QCD
  computations}},
  \href{https://doi.org/10.1146/annurev.nucl.46.1.109}{\emph{Ann. Rev. Nucl.
  Part. Sci.} {\bfseries 46} (1996) 109}
  [\href{https://arxiv.org/abs/hep-ph/9602280}{{\ttfamily hep-ph/9602280}}].

\bibitem{ElvangHuangReview}
H.~Elvang and Y.-t. Huang, \emph{{Scattering amplitudes}},
  \href{https://arxiv.org/abs/1308.1697}{{\ttfamily 1308.1697}}.

\bibitem{Elvang:2015rqa}
H.~Elvang and Y.-t. Huang, \emph{{Scattering Amplitudes in Gauge Theory and
  Gravity}}. Cambridge University Press, 4, 2015.

\bibitem{JJHenrikReview}
J.~J.~M. Carrasco and H.~Johansson, \emph{{Generic multiloop methods and
  application to ${\cal N}=4$ super-Yang-Mills}},
  \href{https://doi.org/10.1088/1751-8113/44/45/454004}{\emph{J. Phys.}
  {\bfseries A44} (2011) 454004}
  [\href{https://arxiv.org/abs/1103.3298}{{\ttfamily 1103.3298}}].

\bibitem{BernHuangReview}
Z.~Bern and Y.-t. Huang, \emph{{Basics of generalized unitarity}},
  \href{https://doi.org/10.1088/1751-8113/44/45/454003}{\emph{J. Phys.}
  {\bfseries A44} (2011) 454003}
  [\href{https://arxiv.org/abs/1103.1869}{{\ttfamily 1103.1869}}].

\bibitem{BCJ}
Z.~Bern, J.~J.~M. Carrasco and H.~Johansson, \emph{{New relations for
  gauge-theory amplitudes}},
  \href{https://doi.org/10.1103/PhysRevD.78.085011}{\emph{Phys. Rev.}
  {\bfseries D78} (2008) 085011}
  [\href{https://arxiv.org/abs/0805.3993}{{\ttfamily 0805.3993}}].

\bibitem{BCJLoop}
Z.~Bern, J.~J.~M. Carrasco and H.~Johansson, \emph{{Perturbative quantum
  gravity as a double copy of gauge theory}},
  \href{https://doi.org/10.1103/PhysRevLett.105.061602}{\emph{Phys. Rev. Lett.}
  {\bfseries 105} (2010) 061602}
  [\href{https://arxiv.org/abs/1004.0476}{{\ttfamily 1004.0476}}].

\bibitem{KLT}
H.~Kawai, D.~C. Lewellen and S.~H.~H. Tye, \emph{{A relation between tree
  amplitudes of closed and open strings}},
  \href{https://doi.org/10.1016/0550-3213(86)90362-7}{\emph{Nucl. Phys.}
  {\bfseries B269} (1986) 1}.

\bibitem{Bern:1993wt}
Z.~Bern, D.~C. Dunbar and T.~Shimada, \emph{{String based methods in
  perturbative gravity}},
  \href{https://doi.org/10.1016/0370-2693(93)91081-W}{\emph{Phys. Lett. B}
  {\bfseries 312} (1993) 277}
  [\href{https://arxiv.org/abs/hep-th/9307001}{{\ttfamily hep-th/9307001}}].

\bibitem{CompactThree}
Z.~Bern, J.~J.~M. Carrasco, L.~J. Dixon, H.~Johansson and R.~Roiban,
  \emph{{Manifest ultraviolet behavior for the three-loop four-point amplitude
  of ${\cal N}=8$ supergravity}},
  \href{https://doi.org/10.1103/PhysRevD.78.105019}{\emph{Phys. Rev.}
  {\bfseries D78} (2008) 105019}
  [\href{https://arxiv.org/abs/0808.4112}{{\ttfamily 0808.4112}}].

\bibitem{Bern:2009kd}
Z.~Bern, J.~J. Carrasco, L.~J. Dixon, H.~Johansson and R.~Roiban, \emph{{The
  Ultraviolet Behavior of N=8 Supergravity at Four Loops}},
  \href{https://doi.org/10.1103/PhysRevLett.103.081301}{\emph{Phys. Rev. Lett.}
  {\bfseries 103} (2009) 081301}
  [\href{https://arxiv.org/abs/0905.2326}{{\ttfamily 0905.2326}}].

\bibitem{GravityThree}
Z.~Bern, J.~J. Carrasco, L.~J. Dixon, H.~Johansson, D.~A. Kosower and
  R.~Roiban, \emph{{Three-loop superfiniteness of ${\cal N}=8$ supergravity}},
  \href{https://doi.org/10.1103/PhysRevLett.98.161303}{\emph{Phys. Rev. Lett.}
  {\bfseries 98} (2007) 161303}
  [\href{https://arxiv.org/abs/hep-th/0702112}{{\ttfamily hep-th/0702112}}].

\bibitem{Bern:2012cd}
Z.~Bern, S.~Davies, T.~Dennen and Y.-t. Huang, \emph{{Absence of three-loop
  four-point divergences in ${\cal N}=4$ supergravity}},
  \href{https://doi.org/10.1103/PhysRevLett.108.201301}{\emph{Phys. Rev. Lett.}
  {\bfseries 108} (2012) 201301}
  [\href{https://arxiv.org/abs/1202.3423}{{\ttfamily 1202.3423}}].

\bibitem{N4GravFourLoop}
Z.~Bern, S.~Davies, T.~Dennen, A.~V. Smirnov and V.~A. Smirnov,
  \emph{{Ultraviolet properties of ${\cal N}=4$ supergravity at four loops}},
  \href{https://doi.org/10.1103/PhysRevLett.111.231302}{\emph{Phys. Rev. Lett.}
  {\bfseries 111} (2013) 231302}
  [\href{https://arxiv.org/abs/1309.2498}{{\ttfamily 1309.2498}}].

\bibitem{N5GravFourLoop}
Z.~Bern, S.~Davies and T.~Dennen, \emph{{Enhanced ultraviolet cancellations in
  ${\cal N}=5$ supergravity at four loops}},
  \href{https://doi.org/10.1103/PhysRevD.90.105011}{\emph{Phys. Rev.}
  {\bfseries D90} (2014) 105011}
  [\href{https://arxiv.org/abs/1409.3089}{{\ttfamily 1409.3089}}].

\bibitem{UVFiveLoops}
Z.~Bern, J.~J. Carrasco, W.-M. Chen, A.~Edison, H.~Johansson, J.~Parra-Martinez
  et~al., \emph{{Ultraviolet properties of $\mathcal N = 8$ supergravity at
  five loops}}, \href{https://doi.org/10.1103/PhysRevD.98.086021}{\emph{Phys.
  Rev.} {\bfseries D98} (2018) 086021}
  [\href{https://arxiv.org/abs/1804.09311}{{\ttfamily 1804.09311}}].

\bibitem{Monteiro2014cda}
R.~Monteiro, D.~O'Connell and C.~D. White, \emph{{Black holes and the double
  copy}}, \href{https://doi.org/10.1007/JHEP12(2014)056}{\emph{JHEP} {\bfseries
  12} (2014) 056} [\href{https://arxiv.org/abs/1410.0239}{{\ttfamily
  1410.0239}}].

\bibitem{BCCJRReview}
Z.~Bern, J.~J. Carrasco, M.~Chiodaroli, H.~Johansson and R.~Roiban, \emph{{The
  Duality Between Color and Kinematics and its Applications}},
  \href{https://arxiv.org/abs/1909.01358}{{\ttfamily 1909.01358}}.

\bibitem{doublecopyWhitePaper2022}
e.~John Joseph~Carrasco, \emph{{Snowmass 2021 Whitepaper: the Double Copy and
  its Applications}},  \href{https://arxiv.org/abs/2203.ikjlm}{{\ttfamily
  2203.ikjlm}}.

\bibitem{Damour:2017zjx}
T.~Damour, \emph{{High-energy gravitational scattering and the general
  relativistic two-body problem}},
  \href{https://doi.org/10.1103/PhysRevD.97.044038}{\emph{Phys. Rev. D}
  {\bfseries 97} (2018) 044038}
  [\href{https://arxiv.org/abs/1710.10599}{{\ttfamily 1710.10599}}].

\bibitem{Einstein:1938yz}
A.~Einstein, L.~Infeld and B.~Hoffmann, \emph{{The Gravitational equations and
  the problem of motion}}, \href{https://doi.org/10.2307/1968714}{\emph{Annals
  Math.} {\bfseries 39} (1938) 65}.

\bibitem{Einstein:1940mt}
A.~Einstein and L.~Infeld, \emph{{The Gravitational equations and the problem
  of motion. 2.}}, \href{https://doi.org/10.2307/1969015}{\emph{Annals Math.}
  {\bfseries 41} (1940) 455}.

\bibitem{Ohta:1973je}
T.~Ohta, H.~Okamura, T.~Kimura and K.~Hiida, \emph{{Physically acceptable
  solution of einstein's equation for many-body system}},
  \href{https://doi.org/10.1143/PTP.50.492}{\emph{Prog. Theor. Phys.}
  {\bfseries 50} (1973) 492}.

\bibitem{Jaranowski:1997ky}
P.~Jaranowski and G.~Schaefer, \emph{{Third postNewtonian higher order ADM
  Hamilton dynamics for two-body point mass systems}},
  \href{https://doi.org/10.1103/PhysRevD.57.7274}{\emph{Phys. Rev. D}
  {\bfseries 57} (1998) 7274}
  [\href{https://arxiv.org/abs/gr-qc/9712075}{{\ttfamily gr-qc/9712075}}].

\bibitem{Damour:1999cr}
T.~Damour, P.~Jaranowski and G.~Schaefer, \emph{{Dynamical invariants for
  general relativistic two-body systems at the third postNewtonian
  approximation}},
  \href{https://doi.org/10.1103/PhysRevD.62.044024}{\emph{Phys. Rev. D}
  {\bfseries 62} (2000) 044024}
  [\href{https://arxiv.org/abs/gr-qc/9912092}{{\ttfamily gr-qc/9912092}}].

\bibitem{Blanchet:2000nv}
L.~Blanchet and G.~Faye, \emph{{Equations of motion of point particle binaries
  at the third postNewtonian order}},
  \href{https://doi.org/10.1016/S0375-9601(00)00360-1}{\emph{Phys. Lett. A}
  {\bfseries 271} (2000) 58}
  [\href{https://arxiv.org/abs/gr-qc/0004009}{{\ttfamily gr-qc/0004009}}].

\bibitem{Damour:2001bu}
T.~Damour, P.~Jaranowski and G.~Schaefer, \emph{{Dimensional regularization of
  the gravitational interaction of point masses}},
  \href{https://doi.org/10.1016/S0370-2693(01)00642-6}{\emph{Phys. Lett. B}
  {\bfseries 513} (2001) 147}
  [\href{https://arxiv.org/abs/gr-qc/0105038}{{\ttfamily gr-qc/0105038}}].

\bibitem{Damour:2014jta}
T.~Damour, P.~Jaranowski and G.~Sch\"afer, \emph{{Nonlocal-in-time action for
  the fourth post-Newtonian conservative dynamics of two-body systems}},
  \href{https://doi.org/10.1103/PhysRevD.89.064058}{\emph{Phys. Rev. D}
  {\bfseries 89} (2014) 064058}
  [\href{https://arxiv.org/abs/1401.4548}{{\ttfamily 1401.4548}}].

\bibitem{Jaranowski:2015lha}
P.~Jaranowski and G.~Sch\"afer, \emph{{Derivation of local-in-time fourth
  post-Newtonian ADM Hamiltonian for spinless compact binaries}},
  \href{https://doi.org/10.1103/PhysRevD.92.124043}{\emph{Phys. Rev. D}
  {\bfseries 92} (2015) 124043}
  [\href{https://arxiv.org/abs/1508.01016}{{\ttfamily 1508.01016}}].

\bibitem{Mino:1996nk}
Y.~Mino, M.~Sasaki and T.~Tanaka, \emph{{Gravitational radiation reaction to a
  particle motion}},
  \href{https://doi.org/10.1103/PhysRevD.55.3457}{\emph{Phys. Rev. D}
  {\bfseries 55} (1997) 3457}
  [\href{https://arxiv.org/abs/gr-qc/9606018}{{\ttfamily gr-qc/9606018}}].

\bibitem{Quinn:1996am}
T.~C. Quinn and R.~M. Wald, \emph{{An Axiomatic approach to electromagnetic and
  gravitational radiation reaction of particles in curved space-time}},
  \href{https://doi.org/10.1103/PhysRevD.56.3381}{\emph{Phys. Rev. D}
  {\bfseries 56} (1997) 3381}
  [\href{https://arxiv.org/abs/gr-qc/9610053}{{\ttfamily gr-qc/9610053}}].

\bibitem{Goldberger:2004jt}
W.~D. Goldberger and I.~Z. Rothstein, \emph{{An Effective field theory of
  gravity for extended objects}},
  \href{https://doi.org/10.1103/PhysRevD.73.104029}{\emph{Phys. Rev. D}
  {\bfseries 73} (2006) 104029}
  [\href{https://arxiv.org/abs/hep-th/0409156}{{\ttfamily hep-th/0409156}}].

\bibitem{Bertotti:1956pxu}
B.~Bertotti, \emph{{On gravitational motion}},
  \href{https://doi.org/10.1007/bf02746175}{\emph{Nuovo Cim.} {\bfseries 4}
  (1956) 898}.

\bibitem{Kerr:1959zlt}
R.~P. Kerr, \emph{{The Lorentz-covariant approximation method in general
  relativity I}}, \href{https://doi.org/10.1007/bf02732767}{\emph{Nuovo Cim.}
  {\bfseries 13} (1959) 469}.

\bibitem{Bertotti:1960wuq}
B.~Bertotti and J.~Plebanski, \emph{{Theory of gravitational perturbations in
  the fast motion approximation}},
  \href{https://doi.org/10.1016/0003-4916(60)90132-9}{\emph{Annals Phys.}
  {\bfseries 11} (1960) 169}.

\bibitem{Portilla:1979xx}
M.~Portilla, \emph{{MOMENTUM AND ANGULAR MOMENTUM OF TWO GRAVITATING
  PARTICLES}}, \href{https://doi.org/10.1088/0305-4470/12/7/025}{\emph{J. Phys.
  A} {\bfseries 12} (1979) 1075}.

\bibitem{Westpfahl:1979gu}
K.~Westpfahl and M.~Goller, \emph{{GRAVITATIONAL SCATTERING OF TWO RELATIVISTIC
  PARTICLES IN POSTLINEAR APPROXIMATION}},
  \href{https://doi.org/10.1007/BF02817047}{\emph{Lett. Nuovo Cim.} {\bfseries
  26} (1979) 573}.

\bibitem{Portilla:1980uz}
M.~Portilla, \emph{{SCATTERING OF TWO GRAVITATING PARTICLES: CLASSICAL
  APPROACH}}, \href{https://doi.org/10.1088/0305-4470/13/12/017}{\emph{J. Phys.
  A} {\bfseries 13} (1980) 3677}.

\bibitem{Bel:1981be}
L.~Bel, T.~Damour, N.~Deruelle, J.~Ibanez and J.~Martin,
  \emph{{Poincar\'e-invariant gravitational field and equations of motion of
  two pointlike objects: The postlinear approximation of general relativity}},
  \href{https://doi.org/10.1007/BF00756073}{\emph{Gen. Rel. Grav.} {\bfseries
  13} (1981) 963}.

\bibitem{Westpfahl:1985tsl}
K.~Westpfahl, \emph{{High-Speed Scattering of Charged and Uncharged Particles
  in General Relativity}},
  \href{https://doi.org/10.1002/prop.2190330802}{\emph{Fortsch. Phys.}
  {\bfseries 33} (1985) 417}.

\bibitem{Damour:2016gwp}
T.~Damour, \emph{{Gravitational scattering, post-Minkowskian approximation and
  Effective One-Body theory}},
  \href{https://doi.org/10.1103/PhysRevD.94.104015}{\emph{Phys. Rev. D}
  {\bfseries 94} (2016) 104015}
  [\href{https://arxiv.org/abs/1609.00354}{{\ttfamily 1609.00354}}].

\bibitem{Levi:2018nxp}
M.~Levi, \emph{{Effective Field Theories of Post-Newtonian Gravity: A
  comprehensive review}},
  \href{https://doi.org/10.1088/1361-6633/ab12bc}{\emph{Rept. Prog. Phys.}
  {\bfseries 83} (2020) 075901}
  [\href{https://arxiv.org/abs/1807.01699}{{\ttfamily 1807.01699}}].

\bibitem{NRGRWhitePaper2022}
e.~Walter~Goldberger, \emph{{Snowmass 2021 Whitepaper: EFT of Gravity and
  NRGR}},  \href{https://arxiv.org/abs/2203.ikjlm}{{\ttfamily 2203.ikjlm}}.

\bibitem{Anastasiou:2015vya}
C.~Anastasiou, C.~Duhr, F.~Dulat, F.~Herzog and B.~Mistlberger, \emph{{Higgs
  Boson Gluon-Fusion Production in QCD at Three Loops}},
  \href{https://doi.org/10.1103/PhysRevLett.114.212001}{\emph{Phys. Rev. Lett.}
  {\bfseries 114} (2015) 212001}
  [\href{https://arxiv.org/abs/1503.06056}{{\ttfamily 1503.06056}}].

\bibitem{Gehrmann-DeRidder:2007vsv}
A.~Gehrmann-De~Ridder, T.~Gehrmann, E.~W.~N. Glover and G.~Heinrich,
  \emph{{NNLO corrections to event shapes in e+ e- annihilation}},
  \href{https://doi.org/10.1088/1126-6708/2007/12/094}{\emph{JHEP} {\bfseries
  12} (2007) 094} [\href{https://arxiv.org/abs/0711.4711}{{\ttfamily
  0711.4711}}].

\bibitem{Henn:2019swt}
J.~M. Henn, G.~P. Korchemsky and B.~Mistlberger, \emph{{The full four-loop cusp
  anomalous dimension in $\mathcal{N}=4$ super Yang-Mills and QCD}},
  \href{https://doi.org/10.1007/JHEP04(2020)018}{\emph{JHEP} {\bfseries 04}
  (2020) 018} [\href{https://arxiv.org/abs/1911.10174}{{\ttfamily
  1911.10174}}].

\bibitem{Laporta:2017okg}
S.~Laporta, \emph{{High-precision calculation of the 4-loop contribution to the
  electron g-2 in QED}},
  \href{https://doi.org/10.1016/j.physletb.2017.06.056}{\emph{Phys. Lett. B}
  {\bfseries 772} (2017) 232}
  [\href{https://arxiv.org/abs/1704.06996}{{\ttfamily 1704.06996}}].

\bibitem{Bern:2018jmv}
Z.~Bern, J.~J. Carrasco, W.-M. Chen, A.~Edison, H.~Johansson, J.~Parra-Martinez
  et~al., \emph{{Ultraviolet Properties of $\mathcal N = 8$ Supergravity at
  Five Loops}}, \href{https://doi.org/10.1103/PhysRevD.98.086021}{\emph{Phys.
  Rev. D} {\bfseries 98} (2018) 086021}
  [\href{https://arxiv.org/abs/1804.09311}{{\ttfamily 1804.09311}}].

\bibitem{Carrasco:2021otn}
J.~J.~M. Carrasco, A.~Edison and H.~Johansson, \emph{{Maximal Super-Yang-Mills
  at Six Loops via Novel Integrand Bootstrap}},
  \href{https://arxiv.org/abs/2112.05178}{{\ttfamily 2112.05178}}.

\bibitem{Chetyrkin:1981qh}
K.~G. Chetyrkin and F.~V. Tkachov, \emph{{Integration by Parts: The Algorithm
  to Calculate beta Functions in 4 Loops}},
  \href{https://doi.org/10.1016/0550-3213(81)90199-1}{\emph{Nucl. Phys. B}
  {\bfseries 192} (1981) 159}.

\bibitem{Laporta:2000dsw}
S.~Laporta, \emph{{High precision calculation of multiloop Feynman integrals by
  difference equations}},
  \href{https://doi.org/10.1142/S0217751X00002159}{\emph{Int. J. Mod. Phys. A}
  {\bfseries 15} (2000) 5087}
  [\href{https://arxiv.org/abs/hep-ph/0102033}{{\ttfamily hep-ph/0102033}}].

\bibitem{Smirnov:2008iw}
A.~V. Smirnov, \emph{{Algorithm FIRE -- Feynman Integral REduction}},
  \href{https://doi.org/10.1088/1126-6708/2008/10/107}{\emph{JHEP} {\bfseries
  10} (2008) 107} [\href{https://arxiv.org/abs/0807.3243}{{\ttfamily
  0807.3243}}].

\bibitem{Kotikov:1990kg}
A.~V. Kotikov, \emph{{Differential equations method: New technique for massive
  Feynman diagrams calculation}},
  \href{https://doi.org/10.1016/0370-2693(91)90413-K}{\emph{Phys. Lett.}
  {\bfseries B254} (1991) 158}.

\bibitem{Bern:1993kr}
Z.~Bern, L.~J. Dixon and D.~A. Kosower, \emph{{Dimensionally regulated pentagon
  integrals}}, \href{https://doi.org/10.1016/0550-3213(94)90398-0}{\emph{Nucl.
  Phys.} {\bfseries B412} (1994) 751}
  [\href{https://arxiv.org/abs/hep-ph/9306240}{{\ttfamily hep-ph/9306240}}].

\bibitem{Remiddi:1997ny}
E.~Remiddi, \emph{{Differential equations for Feynman graph amplitudes}},
  {\emph{Nuovo Cim.} {\bfseries A110} (1997) 1435}
  [\href{https://arxiv.org/abs/hep-th/9711188}{{\ttfamily hep-th/9711188}}].

\bibitem{Gehrmann:1999as}
T.~Gehrmann and E.~Remiddi, \emph{{Differential equations for two loop four
  point functions}},
  \href{https://doi.org/10.1016/S0550-3213(00)00223-6}{\emph{Nucl. Phys.}
  {\bfseries B580} (2000) 485}
  [\href{https://arxiv.org/abs/hep-ph/9912329}{{\ttfamily hep-ph/9912329}}].

\bibitem{Henn:2013pwa}
J.~M. Henn, \emph{{Multiloop integrals in dimensional regularization made
  simple}}, \href{https://doi.org/10.1103/PhysRevLett.110.251601}{\emph{Phys.
  Rev. Lett.} {\bfseries 110} (2013) 251601}
  [\href{https://arxiv.org/abs/1304.1806}{{\ttfamily 1304.1806}}].

\bibitem{Henn:2013nsa}
J.~M. Henn, A.~V. Smirnov and V.~A. Smirnov, \emph{{Evaluating single-scale
  and/or non-planar diagrams by differential equations}},
  \href{https://doi.org/10.1007/JHEP03(2014)088}{\emph{JHEP} {\bfseries 03}
  (2014) 088} [\href{https://arxiv.org/abs/1312.2588}{{\ttfamily 1312.2588}}].

\bibitem{Parra-Martinez:2020dzs}
J.~Parra-Martinez, M.~S. Ruf and M.~Zeng, \emph{{Extremal black hole scattering
  at $\mathcal{O}(G^3)$: graviton dominance, eikonal exponentiation, and
  differential equations}},
  \href{https://doi.org/10.1007/JHEP11(2020)023}{\emph{JHEP} {\bfseries 11}
  (2020) 023} [\href{https://arxiv.org/abs/2005.04236}{{\ttfamily
  2005.04236}}].

\bibitem{Gubser:2002tv}
S.~S. Gubser, I.~R. Klebanov and A.~M. Polyakov, \emph{{A Semiclassical limit
  of the gauge / string correspondence}},
  \href{https://doi.org/10.1016/S0550-3213(02)00373-5}{\emph{Nucl. Phys. B}
  {\bfseries 636} (2002) 99}
  [\href{https://arxiv.org/abs/hep-th/0204051}{{\ttfamily hep-th/0204051}}].

\bibitem{Kruczenski:2004kw}
M.~Kruczenski, A.~V. Ryzhov and A.~A. Tseytlin, \emph{{Large spin limit of
  AdS(5) x S**5 string theory and low-energy expansion of ferromagnetic spin
  chains}}, \href{https://doi.org/10.1016/j.nuclphysb.2004.05.028}{\emph{Nucl.
  Phys. B} {\bfseries 692} (2004) 3}
  [\href{https://arxiv.org/abs/hep-th/0403120}{{\ttfamily hep-th/0403120}}].

\bibitem{Giombi:2009gd}
S.~Giombi, R.~Ricci, R.~Roiban, A.~A. Tseytlin and C.~Vergu, \emph{{Quantum
  AdS(5) x S5 superstring in the AdS light-cone gauge}},
  \href{https://doi.org/10.1007/JHEP03(2010)003}{\emph{JHEP} {\bfseries 03}
  (2010) 003} [\href{https://arxiv.org/abs/0912.5105}{{\ttfamily 0912.5105}}].

\bibitem{Beisert:2010jr}
N.~Beisert et~al., \emph{{Review of AdS/CFT Integrability: An Overview}},
  \href{https://doi.org/10.1007/s11005-011-0529-2}{\emph{Lett. Math. Phys.}
  {\bfseries 99} (2012) 3} [\href{https://arxiv.org/abs/1012.3982}{{\ttfamily
  1012.3982}}].

\bibitem{Kosower:2018adc}
D.~A. Kosower, B.~Maybee and D.~O'Connell, \emph{{Amplitudes, observables, and
  classical scattering}},
  \href{https://doi.org/10.1007/JHEP02(2019)137}{\emph{JHEP} {\bfseries 02}
  (2019) 137} [\href{https://arxiv.org/abs/1811.10950}{{\ttfamily
  1811.10950}}].

\bibitem{Cristofoli:2021jas}
A.~Cristofoli, R.~Gonzo, N.~Moynihan, D.~O'Connell, A.~Ross, M.~Sergola et~al.,
  \emph{{The Uncertainty Principle and Classical Amplitudes}},
  \href{https://arxiv.org/abs/2112.07556}{{\ttfamily 2112.07556}}.

\bibitem{Britto:2021pud}
R.~Britto, R.~Gonzo and G.~R. Jehu, \emph{{Graviton particle statistics and
  coherent states from classical scattering amplitudes}},
  \href{https://arxiv.org/abs/2112.07036}{{\ttfamily 2112.07036}}.

\bibitem{CRS}
C.~Cheung, I.~Z. Rothstein and M.~P. Solon, \emph{{From scattering amplitudes
  to classical potentials in the post-Minkowskian expansion}},
  \href{https://doi.org/10.1103/PhysRevLett.121.251101}{\emph{Phys. Rev. Lett.}
  {\bfseries 121} (2018) 251101}
  [\href{https://arxiv.org/abs/1808.02489}{{\ttfamily 1808.02489}}].

\bibitem{Damour:2019lcq}
T.~Damour, \emph{{Classical and quantum scattering in post-Minkowskian
  gravity}}, \href{https://doi.org/10.1103/PhysRevD.102.024060}{\emph{Phys.
  Rev. D} {\bfseries 102} (2020) 024060}
  [\href{https://arxiv.org/abs/1912.02139}{{\ttfamily 1912.02139}}].

\bibitem{Damgaard:2021rnk}
P.~H. Damgaard and P.~Vanhove, \emph{{Remodeling the effective one-body
  formalism in post-Minkowskian gravity}},
  \href{https://doi.org/10.1103/PhysRevD.104.104029}{\emph{Phys. Rev. D}
  {\bfseries 104} (2021) 104029}
  [\href{https://arxiv.org/abs/2108.11248}{{\ttfamily 2108.11248}}].

\bibitem{Amati:1987wq}
D.~Amati, M.~Ciafaloni and G.~Veneziano, \emph{{Superstring Collisions at
  Planckian Energies}},
  \href{https://doi.org/10.1016/0370-2693(87)90346-7}{\emph{Phys. Lett. B}
  {\bfseries 197} (1987) 81}.

\bibitem{tHooft:1987vrq}
G.~'t~Hooft, \emph{{Graviton Dominance in Ultrahigh-Energy Scattering}},
  \href{https://doi.org/10.1016/0370-2693(87)90159-6}{\emph{Phys. Lett. B}
  {\bfseries 198} (1987) 61}.

\bibitem{Muzinich:1987in}
I.~J. Muzinich and M.~Soldate, \emph{{High-Energy Unitarity of Gravitation and
  Strings}}, \href{https://doi.org/10.1103/PhysRevD.37.359}{\emph{Phys. Rev. D}
  {\bfseries 37} (1988) 359}.

\bibitem{Amati:1987uf}
D.~Amati, M.~Ciafaloni and G.~Veneziano, \emph{{Classical and Quantum Gravity
  Effects from Planckian Energy Superstring Collisions}},
  \href{https://doi.org/10.1142/S0217751X88000710}{\emph{Int. J. Mod. Phys. A}
  {\bfseries 3} (1988) 1615}.

\bibitem{Cho:2018upo}
G.~Cho, A.~Gopakumar, M.~Haney and H.~M. Lee, \emph{{Gravitational waves from
  compact binaries in post-Newtonian accurate hyperbolic orbits}},
  \href{https://doi.org/10.1103/PhysRevD.98.024039}{\emph{Phys. Rev. D}
  {\bfseries 98} (2018) 024039}
  [\href{https://arxiv.org/abs/1807.02380}{{\ttfamily 1807.02380}}].

\bibitem{Kalin:2019rwq}
G.~K\"alin and R.~A. Porto, \emph{{From Boundary Data to Bound States}},
  \href{https://doi.org/10.1007/JHEP01(2020)072}{\emph{JHEP} {\bfseries 01}
  (2020) 072} [\href{https://arxiv.org/abs/1910.03008}{{\ttfamily
  1910.03008}}].

\bibitem{Kalin:2019inp}
G.~K\"alin and R.~A. Porto, \emph{{From boundary data to bound states. Part II.
  Scattering angle to dynamical invariants (with twist)}},
  \href{https://doi.org/10.1007/JHEP02(2020)120}{\emph{JHEP} {\bfseries 02}
  (2020) 120} [\href{https://arxiv.org/abs/1911.09130}{{\ttfamily
  1911.09130}}].

\bibitem{Cho:2021arx}
G.~Cho, G.~K\"alin and R.~A. Porto, \emph{{From Boundary Data to Bound States
  III: Radiative Effects}},  \href{https://arxiv.org/abs/2112.03976}{{\ttfamily
  2112.03976}}.

\bibitem{Thorne:1980ru}
K.~S. Thorne, \emph{{Multipole Expansions of Gravitational Radiation}},
  \href{https://doi.org/10.1103/RevModPhys.52.299}{\emph{Rev. Mod. Phys.}
  {\bfseries 52} (1980) 299}.

\bibitem{Blanchet:1987wq}
L.~Blanchet and T.~Damour, \emph{{Tail Transported Temporal Correlations in the
  Dynamics of a Gravitating System}},
  \href{https://doi.org/10.1103/PhysRevD.37.1410}{\emph{Phys. Rev. D}
  {\bfseries 37} (1988) 1410}.

\bibitem{Blanchet:1993ec}
L.~Blanchet and G.~Schaefer, \emph{{Gravitational wave tails and binary star
  systems}}, \href{https://doi.org/10.1088/0264-9381/10/12/026}{\emph{Class.
  Quant. Grav.} {\bfseries 10} (1993) 2699}.

\bibitem{3PM}
Z.~Bern, C.~Cheung, R.~Roiban, C.-H. Shen, M.~P. Solon and M.~Zeng,
  \emph{{Scattering amplitudes and the conservative Hamiltonian for binary
  systems at third post-Minkowskian order}},
  \href{https://doi.org/10.1103/PhysRevLett.122.201603}{\emph{Phys. Rev. Lett.}
  {\bfseries 122} (2019) 201603}
  [\href{https://arxiv.org/abs/1901.04424}{{\ttfamily 1901.04424}}].

\bibitem{3PMLong}
Z.~Bern, C.~Cheung, R.~Roiban, C.-H. Shen, M.~P. Solon and M.~Zeng,
  \emph{{Black hole binary dynamics from the double copy and effective
  theory}},  \href{https://arxiv.org/abs/1908.01493}{{\ttfamily 1908.01493}}.

\bibitem{Blumlein:2020znm}
J.~Bl\"umlein, A.~Maier, P.~Marquard and G.~Sch\"afer, \emph{{Testing binary
  dynamics in gravity at the sixth post-Newtonian level}},
  \href{https://doi.org/10.1016/j.physletb.2020.135496}{\emph{Phys. Lett. B}
  {\bfseries 807} (2020) 135496}
  [\href{https://arxiv.org/abs/2003.07145}{{\ttfamily 2003.07145}}].

\bibitem{Cheung:2020gyp}
C.~Cheung and M.~P. Solon, \emph{{Classical gravitational scattering at $
  \mathcal{O} $(G$^{3}$) from Feynman diagrams}},
  \href{https://doi.org/10.1007/JHEP06(2020)144}{\emph{JHEP} {\bfseries 06}
  (2020) 144} [\href{https://arxiv.org/abs/2003.08351}{{\ttfamily
  2003.08351}}].

\bibitem{Kalin:2020mvi}
G.~K\"alin and R.~A. Porto, \emph{{Post-Minkowskian Effective Field Theory for
  Conservative Binary Dynamics}},
  \href{https://doi.org/10.1007/JHEP11(2020)106}{\emph{JHEP} {\bfseries 11}
  (2020) 106} [\href{https://arxiv.org/abs/2006.01184}{{\ttfamily
  2006.01184}}].

\bibitem{DiVecchia:2020ymx}
P.~Di~Vecchia, C.~Heissenberg, R.~Russo and G.~Veneziano, \emph{{Universality
  of ultra-relativistic gravitational scattering}},
  \href{https://doi.org/10.1016/j.physletb.2020.135924}{\emph{Phys. Lett. B}
  {\bfseries 811} (2020) 135924}
  [\href{https://arxiv.org/abs/2008.12743}{{\ttfamily 2008.12743}}].

\bibitem{Damour:2020tta}
T.~Damour, \emph{{Radiative contribution to classical gravitational scattering
  at the third order in $G$}},
  \href{https://doi.org/10.1103/PhysRevD.102.124008}{\emph{Phys. Rev. D}
  {\bfseries 102} (2020) 124008}
  [\href{https://arxiv.org/abs/2010.01641}{{\ttfamily 2010.01641}}].

\bibitem{DiVecchia:2021bdo}
P.~Di~Vecchia, C.~Heissenberg, R.~Russo and G.~Veneziano, \emph{{The eikonal
  approach to gravitational scattering and radiation at $ \mathcal{O}
  $(G$^{3}$)}}, \href{https://doi.org/10.1007/JHEP07(2021)169}{\emph{JHEP}
  {\bfseries 07} (2021) 169}
  [\href{https://arxiv.org/abs/2104.03256}{{\ttfamily 2104.03256}}].

\bibitem{Bjerrum-Bohr:2021din}
N.~E.~J. Bjerrum-Bohr, P.~H. Damgaard, L.~Plant\'e and P.~Vanhove, \emph{{The
  amplitude for classical gravitational scattering at third Post-Minkowskian
  order}}, \href{https://doi.org/10.1007/JHEP08(2021)172}{\emph{JHEP}
  {\bfseries 08} (2021) 172}
  [\href{https://arxiv.org/abs/2105.05218}{{\ttfamily 2105.05218}}].

\bibitem{Damgaard:2021ipf}
P.~H. Damgaard, L.~Plante and P.~Vanhove, \emph{{On an exponential
  representation of the gravitational S-matrix}},
  \href{https://doi.org/10.1007/JHEP11(2021)213}{\emph{JHEP} {\bfseries 11}
  (2021) 213} [\href{https://arxiv.org/abs/2107.12891}{{\ttfamily
  2107.12891}}].

\bibitem{Brandhuber:2021eyq}
A.~Brandhuber, G.~Chen, G.~Travaglini and C.~Wen, \emph{{Classical
  gravitational scattering from a gauge-invariant double copy}},
  \href{https://doi.org/10.1007/JHEP10(2021)118}{\emph{JHEP} {\bfseries 10}
  (2021) 118} [\href{https://arxiv.org/abs/2108.04216}{{\ttfamily
  2108.04216}}].

\bibitem{Herrmann:2021lqe}
E.~Herrmann, J.~Parra-Martinez, M.~S. Ruf and M.~Zeng, \emph{{Gravitational
  Bremsstrahlung from Reverse Unitarity}},
  \href{https://doi.org/10.1103/PhysRevLett.126.201602}{\emph{Phys. Rev. Lett.}
  {\bfseries 126} (2021) 201602}
  [\href{https://arxiv.org/abs/2101.07255}{{\ttfamily 2101.07255}}].

\bibitem{Herrmann:2021tct}
E.~Herrmann, J.~Parra-Martinez, M.~S. Ruf and M.~Zeng, \emph{{Radiative
  classical gravitational observables at $ \mathcal{O} $(G$^{3}$) from
  scattering amplitudes}},
  \href{https://doi.org/10.1007/JHEP10(2021)148}{\emph{JHEP} {\bfseries 10}
  (2021) 148} [\href{https://arxiv.org/abs/2104.03957}{{\ttfamily
  2104.03957}}].

\bibitem{DiVecchia:2021ndb}
P.~Di~Vecchia, C.~Heissenberg, R.~Russo and G.~Veneziano, \emph{{Radiation
  Reaction from Soft Theorems}},
  \href{https://doi.org/10.1016/j.physletb.2021.136379}{\emph{Phys. Lett. B}
  {\bfseries 818} (2021) 136379}
  [\href{https://arxiv.org/abs/2101.05772}{{\ttfamily 2101.05772}}].

\bibitem{Heissenberg:2021tzo}
C.~Heissenberg, \emph{{Infrared divergences and the eikonal exponentiation}},
  \href{https://doi.org/10.1103/PhysRevD.104.046016}{\emph{Phys. Rev. D}
  {\bfseries 104} (2021) 046016}
  [\href{https://arxiv.org/abs/2105.04594}{{\ttfamily 2105.04594}}].

\bibitem{Bern:2021dqo}
Z.~Bern, J.~Parra-Martinez, R.~Roiban, M.~S. Ruf, C.-H. Shen, M.~P. Solon
  et~al., \emph{{Scattering Amplitudes and Conservative Binary Dynamics at
  ${\cal O}(G^4)$}},
  \href{https://doi.org/10.1103/PhysRevLett.126.171601}{\emph{Phys. Rev. Lett.}
  {\bfseries 126} (2021) 171601}
  [\href{https://arxiv.org/abs/2101.07254}{{\ttfamily 2101.07254}}].

\bibitem{Bern:2021yeh}
Z.~Bern, J.~Parra-Martinez, R.~Roiban, M.~S. Ruf, C.-H. Shen, M.~P. Solon
  et~al., \emph{{Scattering Amplitudes, the Tail Effect, and Conservative
  Binary Dynamics at $O(G^4)$}},
  \href{https://arxiv.org/abs/2112.10750}{{\ttfamily 2112.10750}}.

\bibitem{Wheeler:1949hn}
J.~A. Wheeler and R.~P. Feynman, \emph{{Classical electrodynamics in terms of
  direct interparticle action}},
  \href{https://doi.org/10.1103/RevModPhys.21.425}{\emph{Rev. Mod. Phys.}
  {\bfseries 21} (1949) 425}.

\bibitem{Damour:1995kt}
T.~Damour and G.~Esposito-Farese, \emph{{Testing gravity to second
  postNewtonian order: A Field theory approach}},
  \href{https://doi.org/10.1103/PhysRevD.53.5541}{\emph{Phys. Rev. D}
  {\bfseries 53} (1996) 5541}
  [\href{https://arxiv.org/abs/gr-qc/9506063}{{\ttfamily gr-qc/9506063}}].

\bibitem{Bini:2021gat}
D.~Bini, T.~Damour and A.~Geralico, \emph{{Radiative contributions to
  gravitational scattering}},
  \href{https://doi.org/10.1103/PhysRevD.104.084031}{\emph{Phys. Rev. D}
  {\bfseries 104} (2021) 084031}
  [\href{https://arxiv.org/abs/2107.08896}{{\ttfamily 2107.08896}}].

\bibitem{Blumlein:2021txe}
J.~Bl\"umlein, A.~Maier, P.~Marquard and G.~Sch\"afer, \emph{{The fifth-order
  post-Newtonian Hamiltonian dynamics of two-body systems from an effective
  field theory approach}},  \href{https://arxiv.org/abs/2110.13822}{{\ttfamily
  2110.13822}}.

\bibitem{Dlapa:2021npj}
C.~Dlapa, G.~K\"alin, Z.~Liu and R.~A. Porto, \emph{{Dynamics of Binary Systems
  to Fourth Post-Minkowskian Order from the Effective Field Theory Approach}},
  \href{https://arxiv.org/abs/2106.08276}{{\ttfamily 2106.08276}}.

\bibitem{Dlapa:2021vgp}
C.~Dlapa, G.~K\"alin, Z.~Liu and R.~A. Porto, \emph{{Conservative Dynamics of
  Binary Systems at Fourth Post-Minkowskian Order in the Large-eccentricity
  Expansion}},  \href{https://arxiv.org/abs/2112.11296}{{\ttfamily
  2112.11296}}.

\bibitem{Shamaly:1972zu}
A.~Shamaly and A.~Z. Capri, \emph{{Propagation of interacting fields}},
  \href{https://doi.org/10.1016/0003-4916(72)90149-2}{\emph{Annals Phys.}
  {\bfseries 74} (1972) 503}.

\bibitem{Hortacsu:1974bm}
M.~Hortacsu, \emph{{Demonstration of noncausality for the rarita-schwinger
  equation}}, \href{https://doi.org/10.1103/PhysRevD.9.928}{\emph{Phys. Rev. D}
  {\bfseries 9} (1974) 928}.

\bibitem{Deser:2000dz}
S.~Deser, V.~Pascalutsa and A.~Waldron, \emph{{Massive spin 3/2
  electrodynamics}},
  \href{https://doi.org/10.1103/PhysRevD.62.105031}{\emph{Phys. Rev. D}
  {\bfseries 62} (2000) 105031}
  [\href{https://arxiv.org/abs/hep-th/0003011}{{\ttfamily hep-th/0003011}}].

\bibitem{Porrati:2008gv}
M.~Porrati and R.~Rahman, \emph{{Intrinsic Cutoff and Acausality for Massive
  Spin 2 Fields Coupled to Electromagnetism}},
  \href{https://doi.org/10.1016/j.nuclphysb.2008.05.013}{\emph{Nucl. Phys. B}
  {\bfseries 801} (2008) 174}
  [\href{https://arxiv.org/abs/0801.2581}{{\ttfamily 0801.2581}}].

\bibitem{Camanho:2014apa}
X.~O. Camanho, J.~D. Edelstein, J.~Maldacena and A.~Zhiboedov, \emph{{Causality
  Constraints on Corrections to the Graviton Three-Point Coupling}},
  \href{https://doi.org/10.1007/JHEP02(2016)020}{\emph{JHEP} {\bfseries 02}
  (2016) 020} [\href{https://arxiv.org/abs/1407.5597}{{\ttfamily 1407.5597}}].

\bibitem{Afkhami-Jeddi:2018apj}
N.~Afkhami-Jeddi, S.~Kundu and A.~Tajdini, \emph{{A Bound on Massive Higher
  Spin Particles}}, \href{https://doi.org/10.1007/JHEP04(2019)056}{\emph{JHEP}
  {\bfseries 04} (2019) 056}
  [\href{https://arxiv.org/abs/1811.01952}{{\ttfamily 1811.01952}}].

\bibitem{Arkani-Hamed:2017jhn}
N.~Arkani-Hamed, T.-C. Huang and Y.-t. Huang, \emph{{Scattering amplitudes for
  all masses and spins}},
  \href{https://doi.org/10.1007/JHEP11(2021)070}{\emph{JHEP} {\bfseries 11}
  (2021) 070} [\href{https://arxiv.org/abs/1709.04891}{{\ttfamily
  1709.04891}}].

\bibitem{Bern:2020buy}
Z.~Bern, A.~Luna, R.~Roiban, C.-H. Shen and M.~Zeng, \emph{{Spinning black hole
  binary dynamics, scattering amplitudes, and effective field theory}},
  \href{https://doi.org/10.1103/PhysRevD.104.065014}{\emph{Phys. Rev. D}
  {\bfseries 104} (2021) 065014}
  [\href{https://arxiv.org/abs/2005.03071}{{\ttfamily 2005.03071}}].

\bibitem{Chiodaroli:2021eug}
M.~Chiodaroli, H.~Johansson and P.~Pichini, \emph{{Compton Black-Hole
  Scattering for $s \leq 5/2$}},
  \href{https://arxiv.org/abs/2107.14779}{{\ttfamily 2107.14779}}.

\bibitem{Aoude:2020onz}
R.~Aoude, K.~Haddad and A.~Helset, \emph{{On-shell heavy particle effective
  theories}}, \href{https://doi.org/10.1007/JHEP05(2020)051}{\emph{JHEP}
  {\bfseries 05} (2020) 051}
  [\href{https://arxiv.org/abs/2001.09164}{{\ttfamily 2001.09164}}].

\bibitem{Jakobsen:2021zvh}
G.~U. Jakobsen, G.~Mogull, J.~Plefka and J.~Steinhoff, \emph{{SUSY in the sky
  with gravitons}}, \href{https://doi.org/10.1007/JHEP01(2022)027}{\emph{JHEP}
  {\bfseries 01} (2022) 027}
  [\href{https://arxiv.org/abs/2109.04465}{{\ttfamily 2109.04465}}].

\bibitem{Vines:2017hyw}
J.~Vines, \emph{{Scattering of two spinning black holes in post-Minkowskian
  gravity, to all orders in spin, and effective-one-body mappings}},
  \href{https://doi.org/10.1088/1361-6382/aaa3a8}{\emph{Class. Quant. Grav.}
  {\bfseries 35} (2018) 084002}
  [\href{https://arxiv.org/abs/1709.06016}{{\ttfamily 1709.06016}}].

\bibitem{Guevara:2019fsj}
A.~Guevara, A.~Ochirov and J.~Vines, \emph{{Black-hole scattering with general
  spin directions from minimal-coupling amplitudes}},
  \href{https://doi.org/10.1103/PhysRevD.100.104024}{\emph{Phys. Rev. D}
  {\bfseries 100} (2019) 104024}
  [\href{https://arxiv.org/abs/1906.10071}{{\ttfamily 1906.10071}}].

\bibitem{Levi:2015msa}
M.~Levi and J.~Steinhoff, \emph{{Spinning gravitating objects in the effective
  field theory in the post-Newtonian scheme}},
  \href{https://doi.org/10.1007/JHEP09(2015)219}{\emph{JHEP} {\bfseries 09}
  (2015) 219} [\href{https://arxiv.org/abs/1501.04956}{{\ttfamily
  1501.04956}}].

\bibitem{Guevara:2018wpp}
A.~Guevara, A.~Ochirov and J.~Vines, \emph{{Scattering of Spinning Black Holes
  from Exponentiated Soft Factors}},
  \href{https://doi.org/10.1007/JHEP09(2019)056}{\emph{JHEP} {\bfseries 09}
  (2019) 056} [\href{https://arxiv.org/abs/1812.06895}{{\ttfamily
  1812.06895}}].

\bibitem{Kosmopoulos:2021zoq}
D.~Kosmopoulos and A.~Luna, \emph{{Quadratic-in-spin Hamiltonian at $
  \mathcal{O} $(G$^{2}$) from scattering amplitudes}},
  \href{https://doi.org/10.1007/JHEP07(2021)037}{\emph{JHEP} {\bfseries 07}
  (2021) 037} [\href{https://arxiv.org/abs/2102.10137}{{\ttfamily
  2102.10137}}].

\bibitem{Chung:2019duq}
M.-Z. Chung, Y.-T. Huang and J.-W. Kim, \emph{{Classical potential for general
  spinning bodies}}, \href{https://doi.org/10.1007/JHEP09(2020)074}{\emph{JHEP}
  {\bfseries 09} (2020) 074}
  [\href{https://arxiv.org/abs/1908.08463}{{\ttfamily 1908.08463}}].

\bibitem{Chung:2020rrz}
M.-Z. Chung, Y.-t. Huang, J.-W. Kim and S.~Lee, \emph{{Complete Hamiltonian for
  spinning binary systems at first post-Minkowskian order}},
  \href{https://doi.org/10.1007/JHEP05(2020)105}{\emph{JHEP} {\bfseries 05}
  (2020) 105} [\href{https://arxiv.org/abs/2003.06600}{{\ttfamily
  2003.06600}}].

\bibitem{Chen:2021qkk}
W.-M. Chen, M.-Z. Chung, Y.-t. Huang and J.-W. Kim, \emph{{The 2PM Hamiltonian
  for binary Kerr to quartic in spin}},
  \href{https://arxiv.org/abs/2111.13639}{{\ttfamily 2111.13639}}.

\bibitem{Jakobsen:2022fcj}
G.~U. Jakobsen and G.~Mogull, \emph{{Conservative and radiative dynamics of
  spinning bodies at third post-Minkowskian order using worldline quantum field
  theory}},  \href{https://arxiv.org/abs/2201.07778}{{\ttfamily 2201.07778}}.

\bibitem{Jakobsen:2021smu}
G.~U. Jakobsen, G.~Mogull, J.~Plefka and J.~Steinhoff, \emph{{Classical
  Gravitational Bremsstrahlung from a Worldline Quantum Field Theory}},
  \href{https://doi.org/10.1103/PhysRevLett.126.201103}{\emph{Phys. Rev. Lett.}
  {\bfseries 126} (2021) 201103}
  [\href{https://arxiv.org/abs/2101.12688}{{\ttfamily 2101.12688}}].

\bibitem{spin5}
Z.~Bern, D.~Kosmopoulos, A.~Luna, R.~Roiban and F.~Teng, \emph{{Binary Dynamics
  Through the Fifth Power of Spin at $\mathcal{O}(G^2)$}},
  \href{https://arxiv.org/abs/2203.06202}{{\ttfamily 2203.06202}}.

\bibitem{spin5competition}
R.~Aoude, K.~Haddad and A.~Helset, \emph{{Searching for Kerr in the 2PM
  amplitude}},  \href{https://arxiv.org/abs/2203.06197}{{\ttfamily
  2203.06197}}.

\bibitem{Luke:1992cs}
M.~E. Luke and A.~V. Manohar, \emph{{Reparametrization invariance constraints
  on heavy particle effective field theories}},
  \href{https://doi.org/10.1016/0370-2693(92)91786-9}{\emph{Phys. Lett. B}
  {\bfseries 286} (1992) 348}
  [\href{https://arxiv.org/abs/hep-ph/9205228}{{\ttfamily hep-ph/9205228}}].

\bibitem{Heinonen:2012km}
J.~Heinonen, R.~J. Hill and M.~P. Solon, \emph{{Lorentz invariance in heavy
  particle effective theories}},
  \href{https://doi.org/10.1103/PhysRevD.86.094020}{\emph{Phys. Rev. D}
  {\bfseries 86} (2012) 094020}
  [\href{https://arxiv.org/abs/1208.0601}{{\ttfamily 1208.0601}}].

\bibitem{delaCruz:2020bbn}
L.~de~la Cruz, B.~Maybee, D.~O'Connell and A.~Ross, \emph{{Classical Yang-Mills
  observables from amplitudes}},
  \href{https://doi.org/10.1007/JHEP12(2020)076}{\emph{JHEP} {\bfseries 12}
  (2020) 076} [\href{https://arxiv.org/abs/2009.03842}{{\ttfamily
  2009.03842}}].

\bibitem{delaCruz:2021gjp}
L.~de~la Cruz, A.~Luna and T.~Scheopner, \emph{{Yang-Mills observables: from
  KMOC to eikonal through EFT}},
  \href{https://doi.org/10.1007/JHEP01(2022)045}{\emph{JHEP} {\bfseries 01}
  (2022) 045} [\href{https://arxiv.org/abs/2108.02178}{{\ttfamily
  2108.02178}}].

\bibitem{LIGOScientific:2020aai}
{\scshape LIGO Scientific, Virgo} collaboration, \emph{{GW190425: Observation
  of a Compact Binary Coalescence with Total Mass $\sim 3.4 M_{\odot}$}},
  \href{https://doi.org/10.3847/2041-8213/ab75f5}{\emph{Astrophys. J. Lett.}
  {\bfseries 892} (2020) L3}
  [\href{https://arxiv.org/abs/2001.01761}{{\ttfamily 2001.01761}}].

\bibitem{Bini:2020flp}
D.~Bini, T.~Damour and A.~Geralico, \emph{{Scattering of tidally interacting
  bodies in post-Minkowskian gravity}},
  \href{https://doi.org/10.1103/PhysRevD.101.044039}{\emph{Phys. Rev. D}
  {\bfseries 101} (2020) 044039}
  [\href{https://arxiv.org/abs/2001.00352}{{\ttfamily 2001.00352}}].

\bibitem{Cheung:2020sdj}
C.~Cheung and M.~P. Solon, \emph{{Tidal Effects in the Post-Minkowskian
  Expansion}},
  \href{https://doi.org/10.1103/PhysRevLett.125.191601}{\emph{Phys. Rev. Lett.}
  {\bfseries 125} (2020) 191601}
  [\href{https://arxiv.org/abs/2006.06665}{{\ttfamily 2006.06665}}].

\bibitem{Kalin:2020lmz}
G.~K\"alin, Z.~Liu and R.~A. Porto, \emph{{Conservative Tidal Effects in
  Compact Binary Systems to Next-to-Leading Post-Minkowskian Order}},
  \href{https://doi.org/10.1103/PhysRevD.102.124025}{\emph{Phys. Rev. D}
  {\bfseries 102} (2020) 124025}
  [\href{https://arxiv.org/abs/2008.06047}{{\ttfamily 2008.06047}}].

\bibitem{Haddad:2020que}
K.~Haddad and A.~Helset, \emph{{Tidal effects in quantum field theory}},
  \href{https://doi.org/10.1007/JHEP12(2020)024}{\emph{JHEP} {\bfseries 12}
  (2020) 024} [\href{https://arxiv.org/abs/2008.04920}{{\ttfamily
  2008.04920}}].

\bibitem{Aoude:2020ygw}
R.~Aoude, K.~Haddad and A.~Helset, \emph{{Tidal effects for spinning
  particles}}, \href{https://doi.org/10.1007/JHEP03(2021)097}{\emph{JHEP}
  {\bfseries 03} (2021) 097}
  [\href{https://arxiv.org/abs/2012.05256}{{\ttfamily 2012.05256}}].

\bibitem{Bern:2020uwk}
Z.~Bern, J.~Parra-Martinez, R.~Roiban, E.~Sawyer and C.-H. Shen, \emph{{Leading
  Nonlinear Tidal Effects and Scattering Amplitudes}},
  \href{https://doi.org/10.1007/JHEP05(2021)188}{\emph{JHEP} {\bfseries 05}
  (2021) 188} [\href{https://arxiv.org/abs/2010.08559}{{\ttfamily
  2010.08559}}].

\bibitem{Cheung:2020gbf}
C.~Cheung, N.~Shah and M.~P. Solon, \emph{{Mining the Geodesic Equation for
  Scattering Data}},
  \href{https://doi.org/10.1103/PhysRevD.103.024030}{\emph{Phys. Rev. D}
  {\bfseries 103} (2021) 024030}
  [\href{https://arxiv.org/abs/2010.08568}{{\ttfamily 2010.08568}}].

\bibitem{Chia:2020yla}
H.~S. Chia, \emph{{Tidal deformation and dissipation of rotating black holes}},
  \href{https://doi.org/10.1103/PhysRevD.104.024013}{\emph{Phys. Rev. D}
  {\bfseries 104} (2021) 024013}
  [\href{https://arxiv.org/abs/2010.07300}{{\ttfamily 2010.07300}}].

\bibitem{Hui:2020xxx}
L.~Hui, A.~Joyce, R.~Penco, L.~Santoni and A.~R. Solomon, \emph{{Static
  response and Love numbers of Schwarzschild black holes}},
  \href{https://doi.org/10.1088/1475-7516/2021/04/052}{\emph{JCAP} {\bfseries
  04} (2021) 052} [\href{https://arxiv.org/abs/2010.00593}{{\ttfamily
  2010.00593}}].

\bibitem{Charalambous:2021mea}
P.~Charalambous, S.~Dubovsky and M.~M. Ivanov, \emph{{On the Vanishing of Love
  Numbers for Kerr Black Holes}},
  \href{https://doi.org/10.1007/JHEP05(2021)038}{\emph{JHEP} {\bfseries 05}
  (2021) 038} [\href{https://arxiv.org/abs/2102.08917}{{\ttfamily
  2102.08917}}].

\bibitem{Charalambous:2021kcz}
P.~Charalambous, S.~Dubovsky and M.~M. Ivanov, \emph{{Hidden Symmetry of
  Vanishing Love Numbers}},
  \href{https://doi.org/10.1103/PhysRevLett.127.101101}{\emph{Phys. Rev. Lett.}
  {\bfseries 127} (2021) 101101}
  [\href{https://arxiv.org/abs/2103.01234}{{\ttfamily 2103.01234}}].

\bibitem{Hui:2021vcv}
L.~Hui, A.~Joyce, R.~Penco, L.~Santoni and A.~R. Solomon, \emph{{Ladder
  symmetries of black holes. Implications for love numbers and no-hair
  theorems}}, \href{https://doi.org/10.1088/1475-7516/2022/01/032}{\emph{JCAP}
  {\bfseries 01} (2022) 032}
  [\href{https://arxiv.org/abs/2105.01069}{{\ttfamily 2105.01069}}].

\bibitem{Emond:2019crr}
W.~T. Emond and N.~Moynihan, \emph{{Scattering Amplitudes, Black Holes and
  Leading Singularities in Cubic Theories of Gravity}},
  \href{https://doi.org/10.1007/JHEP12(2019)019}{\emph{JHEP} {\bfseries 12}
  (2019) 019} [\href{https://arxiv.org/abs/1905.08213}{{\ttfamily
  1905.08213}}].

\bibitem{Cristofoli:2019ewu}
A.~Cristofoli, \emph{{Post-Minkowskian Hamiltonians in Modified Theories of
  Gravity}}, \href{https://doi.org/10.1016/j.physletb.2019.135095}{\emph{Phys.
  Lett. B} {\bfseries 800} (2020) 135095}
  [\href{https://arxiv.org/abs/1906.05209}{{\ttfamily 1906.05209}}].

\bibitem{AccettulliHuber:2019jqo}
M.~Accettulli~Huber, A.~Brandhuber, S.~De~Angelis and G.~Travaglini,
  \emph{{Note on the absence of $R^2$ corrections to Newton\textquoteright{}s
  potential}}, \href{https://doi.org/10.1103/PhysRevD.101.046011}{\emph{Phys.
  Rev. D} {\bfseries 101} (2020) 046011}
  [\href{https://arxiv.org/abs/1911.10108}{{\ttfamily 1911.10108}}].

\bibitem{AccettulliHuber:2020dal}
M.~Accettulli~Huber, A.~Brandhuber, S.~De~Angelis and G.~Travaglini,
  \emph{{From amplitudes to gravitational radiation with cubic interactions and
  tidal effects}},
  \href{https://doi.org/10.1103/PhysRevD.103.045015}{\emph{Phys. Rev. D}
  {\bfseries 103} (2021) 045015}
  [\href{https://arxiv.org/abs/2012.06548}{{\ttfamily 2012.06548}}].

\bibitem{AccettulliHuber:2020oou}
M.~Accettulli~Huber, A.~Brandhuber, S.~De~Angelis and G.~Travaglini,
  \emph{{Eikonal phase matrix, deflection angle and time delay in effective
  field theories of gravity}},
  \href{https://doi.org/10.1103/PhysRevD.102.046014}{\emph{Phys. Rev. D}
  {\bfseries 102} (2020) 046014}
  [\href{https://arxiv.org/abs/2006.02375}{{\ttfamily 2006.02375}}].

\bibitem{Carrillo-Gonzalez:2021mqj}
M.~Carrillo-Gonz\'alez, C.~de~Rham and A.~J. Tolley, \emph{{Scattering
  amplitudes for binary systems beyond GR}},
  \href{https://doi.org/10.1007/JHEP11(2021)087}{\emph{JHEP} {\bfseries 11}
  (2021) 087} [\href{https://arxiv.org/abs/2107.11384}{{\ttfamily
  2107.11384}}].

\bibitem{Riva:2021vnj}
M.~M. Riva and F.~Vernizzi, \emph{{Radiated momentum in the post-Minkowskian
  worldline approach via reverse unitarity}},
  \href{https://doi.org/10.1007/JHEP11(2021)228}{\emph{JHEP} {\bfseries 11}
  (2021) 228} [\href{https://arxiv.org/abs/2110.10140}{{\ttfamily
  2110.10140}}].

\bibitem{Gralla:2021qaf}
S.~E. Gralla and K.~Lobo, \emph{{Self-force effects in post-Minkowskian
  scattering}},  \href{https://arxiv.org/abs/2110.08681}{{\ttfamily
  2110.08681}}.

\bibitem{Jakobsen:2021lvp}
G.~U. Jakobsen, G.~Mogull, J.~Plefka and J.~Steinhoff, \emph{{Gravitational
  Bremsstrahlung and Hidden Supersymmetry of Spinning Bodies}},
  \href{https://doi.org/10.1103/PhysRevLett.128.011101}{\emph{Phys. Rev. Lett.}
  {\bfseries 128} (2022) 011101}
  [\href{https://arxiv.org/abs/2106.10256}{{\ttfamily 2106.10256}}].

\bibitem{Mougiakakos:2021ckm}
S.~Mougiakakos, M.~M. Riva and F.~Vernizzi, \emph{{Gravitational Bremsstrahlung
  in the post-Minkowskian effective field theory}},
  \href{https://doi.org/10.1103/PhysRevD.104.024041}{\emph{Phys. Rev. D}
  {\bfseries 104} (2021) 024041}
  [\href{https://arxiv.org/abs/2102.08339}{{\ttfamily 2102.08339}}].

\bibitem{Manohar:2022dea}
A.~V. Manohar, A.~K. Ridgway and C.-H. Shen, \emph{{Radiated Angular Momentum
  and Dissipative Effects in Classical Scattering}},
  \href{https://arxiv.org/abs/2203.04283}{{\ttfamily 2203.04283}}.

\bibitem{Bini:2021qvf}
D.~Bini and A.~Geralico, \emph{{Higher-order tail contributions to the energy
  and angular momentum fluxes in a two-body scattering process}},
  \href{https://doi.org/10.1103/PhysRevD.104.104020}{\emph{Phys. Rev. D}
  {\bfseries 104} (2021) 104020}
  [\href{https://arxiv.org/abs/2108.05445}{{\ttfamily 2108.05445}}].

\bibitem{Edison:2022cdu}
A.~Edison and M.~Levi, \emph{{A tale of tails through generalized unitarity}},
  \href{https://arxiv.org/abs/2202.04674}{{\ttfamily 2202.04674}}.

\bibitem{Goldberger:2019sya}
W.~D. Goldberger and I.~Z. Rothstein, \emph{{An Effective Field Theory of
  Quantum Mechanical Black Hole Horizons}},
  \href{https://doi.org/10.1007/JHEP04(2020)056}{\emph{JHEP} {\bfseries 04}
  (2020) 056} [\href{https://arxiv.org/abs/1912.13435}{{\ttfamily
  1912.13435}}].

\bibitem{Bloch:1937pw}
F.~Bloch and A.~Nordsieck, \emph{{Note on the Radiation Field of the
  electron}}, \href{https://doi.org/10.1103/PhysRev.52.54}{\emph{Phys. Rev.}
  {\bfseries 52} (1937) 54}.

\bibitem{Kinoshita:1962ur}
T.~Kinoshita, \emph{{Mass singularities of Feynman amplitudes}},
  \href{https://doi.org/10.1063/1.1724268}{\emph{J. Math. Phys.} {\bfseries 3}
  (1962) 650}.

\bibitem{Lee:1964is}
T.~D. Lee and M.~Nauenberg, \emph{{Degenerate Systems and Mass Singularities}},
  \href{https://doi.org/10.1103/PhysRev.133.B1549}{\emph{Phys. Rev.} {\bfseries
  133} (1964) B1549}.

\bibitem{Akhoury:2011kq}
R.~Akhoury, R.~Saotome and G.~Sterman, \emph{{Collinear and Soft Divergences in
  Perturbative Quantum Gravity}},
  \href{https://doi.org/10.1103/PhysRevD.84.104040}{\emph{Phys. Rev. D}
  {\bfseries 84} (2011) 104040}
  [\href{https://arxiv.org/abs/1109.0270}{{\ttfamily 1109.0270}}].

\bibitem{Henn:2012qz}
J.~M. Henn and T.~Huber, \emph{{Systematics of the cusp anomalous dimension}},
  \href{https://doi.org/10.1007/JHEP11(2012)058}{\emph{JHEP} {\bfseries 11}
  (2012) 058} [\href{https://arxiv.org/abs/1207.2161}{{\ttfamily 1207.2161}}].

\bibitem{Manohar:2000dt}
A.~V. Manohar and M.~B. Wise, \emph{{Heavy quark physics}}, vol.~10. 2000.

\bibitem{Levy:1969cr}
M.~Levy and J.~Sucher, \emph{{Eikonal approximation in quantum field theory}},
  \href{https://doi.org/10.1103/PhysRev.186.1656}{\emph{Phys. Rev.} {\bfseries
  186} (1969) 1656}.

\bibitem{Abarbanel:1969ek}
H.~D.~I. Abarbanel and C.~Itzykson, \emph{{Relativistic eikonal expansion}},
  \href{https://doi.org/10.1103/PhysRevLett.23.53}{\emph{Phys. Rev. Lett.}
  {\bfseries 23} (1969) 53}.

\bibitem{Cheng:1969eh}
H.~Cheng and T.~T. Wu, \emph{{High-energy elastic scattering in quantum
  electrodynamics}},
  \href{https://doi.org/10.1103/PhysRevLett.22.666}{\emph{Phys. Rev. Lett.}
  {\bfseries 22} (1969) 666}.

\bibitem{Amati:1990xe}
D.~Amati, M.~Ciafaloni and G.~Veneziano, \emph{{Higher Order Gravitational
  Deflection and Soft Bremsstrahlung in Planckian Energy Superstring
  Collisions}}, \href{https://doi.org/10.1016/0550-3213(90)90375-N}{\emph{Nucl.
  Phys. B} {\bfseries 347} (1990) 550}.

\bibitem{Kabat:1992tb}
D.~N. Kabat and M.~Ortiz, \emph{{Eikonal quantum gravity and Planckian
  scattering}}, \href{https://doi.org/10.1016/0550-3213(92)90627-N}{\emph{Nucl.
  Phys. B} {\bfseries 388} (1992) 570}
  [\href{https://arxiv.org/abs/hep-th/9203082}{{\ttfamily hep-th/9203082}}].

\bibitem{Saotome:2012vy}
R.~Saotome and R.~Akhoury, \emph{{Relationship Between Gravity and Gauge
  Scattering in the High Energy Limit}},
  \href{https://doi.org/10.1007/JHEP01(2013)123}{\emph{JHEP} {\bfseries 01}
  (2013) 123} [\href{https://arxiv.org/abs/1210.8111}{{\ttfamily 1210.8111}}].

\bibitem{Akhoury:2013yua}
R.~Akhoury, R.~Saotome and G.~Sterman, \emph{{High Energy Scattering in
  Perturbative Quantum Gravity at Next to Leading Power}},
  \href{https://doi.org/10.1103/PhysRevD.103.064036}{\emph{Phys. Rev. D}
  {\bfseries 103} (2021) 064036}
  [\href{https://arxiv.org/abs/1308.5204}{{\ttfamily 1308.5204}}].

\bibitem{Bellini:1992eb}
A.~Bellini, M.~Ademollo and M.~Ciafaloni, \emph{{Superstring one loop and
  gravitino contributions to Planckian scattering}},
  \href{https://doi.org/10.1016/0550-3213(93)90238-K}{\emph{Nucl. Phys. B}
  {\bfseries 393} (1993) 79}
  [\href{https://arxiv.org/abs/hep-th/9207113}{{\ttfamily hep-th/9207113}}].

\bibitem{Bern:2020gjj}
Z.~Bern, H.~Ita, J.~Parra-Martinez and M.~S. Ruf, \emph{{Universality in the
  classical limit of massless gravitational scattering}},
  \href{https://doi.org/10.1103/PhysRevLett.125.031601}{\emph{Phys. Rev. Lett.}
  {\bfseries 125} (2020) 031601}
  [\href{https://arxiv.org/abs/2002.02459}{{\ttfamily 2002.02459}}].

\bibitem{Bjerrum-Bohr:2021vuf}
N.~E.~J. Bjerrum-Bohr, P.~H. Damgaard, L.~Plant\'e and P.~Vanhove,
  \emph{{Classical gravity from loop amplitudes}},
  \href{https://doi.org/10.1103/PhysRevD.104.026009}{\emph{Phys. Rev. D}
  {\bfseries 104} (2021) 026009}
  [\href{https://arxiv.org/abs/2104.04510}{{\ttfamily 2104.04510}}].

\bibitem{Amati:2007ak}
D.~Amati, M.~Ciafaloni and G.~Veneziano, \emph{{Towards an S-matrix description
  of gravitational collapse}},
  \href{https://doi.org/10.1088/1126-6708/2008/02/049}{\emph{JHEP} {\bfseries
  02} (2008) 049} [\href{https://arxiv.org/abs/0712.1209}{{\ttfamily
  0712.1209}}].

\bibitem{Haddad:2021znf}
K.~Haddad, \emph{{Exponentiation of the leading eikonal phase with spin}},
  \href{https://doi.org/10.1103/PhysRevD.105.026004}{\emph{Phys. Rev. D}
  {\bfseries 105} (2022) 026004}
  [\href{https://arxiv.org/abs/2109.04427}{{\ttfamily 2109.04427}}].

\bibitem{Kol:2021jjc}
U.~Kol, D.~O'connell and O.~Telem, \emph{{The Radial Action from Probe
  Amplitudes to All Orders}},
  \href{https://arxiv.org/abs/2109.12092}{{\ttfamily 2109.12092}}.

\bibitem{Bjerrum-Bohr:2019kec}
N.~E.~J. Bjerrum-Bohr, A.~Cristofoli and P.~H. Damgaard,
  \emph{{Post-Minkowskian Scattering Angle in Einstein Gravity}},
  \href{https://doi.org/10.1007/JHEP08(2020)038}{\emph{JHEP} {\bfseries 08}
  (2020) 038} [\href{https://arxiv.org/abs/1910.09366}{{\ttfamily
  1910.09366}}].

\bibitem{Newman:1965tw}
E.~T. Newman and A.~I. Janis, \emph{{Note on the Kerr spinning particle
  metric}}, \href{https://doi.org/10.1063/1.1704350}{\emph{J. Math. Phys.}
  {\bfseries 6} (1965) 915}.

\bibitem{Arkani-Hamed:2019ymq}
N.~Arkani-Hamed, Y.-t. Huang and D.~O'Connell, \emph{{Kerr black holes as
  elementary particles}},
  \href{https://doi.org/10.1007/JHEP01(2020)046}{\emph{JHEP} {\bfseries 01}
  (2020) 046} [\href{https://arxiv.org/abs/1906.10100}{{\ttfamily
  1906.10100}}].

\bibitem{Guevara:2020xjx}
A.~Guevara, B.~Maybee, A.~Ochirov, D.~O'connell and J.~Vines, \emph{{A
  worldsheet for Kerr}},
  \href{https://doi.org/10.1007/JHEP03(2021)201}{\emph{JHEP} {\bfseries 03}
  (2021) 201} [\href{https://arxiv.org/abs/2012.11570}{{\ttfamily
  2012.11570}}].

\bibitem{Barack:2010tm}
L.~Barack and N.~Sago, \emph{{Gravitational self-force on a particle in
  eccentric orbit around a Schwarzschild black hole}},
  \href{https://doi.org/10.1103/PhysRevD.81.084021}{\emph{Phys. Rev. D}
  {\bfseries 81} (2010) 084021}
  [\href{https://arxiv.org/abs/1002.2386}{{\ttfamily 1002.2386}}].

\bibitem{vandeMeent:2017bcc}
M.~van~de Meent, \emph{{Gravitational self-force on generic bound geodesics in
  Kerr spacetime}},
  \href{https://doi.org/10.1103/PhysRevD.97.104033}{\emph{Phys. Rev. D}
  {\bfseries 97} (2018) 104033}
  [\href{https://arxiv.org/abs/1711.09607}{{\ttfamily 1711.09607}}].

\bibitem{Pound:2019lzj}
A.~Pound, B.~Wardell, N.~Warburton and J.~Miller, \emph{{Second-Order
  Self-Force Calculation of Gravitational Binding Energy in Compact Binaries}},
  \href{https://doi.org/10.1103/PhysRevLett.124.021101}{\emph{Phys. Rev. Lett.}
  {\bfseries 124} (2020) 021101}
  [\href{https://arxiv.org/abs/1908.07419}{{\ttfamily 1908.07419}}].

\bibitem{BDG1}
D.~Bini, T.~Damour and A.~Geralico, \emph{{Novel approach to binary dynamics:
  application to the fifth post-Newtonian level}},
  \href{https://doi.org/10.1103/PhysRevLett.123.231104}{\emph{Phys. Rev. Lett.}
  {\bfseries 123} (2019) 231104}
  [\href{https://arxiv.org/abs/1909.02375}{{\ttfamily 1909.02375}}].

\bibitem{BDG2}
D.~Bini, T.~Damour and A.~Geralico, \emph{{Binary dynamics at the fifth and
  fifth-and-a-half post-Newtonian orders}},
  \href{https://doi.org/10.1103/PhysRevD.102.024062}{\emph{Phys. Rev. D}
  {\bfseries 102} (2020) 024062}
  [\href{https://arxiv.org/abs/2003.11891}{{\ttfamily 2003.11891}}].

\bibitem{BDG3}
D.~Bini, T.~Damour and A.~Geralico, \emph{{Radiative contributions to
  gravitational scattering}},
  \href{https://doi.org/10.1103/PhysRevD.104.084031}{\emph{Phys. Rev. D}
  {\bfseries 104} (2021) 084031}
  [\href{https://arxiv.org/abs/2107.08896}{{\ttfamily 2107.08896}}].

\bibitem{Antonelli:2020aeb}
A.~Antonelli, C.~Kavanagh, M.~Khalil, J.~Steinhoff and J.~Vines,
  \emph{{Gravitational spin-orbit coupling through third-subleading
  post-Newtonian order: from first-order self-force to arbitrary mass ratios}},
  \href{https://doi.org/10.1103/PhysRevLett.125.011103}{\emph{Phys. Rev. Lett.}
  {\bfseries 125} (2020) 011103}
  [\href{https://arxiv.org/abs/2003.11391}{{\ttfamily 2003.11391}}].

\bibitem{Antonelli:2020ybz}
A.~Antonelli, C.~Kavanagh, M.~Khalil, J.~Steinhoff and J.~Vines,
  \emph{{Gravitational spin-orbit and aligned spin$_1$-spin$_2$ couplings
  through third-subleading post-Newtonian orders}},
  \href{https://doi.org/10.1103/PhysRevD.102.124024}{\emph{Phys. Rev. D}
  {\bfseries 102} (2020) 124024}
  [\href{https://arxiv.org/abs/2010.02018}{{\ttfamily 2010.02018}}].

\bibitem{Khalil:2021fpm}
M.~Khalil, \emph{{Gravitational spin-orbit dynamics at the fifth-and-a-half
  post-Newtonian order}},
  \href{https://doi.org/10.1103/PhysRevD.104.124015}{\emph{Phys. Rev. D}
  {\bfseries 104} (2021) 124015}
  [\href{https://arxiv.org/abs/2110.12813}{{\ttfamily 2110.12813}}].

\bibitem{Cristofoli:2021vyo}
A.~Cristofoli, R.~Gonzo, D.~A. Kosower and D.~O'Connell, \emph{{Waveforms from
  Amplitudes}},  \href{https://arxiv.org/abs/2107.10193}{{\ttfamily
  2107.10193}}.

\bibitem{Aoude:2021oqj}
R.~Aoude and A.~Ochirov, \emph{{Classical observables from coherent-spin
  amplitudes}}, \href{https://doi.org/10.1007/JHEP10(2021)008}{\emph{JHEP}
  {\bfseries 10} (2021) 008}
  [\href{https://arxiv.org/abs/2108.01649}{{\ttfamily 2108.01649}}].

\bibitem{Yaffe:1981vf}
L.~G. Yaffe, \emph{{Large n Limits as Classical Mechanics}},
  \href{https://doi.org/10.1103/RevModPhys.54.407}{\emph{Rev. Mod. Phys.}
  {\bfseries 54} (1982) 407}.

\bibitem{Ciafaloni:2018uwe}
M.~Ciafaloni, D.~Colferai and G.~Veneziano, \emph{{Infrared features of
  gravitational scattering and radiation in the eikonal approach}},
  \href{https://doi.org/10.1103/PhysRevD.99.066008}{\emph{Phys. Rev. D}
  {\bfseries 99} (2019) 066008}
  [\href{https://arxiv.org/abs/1812.08137}{{\ttfamily 1812.08137}}].

\bibitem{Strominger:2014pwa}
A.~Strominger and A.~Zhiboedov, \emph{{Gravitational Memory, BMS
  Supertranslations and Soft Theorems}},
  \href{https://doi.org/10.1007/JHEP01(2016)086}{\emph{JHEP} {\bfseries 01}
  (2016) 086} [\href{https://arxiv.org/abs/1411.5745}{{\ttfamily 1411.5745}}].

\bibitem{Bautista:2019tdr}
Y.~F. Bautista and A.~Guevara, \emph{{From Scattering Amplitudes to Classical
  Physics: Universality, Double Copy and Soft Theorems}},
  \href{https://arxiv.org/abs/1903.12419}{{\ttfamily 1903.12419}}.

\bibitem{Laddha:2018rle}
A.~Laddha and A.~Sen, \emph{{Gravity Waves from Soft Theorem in General
  Dimensions}}, \href{https://doi.org/10.1007/JHEP09(2018)105}{\emph{JHEP}
  {\bfseries 09} (2018) 105}
  [\href{https://arxiv.org/abs/1801.07719}{{\ttfamily 1801.07719}}].

\bibitem{Laddha:2018myi}
A.~Laddha and A.~Sen, \emph{{Logarithmic Terms in the Soft Expansion in Four
  Dimensions}}, \href{https://doi.org/10.1007/JHEP10(2018)056}{\emph{JHEP}
  {\bfseries 10} (2018) 056}
  [\href{https://arxiv.org/abs/1804.09193}{{\ttfamily 1804.09193}}].

\bibitem{Sahoo:2018lxl}
B.~Sahoo and A.~Sen, \emph{{Classical and Quantum Results on Logarithmic Terms
  in the Soft Theorem in Four Dimensions}},
  \href{https://doi.org/10.1007/JHEP02(2019)086}{\emph{JHEP} {\bfseries 02}
  (2019) 086} [\href{https://arxiv.org/abs/1808.03288}{{\ttfamily
  1808.03288}}].

\bibitem{Manu:2020zxl}
A.~Manu, D.~Ghosh, A.~Laddha and P.~V. Athira, \emph{{Soft radiation from
  scattering amplitudes revisited}},
  \href{https://doi.org/10.1007/JHEP05(2021)056}{\emph{JHEP} {\bfseries 05}
  (2021) 056} [\href{https://arxiv.org/abs/2007.02077}{{\ttfamily
  2007.02077}}].

\bibitem{Bautista:2021llr}
Y.~F. Bautista and A.~Laddha, \emph{{Soft Constraints on KMOC Formalism}},
  \href{https://arxiv.org/abs/2111.11642}{{\ttfamily 2111.11642}}.

\bibitem{Caron-Huot:2018ape}
S.~Caron-Huot and Z.~Zahraee, \emph{{Integrability of Black Hole Orbits in
  Maximal Supergravity}},
  \href{https://doi.org/10.1007/JHEP07(2019)179}{\emph{JHEP} {\bfseries 07}
  (2019) 179} [\href{https://arxiv.org/abs/1810.04694}{{\ttfamily
  1810.04694}}].

\bibitem{Bern:2021xze}
Z.~Bern, J.~P. Gatica, E.~Herrmann, A.~Luna and M.~Zeng, \emph{{Scalar QED as a
  toy model for higher-order effects in classical gravitational scattering}},
  \href{https://arxiv.org/abs/2112.12243}{{\ttfamily 2112.12243}}.

\bibitem{Saketh:2021sri}
M.~V.~S. Saketh, J.~Vines, J.~Steinhoff and A.~Buonanno, \emph{{Conservative
  and radiative dynamics in classical relativistic scattering and bound
  systems}},  \href{https://arxiv.org/abs/2109.05994}{{\ttfamily 2109.05994}}.

\bibitem{Buonanno:2000qq}
A.~Buonanno, \emph{{Reduction of the two-body dynamics to a one-body
  description in classical electrodynamics}},
  \href{https://doi.org/10.1103/PhysRevD.62.104022}{\emph{Phys. Rev. D}
  {\bfseries 62} (2000) 104022}
  [\href{https://arxiv.org/abs/hep-th/0004042}{{\ttfamily hep-th/0004042}}].

\bibitem{Caswell:1985ui}
W.~E. Caswell and G.~P. Lepage, \emph{{Effective Lagrangians for Bound State
  Problems in QED, QCD, and Other Field Theories}},
  \href{https://doi.org/10.1016/0370-2693(86)91297-9}{\emph{Phys. Lett. B}
  {\bfseries 167} (1986) 437}.

\bibitem{Isgur:1989vq}
N.~Isgur and M.~B. Wise, \emph{{Weak Decays of Heavy Mesons in the Static Quark
  Approximation}},
  \href{https://doi.org/10.1016/0370-2693(89)90566-2}{\emph{Phys. Lett. B}
  {\bfseries 232} (1989) 113}.

\bibitem{Bodwin:1994jh}
G.~T. Bodwin, E.~Braaten and G.~P. Lepage, \emph{{Rigorous QCD analysis of
  inclusive annihilation and production of heavy quarkonium}},
  \href{https://doi.org/10.1103/PhysRevD.55.5853}{\emph{Phys. Rev. D}
  {\bfseries 51} (1995) 1125}
  [\href{https://arxiv.org/abs/hep-ph/9407339}{{\ttfamily hep-ph/9407339}}].

\bibitem{Beneke:1997zp}
M.~Beneke and V.~A. Smirnov, \emph{{Asymptotic expansion of Feynman integrals
  near threshold}},
  \href{https://doi.org/10.1016/S0550-3213(98)00138-2}{\emph{Nucl. Phys. B}
  {\bfseries 522} (1998) 321}
  [\href{https://arxiv.org/abs/hep-ph/9711391}{{\ttfamily hep-ph/9711391}}].

\bibitem{Manohar:2000hj}
A.~V. Manohar and I.~W. Stewart, \emph{{The QCD heavy quark potential to order
  v**2: One loop matching conditions}},
  \href{https://doi.org/10.1103/PhysRevD.62.074015}{\emph{Phys. Rev. D}
  {\bfseries 62} (2000) 074015}
  [\href{https://arxiv.org/abs/hep-ph/0003032}{{\ttfamily hep-ph/0003032}}].

\bibitem{Manohar:2006nz}
A.~V. Manohar and I.~W. Stewart, \emph{{The Zero-Bin and Mode Factorization in
  Quantum Field Theory}},
  \href{https://doi.org/10.1103/PhysRevD.76.074002}{\emph{Phys. Rev. D}
  {\bfseries 76} (2007) 074002}
  [\href{https://arxiv.org/abs/hep-ph/0605001}{{\ttfamily hep-ph/0605001}}].

\bibitem{Neill:2013wsa}
D.~Neill and I.~Z. Rothstein, \emph{{Classical Space-Times from the S Matrix}},
  \href{https://doi.org/10.1016/j.nuclphysb.2013.09.007}{\emph{Nucl. Phys. B}
  {\bfseries 877} (2013) 177}
  [\href{https://arxiv.org/abs/1304.7263}{{\ttfamily 1304.7263}}].

\bibitem{Vaidya:2014kza}
V.~Vaidya, \emph{{Gravitational spin Hamiltonians from the S matrix}},
  \href{https://doi.org/10.1103/PhysRevD.91.024017}{\emph{Phys. Rev. D}
  {\bfseries 91} (2015) 024017}
  [\href{https://arxiv.org/abs/1410.5348}{{\ttfamily 1410.5348}}].

\bibitem{Cristofoli:2020uzm}
A.~Cristofoli, P.~H. Damgaard, P.~Di~Vecchia and C.~Heissenberg,
  \emph{{Second-order Post-Minkowskian scattering in arbitrary dimensions}},
  \href{https://doi.org/10.1007/JHEP07(2020)122}{\emph{JHEP} {\bfseries 07}
  (2020) 122} [\href{https://arxiv.org/abs/2003.10274}{{\ttfamily
  2003.10274}}].

\bibitem{Anastasiou:2004vj}
C.~Anastasiou and A.~Lazopoulos, \emph{{Automatic integral reduction for higher
  order perturbative calculations}},
  \href{https://doi.org/10.1088/1126-6708/2004/07/046}{\emph{JHEP} {\bfseries
  07} (2004) 046} [\href{https://arxiv.org/abs/hep-ph/0404258}{{\ttfamily
  hep-ph/0404258}}].

\bibitem{Smirnov:2013dia}
A.~V. Smirnov and V.~A. Smirnov, \emph{{FIRE4, LiteRed and accompanying tools
  to solve integration by parts relations}},
  \href{https://doi.org/10.1016/j.cpc.2013.06.016}{\emph{Comput. Phys. Commun.}
  {\bfseries 184} (2013) 2820}
  [\href{https://arxiv.org/abs/1302.5885}{{\ttfamily 1302.5885}}].

\bibitem{Maierhoefer:2017hyi}
P.~Maierhoefer, J.~Usovitsch and P.~Uwer, \emph{{Kira - A Feynman Integral
  Reduction Program}},  \href{https://arxiv.org/abs/1705.05610}{{\ttfamily
  1705.05610}}.

\bibitem{Studerus:2009ye}
C.~Studerus, \emph{{Reduze-Feynman Integral Reduction in C++}},
  \href{https://doi.org/10.1016/j.cpc.2010.03.012}{\emph{Comput. Phys. Commun.}
  {\bfseries 181} (2010) 1293}
  [\href{https://arxiv.org/abs/0912.2546}{{\ttfamily 0912.2546}}].

\bibitem{Lee:2012cn}
R.~N. Lee, \emph{{Presenting LiteRed: a tool for the Loop InTEgrals
  REDuction}},  \href{https://arxiv.org/abs/1212.2685}{{\ttfamily 1212.2685}}.

\bibitem{Lee:2013mka}
R.~N. Lee, \emph{{LiteRed 1.4: a powerful tool for reduction of multiloop
  integrals}}, \href{https://doi.org/10.1088/1742-6596/523/1/012059}{\emph{J.
  Phys. Conf. Ser.} {\bfseries 523} (2014) 012059}
  [\href{https://arxiv.org/abs/1310.1145}{{\ttfamily 1310.1145}}].

\bibitem{Landshoff:1969yyn}
P.~V. Landshoff and J.~C. Polkinghorne, \emph{{Iterations of regge cuts}},
  \href{https://doi.org/10.1103/PhysRev.181.1989}{\emph{Phys. Rev.} {\bfseries
  181} (1969) 1989}.

\bibitem{Bini:2020uiq}
D.~Bini, T.~Damour, A.~Geralico, S.~Laporta and P.~Mastrolia,
  \emph{{Gravitational dynamics at $O(G^6)$: perturbative gravitational
  scattering meets experimental mathematics}},
  \href{https://arxiv.org/abs/2008.09389}{{\ttfamily 2008.09389}}.

\bibitem{Bini:2020rzn}
D.~Bini, T.~Damour, A.~Geralico, S.~Laporta and P.~Mastrolia,
  \emph{{Gravitational scattering at the seventh order in $G$: nonlocal
  contribution at the sixth post-Newtonian accuracy}},
  \href{https://doi.org/10.1103/PhysRevD.103.044038}{\emph{Phys. Rev. D}
  {\bfseries 103} (2021) 044038}
  [\href{https://arxiv.org/abs/2012.12918}{{\ttfamily 2012.12918}}].

\bibitem{Bjerrum-Bohr:2021wwt}
N.~E.~J. Bjerrum-Bohr, L.~Plante and P.~Vanhove, \emph{{Post-Minkowskian Radial
  Action from Soft Limits and Velocity Cuts}},
  \href{https://arxiv.org/abs/2111.02976}{{\ttfamily 2111.02976}}.

\bibitem{Basham:1978bw}
C.~L. Basham, L.~S. Brown, S.~D. Ellis and S.~T. Love, \emph{{Energy
  Correlations in electron - Positron Annihilation: Testing QCD}},
  \href{https://doi.org/10.1103/PhysRevLett.41.1585}{\emph{Phys. Rev. Lett.}
  {\bfseries 41} (1978) 1585}.

\bibitem{Basham:1978zq}
C.~L. Basham, L.~S. Brown, S.~D. Ellis and S.~T. Love, \emph{{Energy
  Correlations in electron-Positron Annihilation in Quantum Chromodynamics:
  Asymptotically Free Perturbation Theory}},
  \href{https://doi.org/10.1103/PhysRevD.19.2018}{\emph{Phys. Rev. D}
  {\bfseries 19} (1979) 2018}.

\bibitem{Richards:1983sr}
D.~G. Richards, W.~J. Stirling and S.~D. Ellis, \emph{{Energy-energy
  Correlations to Second Order in Quantum Chromodynamics}},
  \href{https://doi.org/10.1016/0550-3213(83)90335-8}{\emph{Nucl. Phys. B}
  {\bfseries 229} (1983) 317}.

\bibitem{Korchemsky:1999kt}
G.~P. Korchemsky and G.~F. Sterman, \emph{{Power corrections to event shapes
  and factorization}},
  \href{https://doi.org/10.1016/S0550-3213(99)00308-9}{\emph{Nucl. Phys. B}
  {\bfseries 555} (1999) 335}
  [\href{https://arxiv.org/abs/hep-ph/9902341}{{\ttfamily hep-ph/9902341}}].

\bibitem{Dokshitzer:1999sh}
Y.~L. Dokshitzer, G.~Marchesini and B.~R. Webber, \emph{{Nonperturbative
  effects in the energy energy correlation}},
  \href{https://doi.org/10.1088/1126-6708/1999/07/012}{\emph{JHEP} {\bfseries
  07} (1999) 012} [\href{https://arxiv.org/abs/hep-ph/9905339}{{\ttfamily
  hep-ph/9905339}}].

\bibitem{Lee:2006fn}
C.~Lee and G.~F. Sterman, \emph{{Universality of nonperturbative effects in
  event shapes}}, {\emph{eConf} {\bfseries C0601121} (2006) A001}
  [\href{https://arxiv.org/abs/hep-ph/0603066}{{\ttfamily hep-ph/0603066}}].

\bibitem{Bauer:2008dt}
C.~W. Bauer, S.~P. Fleming, C.~Lee and G.~F. Sterman, \emph{{Factorization of
  e+e- Event Shape Distributions with Hadronic Final States in Soft Collinear
  Effective Theory}},
  \href{https://doi.org/10.1103/PhysRevD.78.034027}{\emph{Phys. Rev. D}
  {\bfseries 78} (2008) 034027}
  [\href{https://arxiv.org/abs/0801.4569}{{\ttfamily 0801.4569}}].

\bibitem{Belitsky:2001ij}
A.~V. Belitsky, G.~P. Korchemsky and G.~F. Sterman, \emph{{Energy flow in QCD
  and event shape functions}},
  \href{https://doi.org/10.1016/S0370-2693(01)00899-1}{\emph{Phys. Lett. B}
  {\bfseries 515} (2001) 297}
  [\href{https://arxiv.org/abs/hep-ph/0106308}{{\ttfamily hep-ph/0106308}}].

\bibitem{Maybee:2019jus}
B.~Maybee, D.~O'Connell and J.~Vines, \emph{{Observables and amplitudes for
  spinning particles and black holes}},
  \href{https://doi.org/10.1007/JHEP12(2019)156}{\emph{JHEP} {\bfseries 12}
  (2019) 156} [\href{https://arxiv.org/abs/1906.09260}{{\ttfamily
  1906.09260}}].

\bibitem{Damgaard:2019lfh}
P.~H. Damgaard, K.~Haddad and A.~Helset, \emph{{Heavy Black Hole Effective
  Theory}}, \href{https://doi.org/10.1007/JHEP11(2019)070}{\emph{JHEP}
  {\bfseries 11} (2019) 070}
  [\href{https://arxiv.org/abs/1908.10308}{{\ttfamily 1908.10308}}].

\bibitem{Brandhuber:2021kpo}
A.~Brandhuber, G.~Chen, G.~Travaglini and C.~Wen, \emph{{A new gauge-invariant
  double copy for heavy-mass effective theory}},
  \href{https://doi.org/10.1007/JHEP07(2021)047}{\emph{JHEP} {\bfseries 07}
  (2021) 047} [\href{https://arxiv.org/abs/2104.11206}{{\ttfamily
  2104.11206}}].

\bibitem{Brandhuber:2021bsf}
A.~Brandhuber, G.~Chen, H.~Johansson, G.~Travaglini and C.~Wen,
  \emph{{Kinematic Hopf Algebra for BCJ Numerators in Heavy-Mass Effective
  Field Theory and Yang-Mills Theory}},
  \href{https://arxiv.org/abs/2111.15649}{{\ttfamily 2111.15649}}.

\bibitem{Chen:2019ywi}
G.~Chen, H.~Johansson, F.~Teng and T.~Wang, \emph{{On the kinematic algebra for
  BCJ numerators beyond the MHV sector}},
  \href{https://doi.org/10.1007/JHEP11(2019)055}{\emph{JHEP} {\bfseries 11}
  (2019) 055} [\href{https://arxiv.org/abs/1906.10683}{{\ttfamily
  1906.10683}}].

\bibitem{Chen:2021chy}
G.~Chen, H.~Johansson, F.~Teng and T.~Wang, \emph{{Next-to-MHV Yang-Mills
  kinematic algebra}},
  \href{https://doi.org/10.1007/JHEP10(2021)042}{\emph{JHEP} {\bfseries 10}
  (2021) 042} [\href{https://arxiv.org/abs/2104.12726}{{\ttfamily
  2104.12726}}].

\bibitem{Carrasco:2020ywq}
J.~J.~M. Carrasco and I.~A. Vazquez-Holm, \emph{{Loop-Level Double-Copy for
  Massive Quantum Particles}},
  \href{https://doi.org/10.1103/PhysRevD.103.045002}{\emph{Phys. Rev. D}
  {\bfseries 103} (2021) 045002}
  [\href{https://arxiv.org/abs/2010.13435}{{\ttfamily 2010.13435}}].

\bibitem{Johansson:2014zca}
H.~Johansson and A.~Ochirov, \emph{{Pure Gravities via Color-Kinematics Duality
  for Fundamental Matter}},
  \href{https://doi.org/10.1007/JHEP11(2015)046}{\emph{JHEP} {\bfseries 11}
  (2015) 046} [\href{https://arxiv.org/abs/1407.4772}{{\ttfamily 1407.4772}}].

\bibitem{Luna:2017dtq}
A.~Luna, I.~Nicholson, D.~O'Connell and C.~D. White, \emph{{Inelastic Black
  Hole Scattering from Charged Scalar Amplitudes}},
  \href{https://doi.org/10.1007/JHEP03(2018)044}{\emph{JHEP} {\bfseries 03}
  (2018) 044} [\href{https://arxiv.org/abs/1711.03901}{{\ttfamily
  1711.03901}}].

\bibitem{Haddad:2020tvs}
K.~Haddad and A.~Helset, \emph{{The double copy for heavy particles}},
  \href{https://doi.org/10.1103/PhysRevLett.125.181603}{\emph{Phys. Rev. Lett.}
  {\bfseries 125} (2020) 181603}
  [\href{https://arxiv.org/abs/2005.13897}{{\ttfamily 2005.13897}}].

\bibitem{Bjerrum-Bohr:2020syg}
N.~E.~J. Bjerrum-Bohr, T.~V. Brown and H.~Gomez, \emph{{Scattering of Gravitons
  and Spinning Massive States from Compact Numerators}},
  \href{https://doi.org/10.1007/JHEP04(2021)234}{\emph{JHEP} {\bfseries 04}
  (2021) 234} [\href{https://arxiv.org/abs/2011.10556}{{\ttfamily
  2011.10556}}].

\bibitem{Bjerrum-Bohr:2019nws}
N.~E.~J. Bjerrum-Bohr, A.~Cristofoli, P.~H. Damgaard and H.~Gomez,
  \emph{{Scalar-Graviton Amplitudes}},
  \href{https://doi.org/10.1007/JHEP11(2019)148}{\emph{JHEP} {\bfseries 11}
  (2019) 148} [\href{https://arxiv.org/abs/1908.09755}{{\ttfamily
  1908.09755}}].

\bibitem{Foffa:2019yfl}
S.~Foffa, R.~A. Porto, I.~Rothstein and R.~Sturani, \emph{{Conservative
  dynamics of binary systems to fourth Post-Newtonian order in the EFT approach
  II: Renormalized Lagrangian}},
  \href{https://doi.org/10.1103/PhysRevD.100.024048}{\emph{Phys. Rev. D}
  {\bfseries 100} (2019) 024048}
  [\href{https://arxiv.org/abs/1903.05118}{{\ttfamily 1903.05118}}].

\bibitem{Foffa:2019rdf}
S.~Foffa and R.~Sturani, \emph{{Conservative dynamics of binary systems to
  fourth Post-Newtonian order in the EFT approach I: Regularized Lagrangian}},
  \href{https://doi.org/10.1103/PhysRevD.100.024047}{\emph{Phys. Rev. D}
  {\bfseries 100} (2019) 024047}
  [\href{https://arxiv.org/abs/1903.05113}{{\ttfamily 1903.05113}}].

\bibitem{Blumlein:2019zku}
J.~Bl\"umlein, A.~Maier and P.~Marquard, \emph{{Five-Loop Static Contribution
  to the Gravitational Interaction Potential of Two Point Masses}},
  \href{https://doi.org/10.1016/j.physletb.2019.135100}{\emph{Phys. Lett. B}
  {\bfseries 800} (2020) 135100}
  [\href{https://arxiv.org/abs/1902.11180}{{\ttfamily 1902.11180}}].

\bibitem{Foffa:2019hrb}
S.~Foffa, P.~Mastrolia, R.~Sturani, C.~Sturm and W.~J. Torres~Bobadilla,
  \emph{{Static two-body potential at fifth post-Newtonian order}},
  \href{https://doi.org/10.1103/PhysRevLett.122.241605}{\emph{Phys. Rev. Lett.}
  {\bfseries 122} (2019) 241605}
  [\href{https://arxiv.org/abs/1902.10571}{{\ttfamily 1902.10571}}].

\bibitem{Foffa:2011ub}
S.~Foffa and R.~Sturani, \emph{{Effective field theory calculation of
  conservative binary dynamics at third post-Newtonian order}},
  \href{https://doi.org/10.1103/PhysRevD.84.044031}{\emph{Phys. Rev. D}
  {\bfseries 84} (2011) 044031}
  [\href{https://arxiv.org/abs/1104.1122}{{\ttfamily 1104.1122}}].

\bibitem{Gilmore:2008gq}
J.~B. Gilmore and A.~Ross, \emph{{Effective field theory calculation of second
  post-Newtonian binary dynamics}},
  \href{https://doi.org/10.1103/PhysRevD.78.124021}{\emph{Phys. Rev. D}
  {\bfseries 78} (2008) 124021}
  [\href{https://arxiv.org/abs/0810.1328}{{\ttfamily 0810.1328}}].

\bibitem{Kol:2007bc}
B.~Kol and M.~Smolkin, \emph{{Non-Relativistic Gravitation: From Newton to
  Einstein and Back}},
  \href{https://doi.org/10.1088/0264-9381/25/14/145011}{\emph{Class. Quant.
  Grav.} {\bfseries 25} (2008) 145011}
  [\href{https://arxiv.org/abs/0712.4116}{{\ttfamily 0712.4116}}].

\bibitem{Mogull:2020sak}
G.~Mogull, J.~Plefka and J.~Steinhoff, \emph{{Classical black hole scattering
  from a worldline quantum field theory}},
  \href{https://doi.org/10.1007/JHEP02(2021)048}{\emph{JHEP} {\bfseries 02}
  (2021) 048} [\href{https://arxiv.org/abs/2010.02865}{{\ttfamily
  2010.02865}}].

\bibitem{Cho:2021mqw}
G.~Cho, B.~Pardo and R.~A. Porto, \emph{{Gravitational radiation from
  inspiralling compact objects: Spin-spin effects completed at the
  next-to-leading post-Newtonian order}},
  \href{https://doi.org/10.1103/PhysRevD.104.024037}{\emph{Phys. Rev. D}
  {\bfseries 104} (2021) 024037}
  [\href{https://arxiv.org/abs/2103.14612}{{\ttfamily 2103.14612}}].

\bibitem{Vines:2018gqi}
J.~Vines, J.~Steinhoff and A.~Buonanno, \emph{{Spinning-black-hole scattering
  and the test-black-hole limit at second post-Minkowskian order}},
  \href{https://doi.org/10.1103/PhysRevD.99.064054}{\emph{Phys. Rev. D}
  {\bfseries 99} (2019) 064054}
  [\href{https://arxiv.org/abs/1812.00956}{{\ttfamily 1812.00956}}].

\bibitem{Foffa:2011np}
S.~Foffa and R.~Sturani, \emph{{Tail terms in gravitational radiation reaction
  via effective field theory}},
  \href{https://doi.org/10.1103/PhysRevD.87.044056}{\emph{Phys. Rev. D}
  {\bfseries 87} (2013) 044056}
  [\href{https://arxiv.org/abs/1111.5488}{{\ttfamily 1111.5488}}].

\bibitem{Goldberger:2016iau}
W.~D. Goldberger and A.~K. Ridgway, \emph{{Radiation and the classical double
  copy for color charges}},
  \href{https://doi.org/10.1103/PhysRevD.95.125010}{\emph{Phys. Rev. D}
  {\bfseries 95} (2017) 125010}
  [\href{https://arxiv.org/abs/1611.03493}{{\ttfamily 1611.03493}}].

\bibitem{Goldberger:2017vcg}
W.~D. Goldberger and A.~K. Ridgway, \emph{{Bound states and the classical
  double copy}}, \href{https://doi.org/10.1103/PhysRevD.97.085019}{\emph{Phys.
  Rev. D} {\bfseries 97} (2018) 085019}
  [\href{https://arxiv.org/abs/1711.09493}{{\ttfamily 1711.09493}}].

\bibitem{Shen:2018ebu}
C.-H. Shen, \emph{{Gravitational Radiation from Color-Kinematics Duality}},
  \href{https://doi.org/10.1007/JHEP11(2018)162}{\emph{JHEP} {\bfseries 11}
  (2018) 162} [\href{https://arxiv.org/abs/1806.07388}{{\ttfamily
  1806.07388}}].

\bibitem{KAGRA:2013rdx}
{\scshape KAGRA, LIGO Scientific, Virgo, VIRGO} collaboration, \emph{{Prospects
  for observing and localizing gravitational-wave transients with Advanced
  LIGO, Advanced Virgo and KAGRA}},
  \href{https://doi.org/10.1007/s41114-020-00026-9}{\emph{Living Rev. Rel.}
  {\bfseries 21} (2018) 3} [\href{https://arxiv.org/abs/1304.0670}{{\ttfamily
  1304.0670}}].

\bibitem{Hinderer:2016eia}
T.~Hinderer, A.~Taracchini, F.~Foucart, A.~Buonanno, J.~Steinhoff et~al.,
  \emph{{Effects of neutron-star dynamic tides on gravitational waveforms
  within the effective-one-body approach}},
  \href{https://doi.org/10.1103/PhysRevLett.116.181101}{\emph{Phys. Rev. Lett.}
  {\bfseries 116} (2016) 181101}
  [\href{https://arxiv.org/abs/1602.00599}{{\ttfamily 1602.00599}}].

\bibitem{Dietrich:2017aum}
T.~Dietrich, S.~Bernuzzi and W.~Tichy, \emph{{Closed-form tidal approximants
  for binary neutron star gravitational waveforms constructed from
  high-resolution numerical relativity simulations}},
  \href{https://doi.org/10.1103/PhysRevD.96.121501}{\emph{Phys. Rev. D}
  {\bfseries 96} (2017) 121501}
  [\href{https://arxiv.org/abs/1706.02969}{{\ttfamily 1706.02969}}].

\bibitem{Nagar:2018plt}
A.~Nagar, F.~Messina, P.~Rettegno, D.~Bini, T.~Damour, A.~Geralico et~al.,
  \emph{{Nonlinear-in-spin effects in effective-one-body waveform models of
  spin-aligned, inspiralling, neutron star binaries}},
  \href{https://doi.org/10.1103/PhysRevD.99.044007}{\emph{Phys. Rev. D}
  {\bfseries 99} (2019) 044007}
  [\href{https://arxiv.org/abs/1812.07923}{{\ttfamily 1812.07923}}].

\bibitem{Nagar:2018zoe}
A.~Nagar et~al., \emph{{Time-domain effective-one-body gravitational waveforms
  for coalescing compact binaries with nonprecessing spins, tides and self-spin
  effects}}, \href{https://doi.org/10.1103/PhysRevD.98.104052}{\emph{Phys. Rev.
  D} {\bfseries 98} (2018) 104052}
  [\href{https://arxiv.org/abs/1806.01772}{{\ttfamily 1806.01772}}].

\bibitem{Cotesta:2018fcv}
R.~Cotesta, A.~Buonanno, A.~Boh\'e, A.~Taracchini, I.~Hinder and S.~Ossokine,
  \emph{{Enriching the Symphony of Gravitational Waves from Binary Black Holes
  by Tuning Higher Harmonics}},
  \href{https://doi.org/10.1103/PhysRevD.98.084028}{\emph{Phys. Rev. D}
  {\bfseries 98} (2018) 084028}
  [\href{https://arxiv.org/abs/1803.10701}{{\ttfamily 1803.10701}}].

\bibitem{Varma:2018mmi}
V.~Varma, S.~E. Field, M.~A. Scheel, J.~Blackman, L.~E. Kidder and H.~P.
  Pfeiffer, \emph{{Surrogate model of hybridized numerical relativity binary
  black hole waveforms}},
  \href{https://doi.org/10.1103/PhysRevD.99.064045}{\emph{Phys. Rev. D}
  {\bfseries 99} (2019) 064045}
  [\href{https://arxiv.org/abs/1812.07865}{{\ttfamily 1812.07865}}].

\bibitem{Dietrich:2019kaq}
T.~Dietrich, A.~Samajdar, S.~Khan, N.~K. Johnson-McDaniel, R.~Dudi and
  W.~Tichy, \emph{{Improving the NRTidal model for binary neutron star
  systems}}, \href{https://doi.org/10.1103/PhysRevD.100.044003}{\emph{Phys.
  Rev. D} {\bfseries 100} (2019) 044003}
  [\href{https://arxiv.org/abs/1905.06011}{{\ttfamily 1905.06011}}].

\bibitem{Varma:2019csw}
V.~Varma, S.~E. Field, M.~A. Scheel, J.~Blackman, D.~Gerosa, L.~C. Stein
  et~al., \emph{{Surrogate models for precessing binary black hole simulations
  with unequal masses}},
  \href{https://doi.org/10.1103/PhysRevResearch.1.033015}{\emph{Phys. Rev.
  Research.} {\bfseries 1} (2019) 033015}
  [\href{https://arxiv.org/abs/1905.09300}{{\ttfamily 1905.09300}}].

\bibitem{Ossokine:2020kjp}
S.~Ossokine, A.~Buonanno, S.~Marsat, R.~Cotesta, S.~Babak et~al.,
  \emph{{Multipolar Effective-One-Body Waveforms for Precessing Binary Black
  Holes: Construction and Validation}},
  \href{https://doi.org/10.1103/PhysRevD.102.044055}{\emph{Phys. Rev. D}
  {\bfseries 102} (2020) 044055}
  [\href{https://arxiv.org/abs/2004.09442}{{\ttfamily 2004.09442}}].

\bibitem{Pratten:2020ceb}
G.~Pratten et~al., \emph{{Computationally efficient models for the dominant and
  subdominant harmonic modes of precessing binary black holes}},
  \href{https://doi.org/10.1103/PhysRevD.103.104056}{\emph{Phys. Rev. D}
  {\bfseries 103} (2021) 104056}
  [\href{https://arxiv.org/abs/2004.06503}{{\ttfamily 2004.06503}}].

\bibitem{Thompson:2020nei}
J.~E. Thompson, E.~Fauchon-Jones, S.~Khan, E.~Nitoglia, F.~Pannarale,
  T.~Dietrich et~al., \emph{{Modeling the gravitational wave signature of
  neutron star black hole coalescences}},
  \href{https://doi.org/10.1103/PhysRevD.101.124059}{\emph{Phys. Rev. D}
  {\bfseries 101} (2020) 124059}
  [\href{https://arxiv.org/abs/2002.08383}{{\ttfamily 2002.08383}}].

\bibitem{Matas:2020wab}
A.~Matas, T.~Dietrich, A.~Buonanno, T.~Hinderer, M.~Puerrer et~al.,
  \emph{{Aligned-spin neutron-star\textendash{}black-hole waveform model based
  on the effective-one-body approach and numerical-relativity simulations}},
  \href{https://doi.org/10.1103/PhysRevD.102.043023}{\emph{Phys. Rev. D}
  {\bfseries 102} (2020) 043023}
  [\href{https://arxiv.org/abs/2004.10001}{{\ttfamily 2004.10001}}].

\bibitem{Estelles:2021gvs}
H.~Estell\'es, M.~Colleoni, C.~Garc\'\i{}a-Quir\'os, S.~Husa, D.~Keitel,
  M.~Mateu-Lucena et~al., \emph{{New twists in compact binary waveform
  modelling: a fast time domain model for precession}},
  \href{https://arxiv.org/abs/2105.05872}{{\ttfamily 2105.05872}}.

\bibitem{Gamba:2021ydi}
R.~Gamba, S.~Ak\c{c}ay, S.~Bernuzzi and J.~Williams, \emph{{Effective-one-body
  waveforms for precessing coalescing compact binaries with post-Newtonian
  Twist}},  \href{https://arxiv.org/abs/2111.03675}{{\ttfamily 2111.03675}}.

\bibitem{Laporta:1996mq}
S.~Laporta and E.~Remiddi, \emph{{The Analytical value of the electron $(g-2)$
  at order $\alpha^3$ in QED}},
  \href{https://doi.org/10.1016/0370-2693(96)00439-X}{\emph{Phys. Lett.}
  {\bfseries B379} (1996) 283}
  [\href{https://arxiv.org/abs/hep-ph/9602417}{{\ttfamily hep-ph/9602417}}].

\bibitem{Gluza:2010ws}
J.~Gluza, K.~Kajda and D.~A. Kosower, \emph{{Towards a Basis for Planar
  Two-Loop Integrals}},
  \href{https://doi.org/10.1103/PhysRevD.83.045012}{\emph{Phys. Rev. D}
  {\bfseries 83} (2011) 045012}
  [\href{https://arxiv.org/abs/1009.0472}{{\ttfamily 1009.0472}}].

\bibitem{Schabinger:2011dz}
R.~M. Schabinger, \emph{{A New Algorithm For The Generation Of
  Unitarity-Compatible Integration By Parts Relations}},
  \href{https://doi.org/10.1007/JHEP01(2012)077}{\emph{JHEP} {\bfseries 01}
  (2012) 077} [\href{https://arxiv.org/abs/1111.4220}{{\ttfamily 1111.4220}}].

\bibitem{Ita:2015tya}
H.~Ita, \emph{{Two-loop Integrand Decomposition into Master Integrals and
  Surface Terms}},
  \href{https://doi.org/10.1103/PhysRevD.94.116015}{\emph{Phys. Rev. D}
  {\bfseries 94} (2016) 116015}
  [\href{https://arxiv.org/abs/1510.05626}{{\ttfamily 1510.05626}}].

\bibitem{Georgoudis:2016wff}
A.~Georgoudis, K.~J. Larsen and Y.~Zhang, \emph{{Azurite: An algebraic geometry
  based package for finding bases of loop integrals}},
  \href{https://doi.org/10.1016/j.cpc.2017.08.013}{\emph{Comput. Phys. Commun.}
  {\bfseries 221} (2017) 203}
  [\href{https://arxiv.org/abs/1612.04252}{{\ttfamily 1612.04252}}].

\bibitem{Ita:2016oar}
H.~Ita, \emph{{Towards a Numerical Unitarity Approach for Two-loop Amplitudes
  in QCD}}, {\emph{PoS} {\bfseries LL2016} (2016) 080}
  [\href{https://arxiv.org/abs/1607.00705}{{\ttfamily 1607.00705}}].

\bibitem{Zhang:2016kfo}
Y.~Zhang, \emph{{Lecture Notes on Multi-loop Integral Reduction and Applied
  Algebraic Geometry}},  2016,
  \href{https://arxiv.org/abs/1612.02249}{{\ttfamily 1612.02249}},
  \href{https://inspirehep.net/record/1502112/files/arXiv:1612.02249.pdf}{https://inspirehep.net/record/1502112/files/arXiv:1612.02249.pdf}.

\bibitem{Bern:2017gdk}
Z.~Bern, M.~Enciso, H.~Ita and M.~Zeng, \emph{{Dual Conformal Symmetry,
  Integration-by-Parts Reduction, Differential Equations and the Nonplanar
  Sector}}, \href{https://doi.org/10.1103/PhysRevD.96.096017}{\emph{Phys. Rev.
  D} {\bfseries 96} (2017) 096017}
  [\href{https://arxiv.org/abs/1709.06055}{{\ttfamily 1709.06055}}].

\bibitem{Mastrolia:2018uzb}
P.~Mastrolia and S.~Mizera, \emph{{Feynman Integrals and Intersection Theory}},
  \href{https://doi.org/10.1007/JHEP02(2019)139}{\emph{JHEP} {\bfseries 02}
  (2019) 139} [\href{https://arxiv.org/abs/1810.03818}{{\ttfamily
  1810.03818}}].

\bibitem{Frellesvig:2019uqt}
H.~Frellesvig, F.~Gasparotto, M.~K. Mandal, P.~Mastrolia, L.~Mattiazzi and
  S.~Mizera, \emph{{Vector Space of Feynman Integrals and Multivariate
  Intersection Numbers}},
  \href{https://doi.org/10.1103/PhysRevLett.123.201602}{\emph{Phys. Rev. Lett.}
  {\bfseries 123} (2019) 201602}
  [\href{https://arxiv.org/abs/1907.02000}{{\ttfamily 1907.02000}}].

\bibitem{Lee:2014ioa}
R.~N. Lee, \emph{{Reducing differential equations for multiloop master
  integrals}}, \href{https://doi.org/10.1007/JHEP04(2015)108}{\emph{JHEP}
  {\bfseries 04} (2015) 108} [\href{https://arxiv.org/abs/1411.0911}{{\ttfamily
  1411.0911}}].

\bibitem{Argeri:2014qva}
M.~Argeri, S.~Di~Vita, P.~Mastrolia, E.~Mirabella, J.~Schlenk, U.~Schubert
  et~al., \emph{{Magnus and Dyson Series for Master Integrals}},
  \href{https://doi.org/10.1007/JHEP03(2014)082}{\emph{JHEP} {\bfseries 03}
  (2014) 082} [\href{https://arxiv.org/abs/1401.2979}{{\ttfamily 1401.2979}}].

\bibitem{Henn:2014qga}
J.~M. Henn, \emph{{Lectures on differential equations for Feynman integrals}},
  \href{https://doi.org/10.1088/1751-8113/48/15/153001}{\emph{J. Phys. A}
  {\bfseries 48} (2015) 153001}
  [\href{https://arxiv.org/abs/1412.2296}{{\ttfamily 1412.2296}}].

\bibitem{Gituliar:2016vfa}
O.~Gituliar and V.~Magerya, \emph{{Fuchsia and master integrals for splitting
  functions from differential equations in QCD}},
  \href{https://doi.org/10.22323/1.260.0030}{\emph{PoS} {\bfseries LL2016}
  (2016) 030} [\href{https://arxiv.org/abs/1607.00759}{{\ttfamily
  1607.00759}}].

\bibitem{Gituliar:2017vzm}
O.~Gituliar and V.~Magerya, \emph{{Fuchsia: a tool for reducing differential
  equations for Feynman master integrals to epsilon form}},
  \href{https://doi.org/10.1016/j.cpc.2017.05.004}{\emph{Comput. Phys. Commun.}
  {\bfseries 219} (2017) 329}
  [\href{https://arxiv.org/abs/1701.04269}{{\ttfamily 1701.04269}}].

\bibitem{Prausa:2017ltv}
M.~Prausa, \emph{{epsilon: A tool to find a canonical basis of master
  integrals}}, \href{https://doi.org/10.1016/j.cpc.2017.05.026}{\emph{Comput.
  Phys. Commun.} {\bfseries 219} (2017) 361}
  [\href{https://arxiv.org/abs/1701.00725}{{\ttfamily 1701.00725}}].

\bibitem{Chicherin:2018old}
D.~Chicherin, T.~Gehrmann, J.~M. Henn, P.~Wasser, Y.~Zhang and S.~Zoia,
  \emph{{All Master Integrals for Three-Jet Production at
  Next-to-Next-to-Leading Order}},
  \href{https://doi.org/10.1103/PhysRevLett.123.041603}{\emph{Phys. Rev. Lett.}
  {\bfseries 123} (2019) 041603}
  [\href{https://arxiv.org/abs/1812.11160}{{\ttfamily 1812.11160}}].

\bibitem{Dlapa:2020cwj}
C.~Dlapa, J.~Henn and K.~Yan, \emph{{Deriving canonical differential equations
  for Feynman integrals from a single uniform weight integral}},
  \href{https://doi.org/10.1007/JHEP05(2020)025}{\emph{JHEP} {\bfseries 05}
  (2020) 025} [\href{https://arxiv.org/abs/2002.02340}{{\ttfamily
  2002.02340}}].

\bibitem{Chen:2020uyk}
J.~Chen, X.~Jiang, X.~Xu and L.~L. Yang, \emph{{Constructing canonical Feynman
  integrals with intersection theory}},
  \href{https://doi.org/10.1016/j.physletb.2021.136085}{\emph{Phys. Lett. B}
  {\bfseries 814} (2021) 136085}
  [\href{https://arxiv.org/abs/2008.03045}{{\ttfamily 2008.03045}}].

\bibitem{Dlapa:2021qsl}
C.~Dlapa, X.~Li and Y.~Zhang, \emph{{Leading singularities in Baikov
  representation and Feynman integrals with uniform transcendental weight}},
  \href{https://doi.org/10.1007/JHEP07(2021)227}{\emph{JHEP} {\bfseries 07}
  (2021) 227} [\href{https://arxiv.org/abs/2103.04638}{{\ttfamily
  2103.04638}}].

\bibitem{Brown:2009ta}
F.~C.~S. Brown, \emph{{On the periods of some Feynman integrals}},
  \href{https://arxiv.org/abs/0910.0114}{{\ttfamily 0910.0114}}.

\bibitem{Brown:2009rc}
F.~Brown and K.~Yeats, \emph{{Spanning forest polynomials and the
  transcendental weight of Feynman graphs}},
  \href{https://doi.org/10.1007/s00220-010-1145-1}{\emph{Commun. Math. Phys.}
  {\bfseries 301} (2011) 357}
  [\href{https://arxiv.org/abs/0910.5429}{{\ttfamily 0910.5429}}].

\bibitem{BrownArxiv}
F.~C.~S. Brown and A.~Levin, \emph{{Multiple Elliptic Polylogarithms}},
  \href{https://arxiv.org/abs/1110.6917}{{\ttfamily 1110.6917}}.

\bibitem{Panzer:2014caa}
E.~Panzer, \emph{{Algorithms for the symbolic integration of hyperlogarithms
  with applications to Feynman integrals}},
  \href{https://doi.org/10.1016/j.cpc.2014.10.019}{\emph{Comput. Phys. Commun.}
  {\bfseries 188} (2015) 148}
  [\href{https://arxiv.org/abs/1403.3385}{{\ttfamily 1403.3385}}].

\bibitem{Broedel:2017kkb}
J.~Broedel, C.~Duhr, F.~Dulat and L.~Tancredi, \emph{{Elliptic polylogarithms
  and iterated integrals on elliptic curves. Part I: general formalism}},
  \href{https://doi.org/10.1007/JHEP05(2018)093}{\emph{JHEP} {\bfseries 05}
  (2018) 093} [\href{https://arxiv.org/abs/1712.07089}{{\ttfamily
  1712.07089}}].

\bibitem{Bourjaily:2018yfy}
J.~L. Bourjaily, A.~J. McLeod, M.~von Hippel and M.~Wilhelm, \emph{{Bounded
  Collection of Feynman Integral Calabi-Yau Geometries}},
  \href{https://doi.org/10.1103/PhysRevLett.122.031601}{\emph{Phys. Rev. Lett.}
  {\bfseries 122} (2019) 031601}
  [\href{https://arxiv.org/abs/1810.07689}{{\ttfamily 1810.07689}}].

\bibitem{integrationWhitePaper2022}
J.~L. Bourjaily et~al., \emph{{Functions Beyond Multiple Polylogarithms for
  Precision Collider Physics}},  3, 2022,
  \href{https://arxiv.org/abs/2203.07088}{{\ttfamily 2203.07088}}.

\bibitem{Smirnov:2004ym}
V.~A. Smirnov, \emph{{Evaluating Feynman integrals}}, {\emph{Springer Tracts
  Mod. Phys.} {\bfseries 211} (2004) 1}.

\bibitem{Bourjaily:2018aeq}
J.~L. Bourjaily, A.~J. McLeod, M.~von Hippel and M.~Wilhelm,
  \emph{{Rationalizing Loop Integration}},
  \href{https://doi.org/10.1007/JHEP08(2018)184}{\emph{JHEP} {\bfseries 08}
  (2018) 184} [\href{https://arxiv.org/abs/1805.10281}{{\ttfamily
  1805.10281}}].

\bibitem{Tarasov:1996br}
O.~V. Tarasov, \emph{{Connection between Feynman integrals having different
  values of the space-time dimension}},
  \href{https://doi.org/10.1103/PhysRevD.54.6479}{\emph{Phys. Rev. D}
  {\bfseries 54} (1996) 6479}
  [\href{https://arxiv.org/abs/hep-th/9606018}{{\ttfamily hep-th/9606018}}].

\bibitem{Lee:2009dh}
R.~N. Lee, \emph{{Space-time dimensionality D as complex variable: Calculating
  loop integrals using dimensional recurrence relation and analytical
  properties with respect to D}},
  \href{https://doi.org/10.1016/j.nuclphysb.2009.12.025}{\emph{Nucl. Phys. B}
  {\bfseries 830} (2010) 474}
  [\href{https://arxiv.org/abs/0911.0252}{{\ttfamily 0911.0252}}].

\bibitem{Lee:2012te}
R.~N. Lee and V.~A. Smirnov, \emph{{The Dimensional Recurrence and Analyticity
  Method for Multicomponent Master Integrals: Using Unitarity Cuts to Construct
  Homogeneous Solutions}},
  \href{https://doi.org/10.1007/JHEP12(2012)104}{\emph{JHEP} {\bfseries 12}
  (2012) 104} [\href{https://arxiv.org/abs/1209.0339}{{\ttfamily 1209.0339}}].

\bibitem{Anastasiou:2002yz}
C.~Anastasiou and K.~Melnikov, \emph{{Higgs boson production at hadron
  colliders in NNLO QCD}},
  \href{https://doi.org/10.1016/S0550-3213(02)00837-4}{\emph{Nucl. Phys. B}
  {\bfseries 646} (2002) 220}
  [\href{https://arxiv.org/abs/hep-ph/0207004}{{\ttfamily hep-ph/0207004}}].

\bibitem{Anastasiou:2002qz}
C.~Anastasiou, L.~J. Dixon and K.~Melnikov, \emph{{NLO Higgs boson rapidity
  distributions at hadron colliders}},
  \href{https://doi.org/10.1016/S0920-5632(03)80168-8}{\emph{Nucl. Phys. B
  Proc. Suppl.} {\bfseries 116} (2003) 193}
  [\href{https://arxiv.org/abs/hep-ph/0211141}{{\ttfamily hep-ph/0211141}}].

\bibitem{Anastasiou:2003yy}
C.~Anastasiou, L.~J. Dixon, K.~Melnikov and F.~Petriello, \emph{{Dilepton
  rapidity distribution in the Drell-Yan process at NNLO in QCD}},
  \href{https://doi.org/10.1103/PhysRevLett.91.182002}{\emph{Phys. Rev. Lett.}
  {\bfseries 91} (2003) 182002}
  [\href{https://arxiv.org/abs/hep-ph/0306192}{{\ttfamily hep-ph/0306192}}].

\bibitem{Anastasiou:2015yha}
C.~Anastasiou, C.~Duhr, F.~Dulat, E.~Furlan, F.~Herzog and B.~Mistlberger,
  \emph{{Soft expansion of double-real-virtual corrections to Higgs production
  at N$^{3}$LO}}, \href{https://doi.org/10.1007/JHEP08(2015)051}{\emph{JHEP}
  {\bfseries 08} (2015) 051}
  [\href{https://arxiv.org/abs/1505.04110}{{\ttfamily 1505.04110}}].

\bibitem{Anastasiou:2003kj}
C.~Anastasiou, Z.~Bern, L.~J. Dixon and D.~A. Kosower, \emph{{Planar amplitudes
  in maximally supersymmetric Yang-Mills theory}},
  \href{https://doi.org/10.1103/PhysRevLett.91.251602}{\emph{Phys. Rev. Lett.}
  {\bfseries 91} (2003) 251602}
  [\href{https://arxiv.org/abs/hep-th/0309040}{{\ttfamily hep-th/0309040}}].

\bibitem{Bern:2005iz}
Z.~Bern, L.~J. Dixon and V.~A. Smirnov, \emph{{Iteration of planar amplitudes
  in maximally supersymmetric Yang-Mills theory at three loops and beyond}},
  \href{https://doi.org/10.1103/PhysRevD.72.085001}{\emph{Phys. Rev. D}
  {\bfseries 72} (2005) 085001}
  [\href{https://arxiv.org/abs/hep-th/0505205}{{\ttfamily hep-th/0505205}}].

\bibitem{Monteiro:2020plf}
R.~Monteiro, D.~O'Connell, D.~Peinador~Veiga and M.~Sergola, \emph{{Classical
  solutions and their double copy in split signature}},
  \href{https://doi.org/10.1007/JHEP05(2021)268}{\emph{JHEP} {\bfseries 05}
  (2021) 268} [\href{https://arxiv.org/abs/2012.11190}{{\ttfamily
  2012.11190}}].

\bibitem{Naoz:2012bx}
S.~Naoz, B.~Kocsis, A.~Loeb and N.~Yunes, \emph{{Resonant Post-Newtonian
  Eccentricity Excitation in Hierarchical Three-body Systems}},
  \href{https://doi.org/10.1088/0004-637X/773/2/187}{\emph{Astrophys. J.}
  {\bfseries 773} (2013) 187}
  [\href{https://arxiv.org/abs/1206.4316}{{\ttfamily 1206.4316}}].

\bibitem{Lim:2020cvm}
H.~Lim and C.~L. Rodriguez, \emph{{Relativistic three-body effects in
  hierarchical triples}},
  \href{https://doi.org/10.1103/PhysRevD.102.064033}{\emph{Phys. Rev. D}
  {\bfseries 102} (2020) 064033}
  [\href{https://arxiv.org/abs/2001.03654}{{\ttfamily 2001.03654}}].

\bibitem{Martinez:2020lzt}
M.~A.~S. Martinez et~al., \emph{{Black Hole Mergers from Hierarchical Triples
  in Dense Star Clusters}},
  \href{https://doi.org/10.3847/1538-4357/abba25}{\emph{Astrophys. J.}
  {\bfseries 903} (2020) 67}
  [\href{https://arxiv.org/abs/2009.08468}{{\ttfamily 2009.08468}}].

\bibitem{Fragione:2020gly}
G.~Fragione, M.~A.~S. Martinez, K.~Kremer, S.~Chatterjee, C.~L. Rodriguez,
  C.~S. Ye et~al., \emph{{Demographics of triple systems in dense star
  clusters}}, \href{https://doi.org/10.3847/1538-4357/aba89b}{\emph{Astrophys.
  J.} {\bfseries 900} (2020) 16}
  [\href{https://arxiv.org/abs/2007.11605}{{\ttfamily 2007.11605}}].

\bibitem{Boekholt:2021ifv}
T.~C.~N. Boekholt, A.~Moerman and S.~F.~P. Zwart, \emph{{Relativistic
  Pythagorean three-body problem}},
  \href{https://doi.org/10.1103/PhysRevD.104.083020}{\emph{Phys. Rev. D}
  {\bfseries 104} (2021) 083020}
  [\href{https://arxiv.org/abs/2109.07013}{{\ttfamily 2109.07013}}].

\bibitem{Zwart:2021qxe}
S.~F.~P. Zwart, T.~C.~N. Boekholt, E.~Por, A.~S. Hamers and S.~L.~W. McMillan,
  \emph{{Chaos in self-gravitating many-body systems: Lyapunov time dependence
  of $N$ and the influence of general relativity}},
  \href{https://arxiv.org/abs/2109.11012}{{\ttfamily 2109.11012}}.

\bibitem{SCHAFER1987336}
G.~Schäfer, \emph{Three-body hamiltonian in general relativity},
  \href{https://doi.org/https://doi.org/10.1016/0375-9601(87)90389-6}{\emph{Physics
  Letters A} {\bfseries 123} (1987) 336}.

\bibitem{Chu:2008xm}
Y.-Z. Chu, \emph{{The n-body problem in General Relativity up to the second
  post-Newtonian order from perturbative field theory}},
  \href{https://doi.org/10.1103/PhysRevD.79.044031}{\emph{Phys. Rev. D}
  {\bfseries 79} (2009) 044031}
  [\href{https://arxiv.org/abs/0812.0012}{{\ttfamily 0812.0012}}].

\bibitem{Will:2013cza}
C.~M. Will, \emph{{Incorporating post-Newtonian effects in $N$-body dynamics}},
  \href{https://doi.org/10.1103/PhysRevD.89.044043}{\emph{Phys. Rev. D}
  {\bfseries 89} (2014) 044043}
  [\href{https://arxiv.org/abs/1312.1289}{{\ttfamily 1312.1289}}].

\bibitem{Loebbert:2020aos}
F.~Loebbert, J.~Plefka, C.~Shi and T.~Wang, \emph{{Three-body effective
  potential in general relativity at second post-Minkowskian order and
  resulting post-Newtonian contributions}},
  \href{https://doi.org/10.1103/PhysRevD.103.064010}{\emph{Phys. Rev. D}
  {\bfseries 103} (2021) 064010}
  [\href{https://arxiv.org/abs/2012.14224}{{\ttfamily 2012.14224}}].

\bibitem{Press:1971wr}
W.~H. Press, \emph{{Long Wave Trains of Gravitational Waves from a Vibrating
  Black Hole}}, \href{https://doi.org/10.1086/180849}{\emph{Astrophys. J.
  Lett.} {\bfseries 170} (1971) L105}.

\bibitem{Kamaretsos:2011um}
I.~Kamaretsos, M.~Hannam, S.~Husa and B.~S. Sathyaprakash, \emph{{Black-hole
  hair loss: learning about binary progenitors from ringdown signals}},
  \href{https://doi.org/10.1103/PhysRevD.85.024018}{\emph{Phys. Rev. D}
  {\bfseries 85} (2012) 024018}
  [\href{https://arxiv.org/abs/1107.0854}{{\ttfamily 1107.0854}}].

\bibitem{Hughes:2019zmt}
S.~A. Hughes, A.~Apte, G.~Khanna and H.~Lim, \emph{{Learning about black hole
  binaries from their ringdown spectra}},
  \href{https://doi.org/10.1103/PhysRevLett.123.161101}{\emph{Phys. Rev. Lett.}
  {\bfseries 123} (2019) 161101}
  [\href{https://arxiv.org/abs/1901.05900}{{\ttfamily 1901.05900}}].

\bibitem{Li:2021wgz}
X.~Li, L.~Sun, R.~K.~L. Lo, E.~Payne and Y.~Chen, \emph{{Angular emission
  patterns of remnant black holes}},
  \href{https://doi.org/10.1103/PhysRevD.105.024016}{\emph{Phys. Rev. D}
  {\bfseries 105} (2022) 024016}
  [\href{https://arxiv.org/abs/2110.03116}{{\ttfamily 2110.03116}}].

\bibitem{Adamo:2017sze}
T.~Adamo, E.~Casali, L.~Mason and S.~Nekovar, \emph{{Amplitudes on plane waves
  from ambitwistor strings}},
  \href{https://doi.org/10.1007/JHEP11(2017)160}{\emph{JHEP} {\bfseries 11}
  (2017) 160} [\href{https://arxiv.org/abs/1708.09249}{{\ttfamily
  1708.09249}}].

\bibitem{Adamo:2018mpq}
T.~Adamo, E.~Casali, L.~Mason and S.~Nekovar, \emph{{Plane wave backgrounds and
  colour-kinematics duality}},
  \href{https://doi.org/10.1007/JHEP02(2019)198}{\emph{JHEP} {\bfseries 02}
  (2019) 198} [\href{https://arxiv.org/abs/1810.05115}{{\ttfamily
  1810.05115}}].

\bibitem{Adamo:2020qru}
T.~Adamo and A.~Ilderton, \emph{{Classical and quantum double copy of
  back-reaction}}, \href{https://doi.org/10.1007/JHEP09(2020)200}{\emph{JHEP}
  {\bfseries 09} (2020) 200}
  [\href{https://arxiv.org/abs/2005.05807}{{\ttfamily 2005.05807}}].

\bibitem{Adamo:2020yzi}
T.~Adamo, L.~Mason and A.~Sharma, \emph{{Gluon scattering on self-dual
  radiative gauge fields}},  \href{https://arxiv.org/abs/2010.14996}{{\ttfamily
  2010.14996}}.

\bibitem{Adamo:2021rfq}
T.~Adamo, A.~Cristofoli and P.~Tourkine, \emph{{Eikonal amplitudes from curved
  backgrounds}},  \href{https://arxiv.org/abs/2112.09113}{{\ttfamily
  2112.09113}}.

\bibitem{Diwakar:2021juk}
P.~Diwakar, A.~Herderschee, R.~Roiban and F.~Teng, \emph{{BCJ amplitude
  relations for Anti-de Sitter boundary correlators in embedding space}},
  \href{https://doi.org/10.1007/JHEP10(2021)141}{\emph{JHEP} {\bfseries 10}
  (2021) 141} [\href{https://arxiv.org/abs/2106.10822}{{\ttfamily
  2106.10822}}].

\bibitem{Cheung:2022pdk}
C.~Cheung, J.~Parra-Martinez and A.~Sivaramakrishnan, \emph{{On-shell
  Correlators and Color-Kinematics Duality in Curved Symmetric Spacetimes}},
  \href{https://arxiv.org/abs/2201.05147}{{\ttfamily 2201.05147}}.

\bibitem{Herderschee:2022ntr}
A.~Herderschee, R.~Roiban and F.~Teng, \emph{{On the Differential
  Representation and Color-Kinematics Duality of AdS Boundary Correlators}},
  \href{https://arxiv.org/abs/2201.05067}{{\ttfamily 2201.05067}}.

\bibitem{Arkani-Hamed:2018kmz}
N.~Arkani-Hamed, D.~Baumann, H.~Lee and G.~L. Pimentel, \emph{{The Cosmological
  Bootstrap: Inflationary Correlators from Symmetries and Singularities}},
  \href{https://doi.org/10.1007/JHEP04(2020)105}{\emph{JHEP} {\bfseries 04}
  (2020) 105} [\href{https://arxiv.org/abs/1811.00024}{{\ttfamily
  1811.00024}}].

\bibitem{Baumann:2020dch}
D.~Baumann, C.~Duaso~Pueyo, A.~Joyce, H.~Lee and G.~L. Pimentel, \emph{{The
  Cosmological Bootstrap: Spinning Correlators from Symmetries and
  Factorization}},
  \href{https://doi.org/10.21468/SciPostPhys.11.3.071}{\emph{SciPost Phys.}
  {\bfseries 11} (2021) 071}
  [\href{https://arxiv.org/abs/2005.04234}{{\ttfamily 2005.04234}}].

\bibitem{Benincasa:2018ssx}
P.~Benincasa, \emph{{From the flat-space S-matrix to the Wavefunction of the
  Universe}},  \href{https://arxiv.org/abs/1811.02515}{{\ttfamily 1811.02515}}.

\bibitem{Baumann:2022jpr}
D.~Baumann, D.~Green, A.~Joyce, E.~Pajer, G.~L. Pimentel, C.~Sleight et~al.,
  \emph{{Snowmass White Paper: The Cosmological Bootstrap}},  in \emph{{2022
  Snowmass Summer Study}}, 3, 2022,
  \href{https://arxiv.org/abs/2203.08121}{{\ttfamily 2203.08121}}.

\bibitem{Luna2015paa}
A.~Luna, R.~Monteiro, D.~O'Connell and C.~D. White, \emph{{The classical double
  copy for Taub--NUT spacetime}},
  \href{https://doi.org/10.1016/j.physletb.2015.09.021}{\emph{Phys. Lett.}
  {\bfseries B750} (2015) 272}
  [\href{https://arxiv.org/abs/1507.01869}{{\ttfamily 1507.01869}}].

\bibitem{Luna2016due}
A.~Luna, R.~Monteiro, I.~Nicholson, D.~O'Connell and C.~D. White, \emph{{The
  double copy: Bremsstrahlung and accelerating black holes}},
  \href{https://doi.org/10.1007/JHEP06(2016)023}{\emph{JHEP} {\bfseries 06}
  (2016) 023} [\href{https://arxiv.org/abs/1603.05737}{{\ttfamily
  1603.05737}}].

\bibitem{Luna2016hge}
A.~Luna, R.~Monteiro, I.~Nicholson, A.~Ochirov, D.~O'Connell, N.~Westerberg
  et~al., \emph{{Perturbative spacetimes from Yang-Mills theory}},
  \href{https://doi.org/10.1007/JHEP04(2017)069}{\emph{JHEP} {\bfseries 04}
  (2017) 069} [\href{https://arxiv.org/abs/1611.07508}{{\ttfamily
  1611.07508}}].

\bibitem{CarrilloGonzalez:2019gof}
M.~Carrillo~Gonz\'alez, B.~Melcher, K.~Ratliff, S.~Watson and C.~D. White,
  \emph{{The classical double copy in three spacetime dimensions}},
  \href{https://doi.org/10.1007/JHEP07(2019)167}{\emph{JHEP} {\bfseries 07}
  (2019) 167} [\href{https://arxiv.org/abs/1904.11001}{{\ttfamily
  1904.11001}}].

\bibitem{CarrilloGonzalez:2018ejf}
M.~Carrillo~Gonz\'alez, R.~Penco and M.~Trodden, \emph{{Radiation of scalar
  modes and the classical double copy}},
  \href{https://doi.org/10.1007/JHEP11(2018)065}{\emph{JHEP} {\bfseries 11}
  (2018) 065} [\href{https://arxiv.org/abs/1809.04611}{{\ttfamily
  1809.04611}}].

\bibitem{Carrillo-Gonzalez:2017iyj}
M.~Carrillo-Gonz\'alez, R.~Penco and M.~Trodden, \emph{{The classical double
  copy in maximally symmetric spacetimes}},
  \href{https://doi.org/10.1007/JHEP04(2018)028}{\emph{JHEP} {\bfseries 04}
  (2018) 028} [\href{https://arxiv.org/abs/1711.01296}{{\ttfamily
  1711.01296}}].

\bibitem{Adamo:2022rmp}
T.~Adamo, A.~Cristofoli and A.~Ilderton, \emph{{Classical physics from
  amplitudes on curved backgrounds}},
  \href{https://arxiv.org/abs/2203.13785}{{\ttfamily 2203.13785}}.

\bibitem{LIGOScientific:2021sio}
{\scshape LIGO Scientific, VIRGO, KAGRA} collaboration, \emph{{Tests of General
  Relativity with GWTC-3}},  \href{https://arxiv.org/abs/2112.06861}{{\ttfamily
  2112.06861}}.

\bibitem{Barack:2018yly}
L.~Barack et~al., \emph{{Black holes, gravitational waves and fundamental
  physics: a roadmap}},
  \href{https://doi.org/10.1088/1361-6382/ab0587}{\emph{Class. Quant. Grav.}
  {\bfseries 36} (2019) 143001}
  [\href{https://arxiv.org/abs/1806.05195}{{\ttfamily 1806.05195}}].

\bibitem{Mathur:2005zp}
S.~D. Mathur, \emph{{The Fuzzball proposal for black holes: An Elementary
  review}}, \href{https://doi.org/10.1002/prop.200410203}{\emph{Fortsch. Phys.}
  {\bfseries 53} (2005) 793}
  [\href{https://arxiv.org/abs/hep-th/0502050}{{\ttfamily hep-th/0502050}}].

\bibitem{Mathur:2008nj}
S.~D. Mathur, \emph{{Fuzzballs and the information paradox: A Summary and
  conjectures}},  \href{https://arxiv.org/abs/0810.4525}{{\ttfamily
  0810.4525}}.

\bibitem{Almheiri:2012rt}
A.~Almheiri, D.~Marolf, J.~Polchinski and J.~Sully, \emph{{Black Holes:
  Complementarity or Firewalls?}},
  \href{https://doi.org/10.1007/JHEP02(2013)062}{\emph{JHEP} {\bfseries 02}
  (2013) 062} [\href{https://arxiv.org/abs/1207.3123}{{\ttfamily 1207.3123}}].

\bibitem{Almheiri:2013hfa}
A.~Almheiri, D.~Marolf, J.~Polchinski, D.~Stanford and J.~Sully, \emph{{An
  Apologia for Firewalls}},
  \href{https://doi.org/10.1007/JHEP09(2013)018}{\emph{JHEP} {\bfseries 09}
  (2013) 018} [\href{https://arxiv.org/abs/1304.6483}{{\ttfamily 1304.6483}}].

\bibitem{Carrasco:2021ptp}
J.~J.~M. Carrasco, L.~Rodina and S.~Zekioglu, \emph{{Composing effective
  prediction at five points}},
  \href{https://doi.org/10.1007/JHEP06(2021)169}{\emph{JHEP} {\bfseries 06}
  (2021) 169} [\href{https://arxiv.org/abs/2104.08370}{{\ttfamily
  2104.08370}}].

\bibitem{Bonnefoy:2021qgu}
Q.~Bonnefoy, G.~Durieux, C.~Grojean, C.~S. Machado and J.~Roosmale~Nepveu,
  \emph{{The seeds of EFT double copy}},
  \href{https://arxiv.org/abs/2112.11453}{{\ttfamily 2112.11453}}.

\bibitem{Chi:2021mio}
H.-H. Chi, H.~Elvang, A.~Herderschee, C.~R.~T. Jones and S.~Paranjape,
  \emph{{Generalizations of the Double-Copy: the KLT Bootstrap}},
  \href{https://arxiv.org/abs/2106.12600}{{\ttfamily 2106.12600}}.

\bibitem{Endlich:2017tqa}
S.~Endlich, V.~Gorbenko, J.~Huang and L.~Senatore, \emph{{An effective
  formalism for testing extensions to General Relativity with gravitational
  waves}}, \href{https://doi.org/10.1007/JHEP09(2017)122}{\emph{JHEP}
  {\bfseries 09} (2017) 122}
  [\href{https://arxiv.org/abs/1704.01590}{{\ttfamily 1704.01590}}].

\bibitem{deRham:2020ejn}
C.~de~Rham, J.~Francfort and J.~Zhang, \emph{{Black Hole Gravitational Waves in
  the Effective Field Theory of Gravity}},
  \href{https://doi.org/10.1103/PhysRevD.102.024079}{\emph{Phys. Rev. D}
  {\bfseries 102} (2020) 024079}
  [\href{https://arxiv.org/abs/2005.13923}{{\ttfamily 2005.13923}}].

\bibitem{Cano:2020cao}
P.~A. Cano, K.~Fransen and T.~Hertog, \emph{{Ringing of rotating black holes in
  higher-derivative gravity}},
  \href{https://doi.org/10.1103/PhysRevD.102.044047}{\emph{Phys. Rev. D}
  {\bfseries 102} (2020) 044047}
  [\href{https://arxiv.org/abs/2005.03671}{{\ttfamily 2005.03671}}].

\bibitem{Cano:2021myl}
P.~A. Cano, K.~Fransen, T.~Hertog and S.~Maenaut, \emph{{Gravitational ringing
  of rotating black holes in higher-derivative gravity}},
  \href{https://doi.org/10.1103/PhysRevD.105.024064}{\emph{Phys. Rev. D}
  {\bfseries 105} (2022) 024064}
  [\href{https://arxiv.org/abs/2110.11378}{{\ttfamily 2110.11378}}].

\bibitem{deRham:2022hpx}
C.~de~Rham, S.~Kundu, M.~Reece, A.~J. Tolley and S.-Y. Zhou, \emph{{Snowmass
  White Paper: UV Constraints on IR Physics}},  in \emph{{2022 Snowmass Summer
  Study}}, 3, 2022, \href{https://arxiv.org/abs/2203.06805}{{\ttfamily
  2203.06805}}.

\end{thebibliography}\endgroup

\end{document}